\documentclass[twocolumn]{aastex631}

\usepackage{graphicx}
\usepackage{amssymb}
\usepackage{longtable}
\usepackage{rotating}

\usepackage{xspace}
\newcommand{\galario}{\texttt{galario}\xspace}
\newcommand{\frank}{\texttt{frank}\xspace}
\newcommand{\discminer}{\texttt{discminer}\xspace}

\begin{document}

\title{exoALMA IV: Substructures, Asymmetries, and the Faint Outer Disk in Continuum Emission}

\correspondingauthor{Pietro Curone}
\email{pcurone@das.uchile.cl}

\author[0000-0003-2045-2154]{Pietro Curone}
\affiliation{Dipartimento di Fisica, Universit\`a degli Studi di Milano, Via Celoria 16, 20133 Milano, Italy}
\affiliation{Departamento de Astronomía, Universidad de Chile, Camino El Observatorio 1515, Las Condes, Santiago, Chile}

\author[0000-0003-4689-2684]{Stefano Facchini}
\affiliation{Dipartimento di Fisica, Universit\`a degli Studi di Milano, Via Celoria 16, 20133 Milano, Italy}

\author[0000-0003-2253-2270]{Sean M. Andrews}
\affiliation{Center for Astrophysics | Harvard \& Smithsonian, Cambridge, MA 02138, USA}

\author[0000-0003-1859-3070]{Leonardo Testi}
\affiliation{Dipartimento di Fisica e Astronomia, Universit\`a di Bologna, I-40190 Bologna, Italy}


\author[0000-0002-7695-7605]{Myriam Benisty}
\affiliation{Universit\'{e} C\^{o}te d'Azur, Observatoire de la C\^{o}te d'Azur, CNRS, Laboratoire Lagrange, France}
\affiliation{Max-Planck Institute for Astronomy (MPIA), Königstuhl 17, 69117 Heidelberg, Germany}

\author[0000-0002-1483-8811]{Ian Czekala}
\affiliation{School of Physics \& Astronomy, University of St. Andrews, North Haugh, St. Andrews KY16 9SS, UK}

\author[0000-0001-6947-6072]{Jane Huang}
\affiliation{Department of Astronomy, Columbia University, 538 W. 120th Street, Pupin Hall, New York, NY 10027, USA}

\author[0000-0003-1008-1142]{John D. Ilee}
\affiliation{School of Physics and Astronomy, University of Leeds, Leeds, LS2 9JT, UK}

\author[0000-0001-8061-2207]{Andrea Isella}
\affiliation{Department of Physics and Astronomy, Rice University, 6100 Main St, Houston, TX 77005, USA}
\affiliation{Rice Space Institute, Rice University, 6100 Main St, Houston, TX 77005, USA}

\author[0000-0002-2357-7692]{Giuseppe Lodato}
\affiliation{Dipartimento di Fisica, Universit\`a degli Studi di Milano, Via Celoria 16, 20133 Milano, Italy}

\author[0000-0002-8932-1219]{Ryan A. Loomis}
\affiliation{National Radio Astronomy Observatory, 520 Edgemont Rd., Charlottesville, VA 22903, USA}

\author[0000-0002-0491-143X]{Jochen Stadler}
\affiliation{Universit\'{e} C\^{o}te d'Azur, Observatoire de la C\^{o}te d'Azur, CNRS, Laboratoire Lagrange, France}

\author[0000-0002-7501-9801]{Andrew J. Winter}
\affiliation{Universit\'{e} C\^{o}te d'Azur, Observatoire de la C\^{o}te d'Azur, CNRS, Laboratoire Lagrange, France}
\affiliation{Max-Planck Institute for Astronomy (MPIA), Königstuhl 17, 69117 Heidelberg, Germany}

\author[0000-0001-7258-770X]{Jaehan Bae}
\affiliation{Department of Astronomy, University of Florida, Gainesville, FL 32611, USA}

\author[0000-0001-6378-7873]{Marcelo Barraza-Alfaro}
\affiliation{Department of Earth, Atmospheric, and Planetary Sciences, Massachusetts Institute of Technology, Cambridge, MA 02139, USA}

\author[0000-0002-2700-9676]{Gianni Cataldi}
\affiliation{National Astronomical Observatory of Japan, 2-21-1 Osawa, Mitaka, Tokyo 181-8588, Japan}

\author[0000-0003-3713-8073]{Nicolás Cuello}
\affiliation{Univ. Grenoble Alpes, CNRS, IPAG, 38000 Grenoble, France}

\author[0000-0003-4679-4072]{Daniele Fasano}
\affiliation{Universit\'{e} C\^{o}te d'Azur, Observatoire de la C\^{o}te d'Azur, CNRS, Laboratoire Lagrange, France}

\author[0000-0002-9298-3029]{Mario Flock}
\affiliation{Max-Planck Institute for Astronomy (MPIA), Königstuhl 17, 69117 Heidelberg, Germany}

\author[0000-0003-1117-9213]{Misato Fukagawa}
\affiliation{National Astronomical Observatory of Japan, 2-21-1 Osawa, Mitaka, Tokyo 181-8588, Japan}

\author[0000-0002-5503-5476]{Maria Galloway-Sprietsma}
\affiliation{Department of Astronomy, University of Florida, Gainesville, FL 32611, USA}

\author[0000-0002-5910-4598]{Himanshi Garg}
\affiliation{School of Physics and Astronomy, Monash University, VIC 3800, Australia}

\author[0000-0002-8138-0425]{Cassandra Hall}
\affiliation{Department of Physics and Astronomy, The University of Georgia, Athens, GA 30602, USA}
\affiliation{Center for Simulational Physics, The University of Georgia, Athens, GA 30602, USA}
\affiliation{Institute for Artificial Intelligence, The University of Georgia, Athens, GA, 30602, USA}

\author[0000-0001-8446-3026]{Andrés F. Izquierdo}
\altaffiliation{NASA Hubble Fellowship Program Sagan Fellow}
\affiliation{Department of Astronomy, University of Florida, Gainesville, FL 32611, USA}
\affiliation{Leiden Observatory, Leiden University, P.O. Box 9513, NL-2300 RA Leiden, The Netherlands}
\affiliation{European Southern Observatory, Karl-Schwarzschild-Str. 2, D-85748 Garching bei München, Germany}

\author[0000-0001-7235-2417]{Kazuhiro Kanagawa}
\affiliation{College of Science, Ibaraki University, 2-1-1 Bunkyo, Mito, Ibaraki 310-8512, Japan}

\author[0000-0002-8896-9435]{Geoffroy Lesur}
\affiliation{Univ. Grenoble Alpes, CNRS, IPAG, 38000 Grenoble, France}

\author[0000-0003-4663-0318]{Cristiano Longarini}
\affiliation{Institute of Astronomy, University of Cambridge, Madingley Rd, CB30HA, Cambridge, UK}
\affiliation{Dipartimento di Fisica, Universit\`a degli Studi di Milano, Via Celoria 16, 20133 Milano, Italy}

\author[0000-0002-1637-7393]{Francois Menard}
\affiliation{Univ. Grenoble Alpes, CNRS, IPAG, 38000 Grenoble, France}

\author[0000-0003-4039-8933]{Ryuta Orihara}
\affiliation{College of Science, Ibaraki University, 2-1-1 Bunkyo, Mito, Ibaraki 310-8512, Japan}

\author[0000-0001-5907-5179]{Christophe Pinte}
\affiliation{Univ. Grenoble Alpes, CNRS, IPAG, 38000 Grenoble, France}
\affiliation{School of Physics and Astronomy, Monash University, VIC 3800, Australia}

\author[0000-0002-4716-4235]{Daniel J. Price}
\affiliation{School of Physics and Astronomy, Monash University, VIC 3800, Australia}

\author[0000-0003-4853-5736]{Giovanni Rosotti}
\affiliation{Dipartimento di Fisica, Universit\`a degli Studi di Milano, Via Celoria 16, 20133 Milano, Italy}

\author[0000-0003-1534-5186]{Richard Teague}
\affiliation{Department of Earth, Atmospheric, and Planetary Sciences, Massachusetts Institute of Technology, Cambridge, MA 02139, USA}

\author[0000-0002-3468-9577]{Gaylor Wafflard-Fernandez}
\affiliation{Univ. Grenoble Alpes, CNRS, IPAG, 38000 Grenoble, France}

\author[0000-0003-1526-7587]{David J. Wilner}
\affiliation{Center for Astrophysics | Harvard \& Smithsonian, Cambridge, MA 02138, USA}

\author[0000-0002-7212-2416]{Lisa Wölfer}
\affiliation{Department of Earth, Atmospheric, and Planetary Sciences, Massachusetts Institute of Technology, Cambridge, MA 02139, USA}

\author[0000-0003-1412-893X]{Hsi-Wei Yen}
\affiliation{Academia Sinica Institute of Astronomy \& Astrophysics, 11F of Astronomy-Mathematics Building, AS/NTU, No.1, Sec. 4, Roosevelt Rd, Taipei 106216, Taiwan}

\author[0000-0001-8002-8473]{Tomohiro C. Yoshida}
\affiliation{National Astronomical Observatory of Japan, 2-21-1 Osawa, Mitaka, Tokyo 181-8588, Japan}
\affiliation{Department of Astronomical Science, The Graduate University for Advanced Studies, SOKENDAI, 2-21-1 Osawa, Mitaka, Tokyo 181-8588, Japan}

\author[0000-0001-9319-1296]{Brianna Zawadzki}
\affiliation{Department of Astronomy, Van Vleck Observatory, Wesleyan University, 96 Foss Hill Drive, Middletown, CT 06459, USA}
\affiliation{Department of Astronomy \& Astrophysics, 525 Davey Laboratory, The Pennsylvania State University, University Park, PA 16802, USA}

\begin{abstract}

The exoALMA Large Program targeted a sample of 15 disks to study gas dynamics within these systems, and these observations simultaneously produced continuum data at 0.9~mm (331.6~GHz) with exceptional surface brightness sensitivity at high angular resolution. To provide a robust characterization of the observed substructures, we performed a visibility space analysis of the continuum emission from the exoALMA data, characterizing axisymmetric substructures and nonaxisymmetric residuals obtained by subtracting an axisymmetric model from the observed data. We defined a nonaxisymmetry index and found that the most asymmetric disks predominantly show an inner cavity and consistently present higher values of mass accretion rate and near-infrared excess. This suggests a connection between outer disk dust substructures and inner disk properties. The depth of the data allowed us to describe the azimuthally averaged continuum emission in the outer disk, revealing that larger disks (both in dust and gas) in our sample tend to be gradually tapered compared to the sharper outer edge of more compact sources. Additionally, the data quality revealed peculiar features in various sources, such as shadows, inner disk offsets, tentative external substructures, and a possible dust cavity wall. 
\end{abstract}


\section{Introduction} \label{sect:introduction}

Over the last decade, the capabilities of the Atacama Large Millimeter/submillimeter Array (ALMA) allowed us to reveal and extensively explore substructures in protoplanetary disks. This effort began with the dust continuum observation of HL~Tau by \cite{ALMAPartnership_2015} and has continued with numerous other high-resolution observations (see, e.g., \citealt{Andrews2020_review} for a review). Substructures in disks have also been detected using other tracers and wavelengths, such as in the gas line emission (e.g., \citealt{Law_2021_MAPS_III_substructures}) and the near-infrared (NIR) scattered light (review by \citealt{Benisty_2023_PPVII}). These substructures include rings and gaps \citep{Andrews2016, Long2018, Andrews18_DSHARP, Perez2019}, cavities \citep{Francis2020, Facchini2020, Sierra2024}, crescents \citep{Casassus2013,vanderMarel2013,Perez2014}, and spirals \citep{Benisty2015, Perez2016, Speedie2024}

Different physical mechanisms have been proposed to explain the formation of such substructures. They encompass a range of hydrodynamic and magnetohydrodynamic processes (e.g., Rossby wave instability, vertical shear instability, gravitational instability, zonal flows, dead zones), photoevaporative and magnetic winds, dust accumulation and growth at ice lines along with dust concentration driven by streaming instability, as well as tidal interactions with a stellar companion and stellar flyby events (\citealt{Bae_2023_PPVII}, \citealt{Lesur2023, Pascucci2023, Cuello2023, Kurtovic2018}). Among these mechanisms, the observed substructures are often interpreted as resulting from interactions between the disk and one or more planets (e.g., \citealt{Ayliffe2012}, \citealt{Dipierro2015}, \citealt{Bae2018}, \citealt{Lodato_2019}, \citealt{Ruzza2024}).

\begin{deluxetable*}{lccccccr}
\tabletypesize{\footnotesize}
\tablewidth{1\textwidth} 
\tablecaption{Continuum Image Properties}
\tablehead{
\colhead{Source} & \colhead{$\theta_\mathrm{b}$} & \colhead{PA${_\mathrm{b}}$} &\colhead{rms Noise} & \colhead{Peak $I_\nu$, $T_\mathrm{b}$} & \colhead{$F_\nu$} &   \colhead{$d$} &  \colhead{$M_\mathrm{d}$}\\
\colhead{} & \colhead{(mas,$\;\;$ au)} & \colhead{(deg)} & \colhead{($\mu$Jy beam$^{-1}$,$\;\;$ K)} & \colhead{(mJy beam$^{-1}$,$\;\;$ K)} & \colhead{(mJy)} &   \colhead{(pc)} &  \colhead{($M_\oplus$,$\;\;$ $M_\mathrm{Jup}$)}} 
\colnumbers
\startdata 
AA Tau      &       $70\times57, \;\; 9\times8$      &      168      &    45$, \;\;$  3.3    &     2.71$, \;\;$ 14.0    &     $189.4\pm0.3$    &  135\tablenotemark{a}   &   37,$\;\;$ 0.12 \\
CQ Tau      &       $82\times59, \;\; 12\times9$      &      153     &    40,$\;\;$  3.1    &     8.10,$\;\;$ 25.8    &     $431.9\pm0.3$    &  149\tablenotemark{a}   &   103,$\;\;$ 0.33 \\
DM Tau      &       $68\times58, \;\; 10\times8$      &      162      &    39,$\;\;$  3.2    &     2.98,$\;\;$ 14.8    &     $226.5\pm0.6$    &   144 &  50,$\;\;$ 0.16 \\
HD 135344B      &       $90\times76, \;\;12\times10$      &      85      &    43,$\;\;$  2.9    &     6.44,$\;\;$ 17.2    &     $424.7\pm0.2$    &  135 &   83,$\;\;$ 0.26 \\
HD 143006      &       $94\times68, \;\;16\times11$      &      97      &    45,$\;\;$  3.0    &     3.44,$\;\;$ 12.3    &     $155.5\pm0.2$    &  167  &   47,$\;\;$ 0.15 \\
HD 34282      &       $67\times54, \;\;21\times17$      &      95      &    40,$\;\;$  3.3    &     3.80,$\;\;$ 18.5    &     $343.4\pm0.3$    &  309  &   351,$\;\;$ 1.10 \\
J1604      &       $95\times73, \;\;14\times11$      &      91      &    43,$\;\;$  2.9    &     2.72,$\;\;$ 10.4    &     $198.4\pm0.3$    &  145  &   44,$\;\;$ 0.14 \\
J1615      &       $97\times83, \;\;15\times13$      &      85      &    38,$\;\;$  2.8    &     5.83,$\;\;$ 14.5    &     $386.0\pm0.7$    &  156  &  100,$\;\;$ 0.32 \\
J1842      &       $97\times72, \;\;15\times11$      &      78      &    43,$\;\;$  2.9    &     3.55,$\;\;$ 12.0    &     $141.5\pm0.2$    &   151  &   35,$\;\;$ 0.11 \\
J1852      &       $99\times71, \;\;15\times10$      &      67      &    37,$\;\;$  2.8    &     4.52,$\;\;$ 13.6    &     $150.9\pm0.1$    &   147  &  35,$\;\;$ 0.11 \\
LkCa 15      &       $75\times59, \;\;12\times9$      &      150      &    34,$\;\;$  3.1    &     2.75,$\;\;$ 13.3    &     $407.1\pm0.4$    &  156  &   108,$\;\;$ 0.34 \\
MWC 758      &       $101\times75, \;\;16\times12$      &      130      &    56,$\;\;$  3.0    &     6.85,$\;\;$ 16.7    &     $214.5\pm0.2$    &    156  &  56,$\;\;$ 0.18 \\
PDS 66      &       $93\times73, \;\;9\times7$      &      19      &    47,$\;\;$  3.0    &     16.28,$\;\;$ 33.9    &     $336.1\pm0.2$    &   98  &  35,$\;\;$ 0.11 \\
SY Cha      &       $91\times68, \;\;16\times12$      &      171      &    56,$\;\;$  3.1    &     1.54,$\;\;$ 8.3    &     $158.4\pm0.5$    &  182  &   55,$\;\;$ 0.17 \\
V4046 Sgr      &       $89\times72, \;\;6\times5$      &      88      &    37,$\;\;$  2.9    &     5.20,$\;\;$ 15.7  &     $668.4\pm1.0$    &   72 &   37,$\;\;$ 0.12 \\
\enddata
\tablecomments{All properties were obtained from fiducial CLEAN images with robust -0.5. The mean frequency is 331.6~GHz for each image. Column (1): target name. Column (2): synthesized beam FWHM major and minor axes. Column (3): synthesized beam PA. Column (4): image rms noise. Column (5): image peak intensity. Note that the noise and peak brightness temperature were computed using the full Planck law. 
Column(6): integrated flux density with statistical uncertainty, excluding the $10\%$ absolute flux calibration. Column (7): source distance as measured by \textit{Gaia} DR3 \citep{GaiaDR3}. Column (8): estimated dust mass.}
\tablenotetext{a}{As reported in \cite{Teague_exoALMA}, the renormalized unit weight error (RUWE) values from Gaia \citep{GaiaDR3} for these sources are high, indicating that their distances should be interpreted with caution.}
\label{tab:image_properties}
\end{deluxetable*}

Among the different tracers used to study disk substructures, dust continuum emission at submillimeter wavelengths holds particular importance. Dust particles in disks constitute the fundamental building blocks of planets, and their thermal emission allows us to trace the distribution and properties of millimeter-sized grains concentrated in the disk midplane, where planet formation is thought to occur \citep{Drazkowska2023}. By studying dust continuum emission, we gain insights into the processes that shape dust distribution, growth, concentration, and evolution, all of which are essential for understanding the early stages of planet formation \citep{Testi2014}. 

However, what governs the morphology of dust continuum emission in protoplanetary disks remains an open question. In this paper, we aim to bring new insights to this question by analyzing the homogeneous, deep observations at high angular resolution of dust continuum emission from the exoALMA Large Program\footnote{\url{https://www.exoalma.com}} (2021.1.01123.L; \citealt{Teague_exoALMA}). The continuum emission features observed in the exoALMA sample are then connected to properties derived from gas emission observations and model predictions in other papers of this series \citep{Galloway_exoALMA, Stadler_exoALMA, Gardner_exoALMA, Longarini_exoALMA, Yoshida_exoALMA, Wölfer_exoALMA}. Additionally, we introduce two new metrics: one to quantify the level of nonaxisymmetry in disks, used to explore connections between observed dust substructures and inner disk properties, and another to investigate the falloff of the outer disk emission.

Section~\ref{sect:data} presents the exoALMA data. Section~\ref{sect:methods} describes the pipeline adopted to characterize the observed substructures in the visibility space. Section~\ref{sect:results} presents the results of the analysis, including axisymmetric substructures and nonaxisymmetric residuals obtained by subtracting an axisymmetric model from the data.  Section~\ref{sect:discussion} discusses the results by examining the observed substructures in the context of what is already known for each disk.  We also discuss nonaxisymmetries, the faint outer disk, and hints of the presence of companions in the disks in our sample, comparing our findings with previous studies and with the velocity kink results presented by \cite{Pinte_exoALMA}. Section~\ref{sect:conlusion} summarizes the main results.

\section{Data}
\label{sect:data}

\begin{figure*}[t]
\centering
\includegraphics[width=1\hsize]{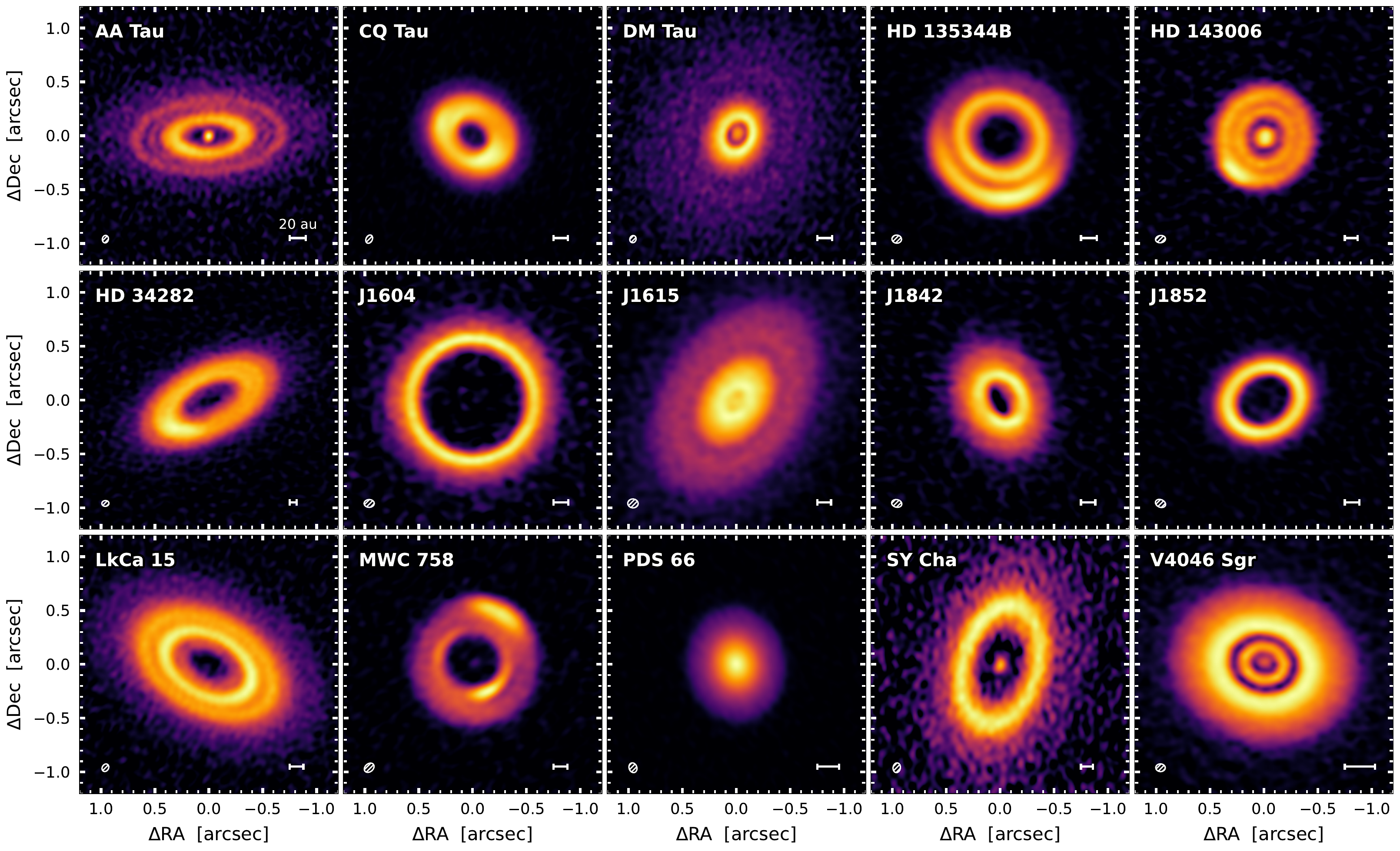}
\caption{Gallery of fiducial continuum images at 0.9~mm (331.6~GHz) of the exoALMA sample, obtained with the CLEAN algorithm and robust of -0.5. The source order is alphabetical. All images are shown on the same angular scales. The FWHM of the synthesized beams and the 20~au scale bars are indicated in the lower left and right corners of each plot, respectively. The color scale ranges from null to peak intensity for each disk. An asinh stretch is applied to the color scale to visually enhance the fainter emission. A linear stretch version is presented in Figure~\ref{fig:gallery_images_linear}.}
\label{fig:gallery_images_asinh}
\end{figure*}

exoALMA targeted 15 protoplanetary disks with deep observations at high angular and spectral resolution. The primary objective was to study the physical and dynamical structure of the gas in these disks and to reveal perturbations that may be produced by embedded planets \citep{Teague_exoALMA}. For this aim, as detailed in \cite{Teague_exoALMA}, the selection criteria focused on sources that were extended in gas (at least~1\arcsec) and had favorable inclinations (between ${\sim}5^\circ$ and $60^\circ$), and whose gas emission was free from absorption or contamination by large-scale emission. Preference was given to brighter sources in $^{12}$CO, with a distribution in R.A. to facilitate scheduling observations. This resulted in an intentionally biased sample toward bright and extended disks from different star-forming regions, most of which have already been observed in continuum by ALMA at high angular resolution, revealing a variety of dust substructures indicative of planet-disk interactions. 

The exoALMA observations also produced extremely deep dust continuum data at high angular resolution. Observations were carried out in ALMA Band~7, combining configurations C-6 and C-3 (and the Atacama Compact Array, ACA, for the most extended sources), having one spectral window with a bandwidth of 1875~MHz centered at 331.57~GHz (0.904~mm) dedicated to the continuum emission.  This resulted in continuum images with an angular resolution of ${\sim}0.09\arcsec$, maximum recoverable scale of $4.7\arcsec$ ($19.3\arcsec$ for sources with ACA observations), and a sensitivity of ${\sim}40\, \mu\mathrm{Jy\,beam}^{-1}$.  The exoALMA observations provide deep surface brightness sensitivity at high angular resolution, achieving a noise level of ${\sim}0.05$~K. For comparison, the DSHARP program \citep{Andrews18_DSHARP} reached a noise level of 0.25~K at a wavelength of 1.3~mm and a resolution of ${\sim}0.035\arcsec$ (both noise levels calculated using the Rayleigh-Jeans approximation).

Pipeline calibration and self-calibration have been applied to all sources. A dedicated description of the calibration and imaging pipeline is presented in \citet{Loomis_exoALMA}. For the analyses in this paper, we considered only the spectral window dedicated to the continuum to maintain consistency in frequency coverage, as including the additional continuum data in spectral windows dedicated to line emission would have provided only a marginal 3\% increase in sensitivity. The visibilities in the continuum spectral windows were spectrally averaged down to one channel for each execution block and averaged in time down to 30~s bins to reduce file size and improve processing efficiency. We verified that this averaging did not affect the continuum analysis by comparing images produced with and without the averaging; the resulting images showed no significant differences \citep{Loomis_exoALMA}. All the manipulations of the visibilities were conducted with the software CASA, version~6.2 \citep{CASA2022}.

We present a gallery of the fiducial continuum images of the exoALMA disks in Fig.~\ref{fig:gallery_images_asinh} with an asinh stretch on the color scale (meaning the asinh function has been applied to the observed intensity) and in Fig.~\ref{fig:gallery_images_linear} with a linear stretch. Table~\ref{tab:image_properties} reports the properties of the observed continuum images for each target. We calculated the rms noise in an annulus between $3\arcsec$ and $4\arcsec$ centered on the disk, where no emission from the target was present. Integrated flux density was measured within a mask defined as an ellipse with the same center, position angle (PA), and aspect ratio as the target. The semimajor axis of this ellipse is 1.5 times the outer extent of the observed emission ($R_{\rm out}$), determined by the intersection of the contour reaching twice the rms noise level in the image with the disk major axis. The associated uncertainty reflects only the statistical error and does not include the $10\%$ absolute flux error~(2$\sigma$) in ALMA Band~7 observations\footnote{see Sect.~10.2.6 in the ALMA Technical Handbook \url{https://almascience.nrao.edu/proposing/technical-handbook/}}. To estimate the statistical error, we followed a procedure similar to that of \citet{Rampinelli2024}. We computed the uncertainty as the standard deviation of the integrated flux density measured in 24 nonoverlapping elliptical masks identical to the one used for the disk’s flux measurement placed within the field of view (FOV) outside the disk’s emitting area. Since the continuum emission is always well within the primary beam, we used images without primary beam correction to yield uniform noise.

We derived an estimate of the total dust mass in each disk using the integrated flux density and the relation from \cite{Hildebrand1983}, which is based on the assumption of optically thin dust emission,
\begin{equation}
\label{eq:dust_mass}
M_\mathrm{d} = \frac{F_\nu d^2}{B_\nu(T) k_\nu}, 
\end{equation}
where $d$ is the distance, $B_\nu (T)$ is the blackbody surface brightness at a given temperature, and $k_\nu$ is the dust opacity. We assumed a temperature of 20~K (as in, e.g., \citealt{Ansdell2016}) and an opacity $k_\nu = 3.5 \, \mathrm{cm}^2/\mathrm{g} \times 870  \,\mu\mathrm{m}/\lambda$ \citep{Beckwith1990}. \cite{Longarini_exoALMA} provide a comparison between the masses derived from the continuum and the ones computed from the gas rotation curves. They obtain gas-to-dust mass ratios above the usual value of 100, with an average of ${\sim}400$. These high values indicate that the dust masses we compute are underestimated due to the assumption of optically thin emission when using Equation~\ref{eq:dust_mass}.

\section{Methods for the Continuum Analysis} \label{sect:methods}

The first aim of our continuum data analysis is to perform a morphological characterization of the observed substructures in each disk. To do so, we rely on a two-step visibility-fitting pipeline\footnote{The pipeline is accessible at \url{https://github.com/pcurone/exoALMA_continuum_pipeline}}. 

First, we use the code \galario \citep{Tazzari2018_galario} to recover the disk’s geometric parameters: inclination ($i$), PA, and the offsets in R.A. and decl. between the disk center and the phase center ($\Delta$R.A. and $\Delta$decl.) (Sect.~\ref{subsect:galario_fit}). \galario\ uses a parametric intensity model and a Markov Chain Monte Carlo (MCMC) approach, which ensures accurate estimation of the disk geometry, as demonstrated in several previous studies (e.g., \citealt{Fedele2018,Long2018, Facchini2020}). 

We then use these geometric parameters as input in the second step, where, to obtain a model of the intensity radial profile, we employ \texttt{frankenstein} (hereafter \frank; \citealt{Jennings2020_frank}). Unlike \galario, \frank\ uses a nonparametric approach, offering more flexibility in fitting the observed visibilities without requiring a predefined intensity model. This nonparametric method allows us to reconstruct the intensity radial profile with subbeam resolution, providing a more detailed representation of the disk structure (Sect.~\ref{subsect:frank_fit}).

\begin{deluxetable*}{llllllll}
\tabletypesize{\footnotesize}
\tablewidth{1\textwidth} 
\tablecaption{Dust Disk Geometries}
\tablehead{
\colhead{Source} & \colhead{$i$} & \colhead{PA} &\colhead{$\Delta$R.A.} & \colhead{$\Delta$decl.}  & \colhead{$R_{68}$} &  \colhead{$R_{90}$}  &   \colhead{$R_{95}$}\\
\colhead{} & \colhead{(deg)} & \colhead{(deg)} &\colhead{(mas)} & \colhead{(mas)}    &  \colhead{(au, mas)} &  \colhead{(au, mas)}  &   \colhead{(au, mas)}
} 
\colnumbers
\startdata 
AA Tau     &     $58.54^{+0.02}_{-0.02}$     &    $93.77^{+0.02}_{-0.03}$      &      $-5.46^{+0.11}_{-0.11}$     &     $4.83^{+0.07}_{-0.08}$   &   $92.2_{-0.6}^{+1.2}$, $685_{-5}^{+9}$  &  $139.4_{-1.2}^{+1.2}$, $1035_{-9}^{+9}$  &  $158.6_{-1.2}^{+2.2}$, $1177_{-9}^{+16}$  \\
CQ Tau     &     $35.24^{+0.02}_{-0.02}$     &    $53.87^{+0.02}_{-0.02}$      &      $-8.71^{+0.05}_{-0.05}$     &     $0.99^{+0.04}_{-0.04}$   & $55.8_{-0.1}^{+0.6}$, $373_{-1}^{+4}$  &  $73.1_{-0.6}^{+0.6}$, $489_{-4}^{+4}$  &  $85.4_{-0.1}^{+0.7}$, $572_{-1}^{+5}$  \\
DM Tau     &     $35.97^{+0.05}_{-0.05}$     &    $155.60^{+0.08}_{-0.07}$      &      $-5.51^{+0.07}_{-0.07}$     &     $-6.59^{+0.09}_{-0.09}$      & $118.6_{-0.8}^{+0.8}$, $824_{-6}^{+6}$  &  $201.9_{-0.8}^{+1.6}$, $1402_{-6}^{+11}$  &  $244.8_{-1.6}^{+1.6}$, $1700_{-11}^{+11}$  \\
HD 135344B     &     $20.73^{+0.02}_{-0.02}$     &    $28.92^{+0.09}_{-0.06}$      &      $0.80^{+0.05}_{-0.05}$     &     $-3.21^{+0.05}_{-0.05}$     & $78.7_{-1.2}^{+1.2}$, $583_{-9}^{+9}$  &  $90.2_{-1.2}^{+1.2}$, $668_{-9}^{+9}$  &  $94.4_{-0.6}^{+1.2}$, $700_{-4}^{+9}$  \\
HD 143006     &     $18.69^{+0.09}_{-0.09}$     &    $7.53^{+0.35}_{-0.32}$      &      $8.27^{+0.14}_{-0.13}$     &     $26.42^{+0.16}_{-0.16}$     & 
$69.4_{-0.4}^{+0.4}$, $415_{-2}^{+2}$  &  $79.9_{-0.8}^{+1.1}$, $478_{-4}^{+7}$  &  $84.8_{-0.8}^{+1.5}$, $507_{-4}^{+9}$  \\
HD 34282     &     $59.09^{+0.01}_{-0.01}$     &    $117.15^{+0.01}_{-0.01}$      &      $13.00^{+0.07}_{-0.08}$     &     $15.49^{+0.06}_{-0.06}$      & $179.8_{-1.4}^{+2.8}$, $583_{-4}^{+9}$  &  $239.4_{-2.8}^{+2.8}$, $776_{-9}^{+9}$  &  $289.3_{-2.8}^{+3.2}$, $938_{-9}^{+10}$  \\
J1604     &     $8.72^{+0.09}_{-0.07}$     &    $123.24^{+0.07}_{-0.15}$      &      $-74.82^{+0.07}_{-0.07}$     &     $-16.67^{+0.06}_{-0.06}$      & $94.0_{-0.1}^{+0.5}$, $650_{-1}^{+4}$  &  $112.4_{-0.1}^{+0.5}$, $778_{-1}^{+4}$  &  $122.2_{-0.5}^{+0.1}$, $845_{-4}^{+1}$  \\
J1615     &     $47.10^{+0.01}_{-0.01}$     &    $146.14^{+0.02}_{-0.02}$      &      $-44.32^{+0.05}_{-0.04}$     &     $-5.88^{+0.04}_{-0.05}$      & $116.1_{-0.1}^{+1.0}$, $746_{-1}^{+7}$  &  $169.6_{-1.0}^{+2.1}$, $1090_{-7}^{+13}$  &  $204.2_{-1.0}^{+2.1}$, $1312_{-7}^{+13}$  \\
J1842     &     $39.22^{+0.03}_{-0.04}$     &    $26.35^{+0.06}_{-0.06}$      &      $-3.16^{+0.07}_{-0.07}$     &     $-30.69^{+0.07}_{-0.07}$   & $62.7_{-0.1}^{+0.5}$, $415_{-1}^{+4}$  &  $85.2_{-0.5}^{+1.1}$, $564_{-4}^{+7}$  &  $100.8_{-1.1}^{+0.5}$, $668_{-7}^{+4}$ \\
J1852     &     $32.50^{+0.03}_{-0.05}$     &    $117.61^{+0.03}_{-0.03}$      &      $-23.41^{+0.04}_{-0.04}$     &     $1.91^{+0.04}_{-0.04}$     & $58.0_{-0.1}^{+0.7}$, $394_{-1}^{+4}$  &  $69.9_{-0.7}^{+0.7}$, $475_{-4}^{+4}$  &  $79.8_{-0.7}^{+0.7}$, $542_{-4}^{+4}$  \\
LkCa 15     &     $50.59^{+0.01}_{-0.02}$     &    $61.57^{+0.01}_{-0.01}$      &      $-16.84^{+0.05}_{-0.05}$     &     $20.83^{+0.05}_{-0.05}$      &  $111.0_{-0.8}^{+1.6}$, $706_{-5}^{+10}$  &  $156.3_{-1.6}^{+2.5}$, $994_{-10}^{+16}$  &  $181.0_{-2.5}^{+2.5}$, $1152_{-16}^{+16}$
  \\
MWC 758     &     $7.27^{+0.23}_{-0.17}$     &    $76.17^{+0.13}_{-0.10}$      &      $25.48^{+0.13}_{-0.13}$     &     $18.42^{+0.12}_{-0.12}$      & $79.5_{-0.8}^{+0.4}$, $510_{-5}^{+3}$  &  $91.4_{-1.2}^{+1.2}$, $586_{-8}^{+8}$  &  $96.3_{-1.2}^{+1.2}$, $618_{-8}^{+8}$  \\
PDS 66     &     $32.02^{+0.03}_{-0.03}$     &    $8.91^{+0.05}_{-0.05}$      &      $-3.59^{+0.02}_{-0.02}$     &     $6.68^{+0.03}_{-0.03}$    & $31.7_{-0.1}^{+0.3}$, $324_{-1}^{+3}$  &  $46.9_{-0.3}^{+0.3}$, $479_{-3}^{+3}$  &  $51.5_{-0.3}^{+0.5}$, $526_{-3}^{+5}$ \\
SY Cha     &     $51.65^{+0.03}_{-0.02}$     &    $165.77^{+0.04}_{-0.04}$      &      $-12.66^{+0.12}_{-0.13}$     &     $28.16^{+0.18}_{-0.18}$     &  $132.3_{-1.6}^{+1.6}$, $732_{-9}^{+9}$  &  $197.7_{-1.6}^{+2.4}$, $1094_{-9}^{+13}$  &  $228.1_{-1.6}^{+1.6}$, $1262_{-9}^{+9}$  \\
V4046 Sgr     &     $33.36^{+0.01}_{-0.01}$     &    $76.02^{+0.02}_{-0.01}$      &      $-50.94^{+0.03}_{-0.02}$     &     $-45.18^{+0.02}_{-0.02}$      & $46.5_{-0.4}^{+0.1}$, $650_{-6}^{+1}$  &  $60.9_{-0.4}^{+0.4}$, $852_{-6}^{+6}$  &  $71.8_{-0.4}^{+0.8}$, $1004_{-6}^{+11}$  \\
\enddata
\tablecomments{Column (1): target name. Column (2): disk inclination. Column (3): disk PA. Columns (4) and (5): offsets in R.A. and decl. between the disk center and the phase center. Geometrical parameters in columns (2) - (5) were obtained from \galario fits (see Sect.~\ref{subsect:galario_fit}), and the associated statistical uncertainties represent the 16th and 84th percentiles of the MCMC marginalized distribution.  These uncertainties should not be considered as actual observational errors but rather as uncertainties on the fit given the assumed model. Columns (6), (7), and (8): radial extent of the continuum emission enclosing 68\%, 90\%, and 95\% of the continuum intensity, respectively. Values were computed from \frank model intensity profiles, and 16th and 84th percentiles are derived via bootstrapping varying the geometrical parameters (see Sect~\ref{subsect:frank_fit}).}
\label{tab:disk_geometrical_params}
\end{deluxetable*}

\subsection{\galario Fit}\label{subsect:galario_fit}

The code \galario assumes a 1D or 2D model representing the emission in the image plane and performs a Fourier transform to derive the synthetic visibilities at the same $uv$-points as the observation \citep{Tazzari2018_galario}. The best-fit model is determined by minimizing the $\chi^2$ through an MCMC approach, utilizing the \texttt{emcee} package for parameter sampling \citep{ForemanMackey2013_emcee}. 
In employing this methodology, our primary focus was not an exhaustive characterization of the substructures, a task reserved for the application of \frank. Instead, our objective was to derive robust estimates of the geometrical parameters of each disk, specifically inclination, PA and offsets in R.A. and decl. between the disk center and the phase center. This is reflected by our choices of the parametric models, selected so that they could globally represent the disk observed morphology. 

Of the 15 sources, 10 were characterized using 1D axisymmetric intensity profiles.  Each profile includes one or more Gaussian rings, 
\begin{equation}
    I(R) = f_0 \,\exp{\left[-\frac{(R-R_0)^2}{2\sigma^2}\right]},
\end{equation}
where $R$ is the radial coordinate, $f_0$ is a normalization term, $R_0$ denotes the radial location of the Gaussian peak, and $\sigma$ is the standard deviation. For sources displaying inner emission, we added either a central Gaussian, 
\begin{equation}
    I(R) = f_0 \,\exp{\left[-\frac{R^2}{2\sigma^2}\right]},
\end{equation}
or, in the case of unresolved emission, a central point source,
\begin{equation}
    I(R) = f_0 \,\delta (R),
\end{equation}
where $\delta(R)$ is the Dirac delta function. 

For the five disks showing strong asymmetries (CQ~Tau, HD~135344B, HD~143006, HD~34282, and MWC~758), we employed 2D models. These models combined axisymmetric rings with one or more arcs (as done by, e.g., \citealt{Cazzoletti2018} and \citealt{Perez2018}), defined as Gaussian rings with azimuthal tapering,
\begin{equation}
    I(R, \phi) = f_0 \,\exp{\left[-\frac{(R-R_0)^2}{2\sigma^2}\right]} \, \exp{\left[-\frac{(\phi-\phi_0)^2}{2\sigma_\phi^2}\right]},
\end{equation}
where $\phi$ is the azimuthal coordinate, $\phi_0$ is the azimuthal center of the arc, and $\sigma_\phi$ is its azimuthal extent. 

Uniform priors were applied, and the intensity normalization factor $f_0$ was logarithmically sampled. For each 1D calculation, we used ${\sim}100$ walkers that converged after ${\sim}10^4$ steps, while the 2D runs required a higher number of steps to converge, between  ${\sim}3\times10^4$ and ${\sim}10^5$. The estimates of the geometrical parameters for each disk are reported in Table~\ref{tab:disk_geometrical_params}, while the chosen \galario models along with the best-fit value for each parameter are presented in Tables~\ref{tab:galario_results_1D} and~\ref{tab:galario_results_2D} for 1D and 2D models, respectively. Appendix~\ref{sect:comparing_gas_dust_geom_param} compares the continuum geometrical parameters obtained with \galario with the estimates from the gas data retrieved with \discminer \citep{Izquierdo_exoALMA}, showing a generally good agreement (within 5~deg for $i$ and PA, and within 50~mas in $\Delta$R.A. and $\Delta$decl. in most sources).

\subsection{\frank Fit}\label{subsect:frank_fit}

For a thorough characterization of the intensity profiles as a function of disk radius, we used the code \frank. It reconstructs the protoplanetary disk intensity radial profile by modeling it as a Fourier-Bessel series, then using a discrete Hankel transform to compute synthetic visibilities. These synthetic visibilities are subsequently fitted directly to the observed visibilities within a Bayesian framework, employing a Gaussian process for regularization \citep{Jennings2020_frank}. This method is applied to visibilities that have been deprojected, shifted so that the center of the disk is at phase center, and left unbinned. The fit is nonparametric and 1D, assuming the axisymmetry of the source. Moreover, the disk emission is treated as geometrically flat and optically thick, since visibility deprojection based on inclination scales the total flux. \frank enables the recovery of subbeam resolution features that remain undetected in both the CLEAN image and its azimuthally averaged intensity profile while exploiting the full data sensitivity (\citealt{Jennings2022_DSHARP, Andrews2021, Ilee2022}). 

We performed the \frank fit in logarithmic intensity space, which intrinsically guarantees the intensity to be nonnegative and largely reduces the high-frequency oscillations in the reconstructed intensity profile when compared to the fit in linear space. We verified that the choice of the five hyperparameters $\alpha$, $w_\mathrm{smooth}$, $R_\mathrm{max}$, $N$, $p_0$) had minimal impact on the resulting fit, given the high sensitivity of our data. We selected, nonetheless, conservative values to minimize the chance of artifacts generated by fitting low signal-to-noise features and set $\alpha = 1.3$, and $w_\mathrm{smooth}= 0.01$, with $\alpha$ determining the signal-to-noise ratio (SNR) threshold at which the model stops fitting the data and $w_\mathrm{smooth}$ helping to suppress noisy oscillations. The hyperparameter $R_\mathrm{max}$, indicating the point beyond which \frank assumes zero emission, was established at $1.5R_{\rm out}$ (see Sect.~\ref{sect:data} for the definition of $R_{\rm out}$). The $N$ hyperparameter, determining the radial gridding, was set to 400, and $p_0$, acting in the regularization of the emission power spectrum, was fixed to $10^{-35}$, the standard value for logarithmic intensity space fitting. A comparison of the observed visibility profiles as a function of deprojected baseline with the \galario and \frank fits is presented in Fig.~\ref{fig:uv_profiles}. 

As explained by \cite{Jennings2020_frank, Jennings2022_DSHARP}, obtaining an accurate estimate of the uncertainty associated with the \frank fit is not feasible. This limitation arises from the inherently ill-posed nature of reconstructing brightness from Fourier data. Specifically, there is no robust method to accurately extrapolate visibility amplitudes in a given dataset beyond the longest baseline fitted by \frank. Therefore, to obtain a reasonable uncertainty for the reconstructed intensity radial profile, we bootstrapped the \frank fit by randomly varying the geometrical parameters, similar to what was done by \cite{Carvalho2024}. We ran the \frank fit 500 times for each disk (after testing that 500 iterations produced the same effect as 5000 iterations), randomly picking the $i$, PA, $\Delta$R.A., and $\Delta$decl. from a Gaussian distribution centered on the best-fit values from \galario. Since the uncertainties on the geometric parameters from \galario are considerably underestimated (as is often the case with MCMC methods), we assumed broader ranges for the bootstrapping. The standard deviation was set to 1~deg for inclination and PA and to one-third of the $\sigma$ of the synthesized beam major axis for the R.A. and decl. offsets, resulting in a ${\sim}10$ mas centering accuracy. We then fitted the distribution of the intensity for each radial bin with a Gaussian and took $1\sigma$ as the uncertainty associated with the intensity. We also used this bootstrapping method to assign an uncertainty to the values of the dust disk extent ($R_{68}$, $R_{90}$, and $R_{95}$) reported in Table \ref{tab:disk_geometrical_params}. These values were calculated for each iteration of the bootstrap, and the uncertainties were taken as the 16th and 84th percentiles.

\section{Results} \label{sect:results}

\subsection{Axisymmetric Substructures} \label{sect:axisymm_subs}

\begin{figure*}
\centering
\includegraphics[width=0.94\hsize]{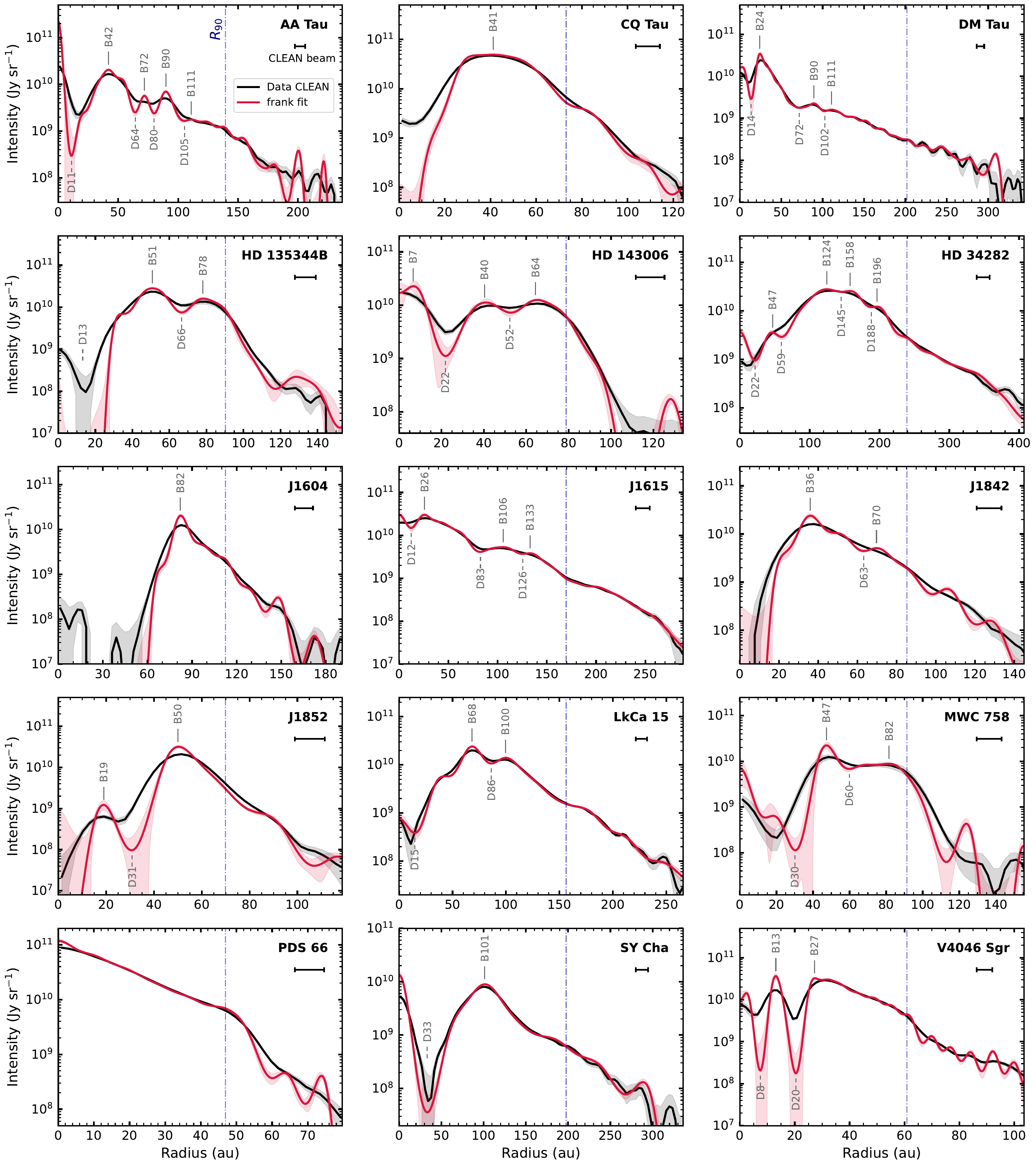}
\caption{Gallery of radial intensity profiles on a log-linear scale of the deprojected and azimuthally averaged CLEAN data (black solid line) and the \frank model (red solid line). Sources are arranged alphabetically. The gray shading represents CLEAN data uncertainty, calculated as the 1$\sigma$ scatter per radial bin, divided by the square root of the number of beams in the associated annulus. The red shading indicates the 1$\sigma$ uncertainty of the \frank model, estimated via bootstrapping, considering small variations of the disk geometrical parameters (see Sect.~\ref{subsect:frank_fit}). The black scale bar in the top right corner of each panel shows the FWHM, averaged between the major and minor axes of the CLEAN synthesized beam.
Radial positions of rings and gaps are marked, with each gap labeled with a dashed line and a “D” (for dark) followed by the distance from the central star in au. Solid lines with “B” labels (for bright) denote the rings. A vertical blue dashed-dotted line indicates $R_{90}$, the radial location within which axisymmetric substructures are defined.}
\label{fig:gallery_intensity_profiles}
\end{figure*}

We employed the intensity profile from the \frank fit to define the annular axisymmetric features, that is, rings and gaps. Figure~\ref{fig:gallery_intensity_profiles} presents the intensity profiles as a function of disk radius of the deprojected and azimuthally averaged data CLEAN image compared to the \frank model. The deprojection and azimuthal averaging of the observed CLEAN image were performed with the package \texttt{GoFish} \citep{GoFish}. The uncertainty was determined by dividing the $1\sigma$ scatter at each intensity radial bin by the square root of the number of beams within the corresponding radial annulus.

Considering the subbeam resolution \frank model of the intensity radial profile, we aim to define annular substructures, that is, rings and gaps that appear as peaks and troughs in the intensity, respectively. Following the nomenclature of \cite{Huang2018_DSHARPII}, we label rings as “B” (for \textit{bright}) and gaps as “D” (for \textit{dark}). To ensure that these features are not simply noise or artifacts from the \frank\ model, we establish four criteria to assess what can be robustly defined as a substructure and determine its radial location ($R_\mathrm{B}$ and $R_\mathrm{D}$, respectively). First, focusing only on the best-fit \frank model intensity radial profile (and not the bootstrapped uncertainties), rings and gaps must correspond to local maxima and minima, respectively. Second, their radial location should fall within the radius enclosing 90\% of the source flux ($R_{90}$) to exclude low-SNR oscillations at larger radii. Third, the peak intensity of each ring must be higher than the rms noise to avoid low-SNR fluctuations within inner cavities. Finally, defining the gap depth as $I_\mathrm{D} / I_\mathrm{B}$, where $I_\mathrm{D}$ is the intensity of the gap minimum at $R_\mathrm{D}$ and  $I_\mathrm{B}$ is the ring peak intensity at $R_\mathrm{B}$ (following the definition in \citealt{Huang2018_DSHARPII}), we accept a pair of gap-ring if it meets the gap depth condition $I_\mathrm{D} / I_\mathrm{B} \le 0.97$ to ensure sufficient contrast.

We adopted the procedure of \cite{Huang2018_DSHARPII} (see their Section~3.2 and Appendix~B for more details) to determine the substructure width. Briefly, it involves deriving the width based on the inner and outer edges of a substructure rather than employing a Gaussian fit, a more suitable method for structures deviating from a Gaussian shape. Applying these criteria to our \frank model intensity profiles, for a gap-ring pair, the dividing point between the outer edge of the gap and the inner edge of the ring, denoted as $R_\mathrm{D,out}\equiv R_\mathrm{B,in}$, is defined as the radius at which the intensity equals $I_\mathrm{mean}=(I_\mathrm{D}+I_\mathrm{B})/2$. The radius of the gap inner edge $R_\mathrm{D,in}$ is the largest radius with $R<R_\mathrm{D}$ and $I(R)=I_\mathrm{mean}$. The radius of the ring outer edge $R_\mathrm{B,out}$ is the smallest radius with $R>R_\mathrm{B}$ and $I(R)=I_\mathrm{mean}$. Consequently, the gap width is given by $R_\mathrm{D,out} - R_\mathrm{D,in}$ and the ring width is $R_\mathrm{B,out} - R_\mathrm{B,in}$. With this approach, we automatically obtain the width of the inner cavities as well. If the first substructure in a disk (counting from the center) is a ring (CQ~Tau, HD~143006, J1604, J1842, J1852), the outer radius of the cavity corresponds to the $R_\mathrm{B,in}$ of that first ring. Conversely, if the first substructure is a gap, occurring when there is an inner disk (AA~Tau, DM~Tau, HD~135344B, HD~34282, J1615, LkCa~15, MWC~758, SY~Cha, V4046~Sgr), the outer radius of the cavity corresponds to the $R_\mathrm{D,out}$ of that first gap. 

All the substructure properties for each disk are presented in Table~\ref{tab:substructures}. PDS~66 is the only source where no annular substructures were detected.

\subsection{Nonaxisymmetric Substructures} \label{subsect:non-axisymm_subs}

\begin{figure*}[t]
\centering
\includegraphics[width=0.77\hsize]{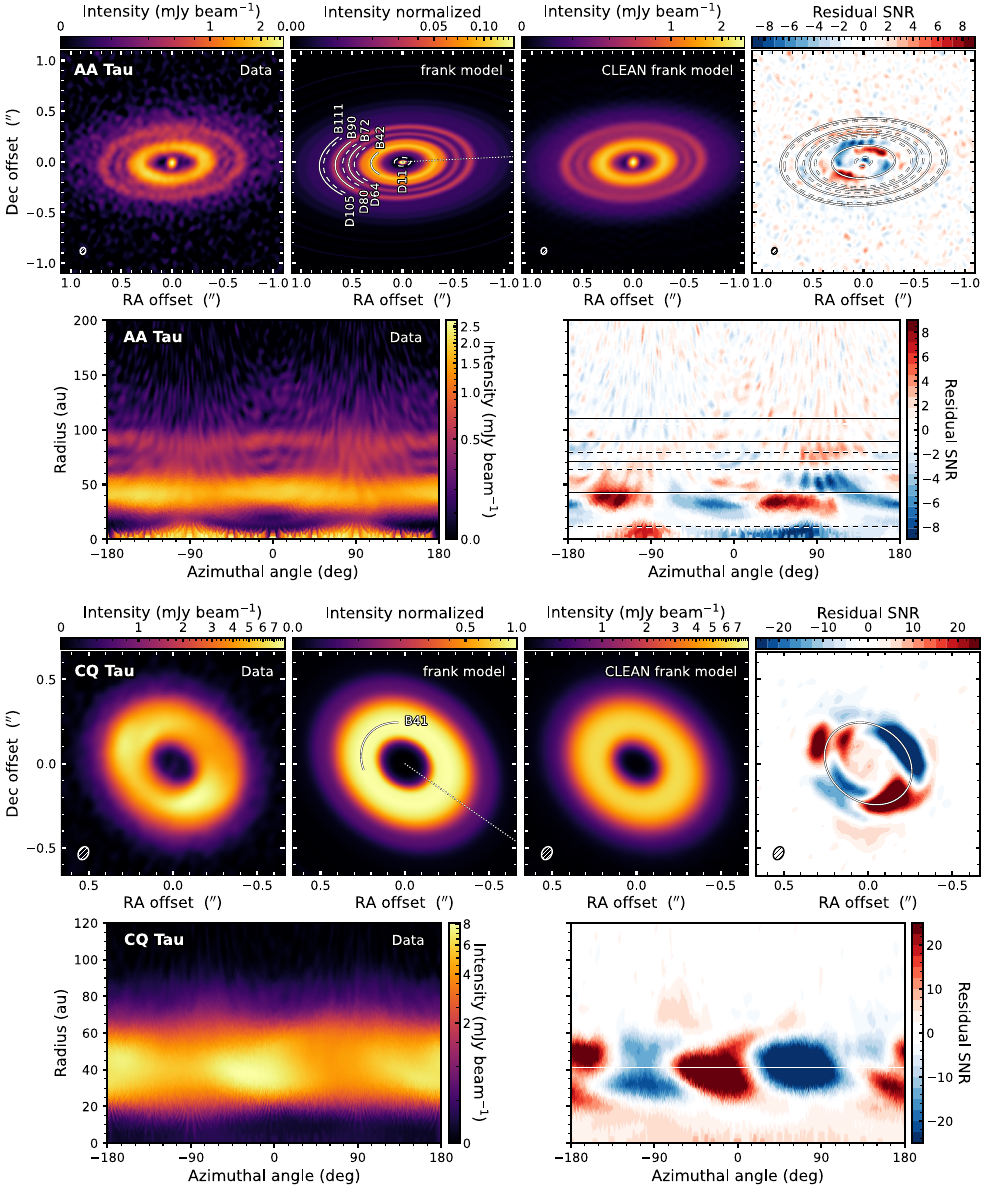}
\caption{Comparison of data, \frank model, CLEAN-imaged \frank model, residuals, and polar plots for each disk (here showing AA~Tau and CQ~Tau and continued in Appendix~\ref{sect:Appendix_Galleries}). (Top to bottom, left to right) First panel: fiducial continuum image of the observed data obtained with robust -0.5, with the synthesized beam’s FWHM shown as an ellipse in the lower left corner. The asinh function was applied to the color scale to visually enhance the fainter emission. Second panel: image of the frank model swept over $2\pi$ and reprojected, with normalized intensity and an asinh stretch. Each gap is marked by a dashed arc labeled “D" with its distance from the star in au, while solid arcs labeled “B"indicate the rings. The dotted line corresponds with the PA measured on gas data with \discminer \citep{Izquierdo_exoALMA} and defines the $\phi=0$ angle of the polar plots, increasing counterclockwise (note possible PA differences with the one measured from \galario and used in \frank, see Appendix~\ref{sect:comparing_gas_dust_geom_param}). Third panel: \frank model sampled at the same $uv$-points of the observation and imaged with CLEAN as the observed data. The color scale is the same as the data panel. Fourth panel: residuals obtained subtracting the \texttt{frank model} from the data. The residual visibilities were calculated at the same $uv$-points of the ALMA observations and imaged with CLEAN as the observed data. The color scale shows the residuals in units of the observed noise ($\sigma_\mathrm{rms}$). Rings and gaps are marked with solid and dashed ellipses, respectively. Fifth panel: polar plot of the data continuum image. Sixth panel: polar plot of the nonaxisymmetric residuals. The locations of rings and gaps are marked by solid and dashed horizontal lines, respectively.}
\label{fig:gallery_single_sources_maintext}
\end{figure*}

We extracted the nonaxisymmetric substructures by computing the residuals between the observed data and the axisymmetric \frank fit. Initially, we sampled the \frank model at the same $uv$-coordinates as the observed data, generating the synthetic visibilities for the fit. Then, we calculated the residual visibilities by subtracting these synthetic visibilities from the corresponding observed ones at each $uv$-location. We imaged the residual visibilities using the CASA \texttt{tclean} algorithm. 

We present in Fig.~\ref{fig:gallery_single_sources_maintext} and in Figs.~\ref{fig:gallery_single_sources_appendix1}, \ref{fig:gallery_single_sources_appendix2}, \ref{fig:gallery_single_sources_appendix3}, \ref{fig:gallery_single_sources_appendix4}, \ref{fig:gallery_single_sources_appendix5}, \ref{fig:gallery_single_sources_appendix6}, \ref{fig:gallery_single_sources_appendix7} in Appendix~\ref{sect:Appendix_Galleries} a gallery for each disk displaying the image of the observed data, the \frank profile swept over $2\pi$, the \frank model imaged with CLEAN, the nonaxisymmetric residuals, and the polar plots of the data and the nonaxisymmetric residuals. The residuals are expressed in terms of the observed rms noise, indicating the nonaxisymmetry signal-to-noise ratio (SNR). Both the data and the residuals are imaged with robust -0.5, which gave the best compromise between angular resolution and SNR \citep{Loomis_exoALMA}. The polar plots were computed by deprojecting and then mapping the intensity distribution onto a radius-azimuthal angle grid. For consistency with the other papers in the exoALMA series, we adopted the same convention used by \discminer for the azimuthal angle in polar plots \citep{Izquierdo_exoALMA}. Specifically, $\phi=0^\circ$ coincides with the PA measured on gas data, corresponding to the direction along the disk's semimajor axis on the redshifted side, with the azimuthal angle increasing counterclockwise. Note that the PA measured on gas data by \discminer may differ from the one measured on continuum data by \galario and employed in the \frank model (see Appendix~\ref{sect:comparing_gas_dust_geom_param}).

\begin{figure*}[t]
\centering
\includegraphics[width=0.78\hsize]{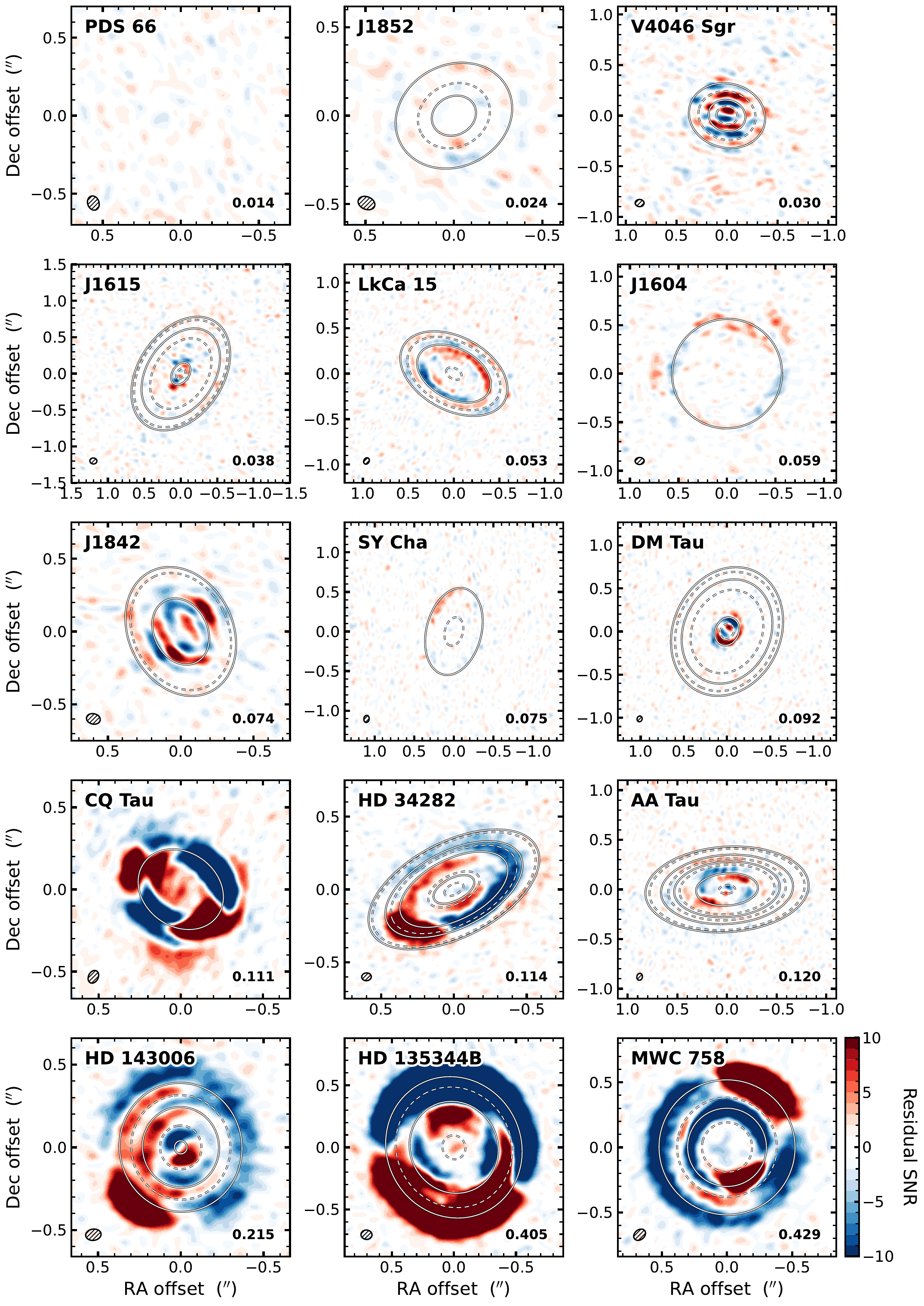}
\caption{Gallery of residuals plots generated subtracting the \frank model (sampled at the same $uv$-points of the observations) from the data and then imaged with CLEAN and robust -0.5. The source order is from the least to the most nonaxisymmetric, according to the NAI presented in Sect.~\ref{subsect:non-axisymm_subs} and reported in the lower right corner of each panel and in Table~\ref{tab:substructures}. The color scale represents the values of residual SNR in units of the observed noise ($\sigma_\mathrm{rms}$) for the respective observation, and the same extremes are applied to each plot. Note that the spatial scales are different for every disk. Rings and gaps are marked with solid and dashed ellipses, respectively, while the synthesized beam is indicated by the ellipse in the lower left corner.}
\label{fig:gallery_residuals}
\end{figure*}

We quantify the level of nonaxisymmetry by evaluating the \frank residuals normalized by the flux in the CLEAN image of the \frank model. We define the nonaxisymmetry index (NAI) as
\begin{equation}
    \mathrm{NAI} = \frac{\sum_{i,j} |I_{\mathrm{res}\, i,j}|}{\sum_{i,j}  | I_{\mathrm{mod}\, i,j}   |} \quad \mathrm{for \; SNR}\geq 5,
\end{equation}
where $I_{\mathrm{res}\, i,j}$ and $I_{\mathrm{mod}\, i,j}$ are the intensity of pixel $i,j$ of the CLEAN images of the \frank residuals and the \frank model. The sums are taken over all pixels within a mask defined by pixels having SNR${\geq}5$ in the CLEAN image of the data. CLEAN images of the data, \frank residuals, and \frank model must be computed with the same \texttt{tclean} parameters, particularly the same pixel size. This index represents a global deviation in flux between observed data and the \frank axisymmetric model. A similar yet distinct approach has been applied to quantify the asymmetries in the gas emission of nearby galaxies by \cite{Davis2022_galaxies} (see their Section~3.1). The NAI values we obtain are provided in Table~\ref{tab:substructures} and each panel of Fig.~\ref{fig:gallery_residuals}, presenting a gallery of \frank residual images with the same SNR scale, ordered by increasing NAI

The disk geometrical parameters, derived from \galario and subsequently employed in the \frank fit, are obtained optimizing a parametric model that assumes fixed values for inclination, PA, and R.A. and decl. offsets for each disk. As \galario minimizes the difference between observed and model intensity at each sampled $uv$-point, the derived geometrical parameters primarily reflect the geometry of the disk region dominating the flux output, namely, the extended disk rather than the inner regions. This is evident in Fig.~\ref{fig:gallery_residuals}, where the most pronounced residuals tend to be concentrated in the inner regions for the majority of sources, leaving larger radii with residuals exhibiting $|\mathrm{SNR}|<3$. 

\section{Discussion} \label{sect:discussion}

\subsection{Source-specific Analysis} \label{subsect:each_source}

In this section, we discuss the substructures observed in the continuum emission of each disk in the exoALMA sample. All features are summarized in Table~\ref{tab:summary_substructures}. For a detailed comparison between the locations of our annular substructures and gradients in the azimuthal velocity deviations from Keplerian rotation, we refer to \citet{Stadler_exoALMA}, who investigate whether the origins of the dust rings are linked to pressure variations. Furthermore, we refer to \citet{Wölfer_exoALMA} for a comprehensive analysis of the dust asymmetries observed in HD~135344B, HD~143006, HD~34282, and MWC~758, comparing them to gas kinematics to explore whether a vortex could be the underlying cause. In addition, we note that the emission from the observed inner disks may originate from nonthermal components, such as free-free or gyrosynchrotron emission of ionized gas in the proximity of the star (see, e.g., \citealt{Rota2024} for HD~135344B and MWC~758, and \citealt{Sierra2025} for LkCa~15).

\hyperref[fig:gallery_single_sources_maintext]{\textit{AA Tau}}. We distinguish three distinct pairs of gaps and rings, along with one fainter outer pair (D105-B111). The nonaxisymmetric features and shadows observed in the first ring (B42) align with the findings of \cite{Loomis2017}, who presented $0.2''$ angular resolution observations. Notably, we detect residuals in the inner disk, possibly indicating a misalignment between the inner disk and the B42 ring. There is no sign of the dust inner streamers proposed by \cite{Loomis2017}. Based on our data and residuals, a plausible explanation involves a misaligned inner disk, casting shadows onto the B42 ring. This results in emission coming from shadowed and geometrically flatter regions, while the brighter areas, receiving more illumination from the central star, exhibit a greater vertical extent.

\hyperref[fig:gallery_single_sources_maintext]{\textit{CQ Tau}}. Prominent spiral-like nonaxisymmetric features are evident, with two on the northeast side and one on the southwest side. It should be noted that in these cases, the \frank model, designed for axisymmetric emission, computes an intermediate intensity between the bright nonaxisymmetric structures and the underlying fainter ring emission. This accounts for the pronounced red-blue patterns observed in the residuals. Our data also reveal a faint inner disk with an integrated intensity of ${\sim}200\,\mu$Jy. This source was previously studied by \cite{UbeiraGabellini2019} with lower resolution $0.15''$ ALMA 1.3~mm (Band~6) observations. By comparing these data to hydrodynamical and radiative transfer simulations, the authors concluded that a a massive planet with a minimum mass of $6-9\,M_\mathrm{Jup}$, located at a distance of $20\,\mathrm{au}$ from the central star, can qualitatively reproduce the continuum intensity radial profile. More recently, \cite{Hammond2022} found prominent spirals in SPHERE scattered-light images, aligning with the nonaxisymmetries in our images, and proposed the presence of an inner companion responsible for inducing such spirals.

\hyperref[fig:gallery_single_sources_appendix1]{\textit{DM Tau}}. The disk of DM~Tau is characterized by a very extended faint emission, reaching $R_{95}=245$~au. Strong residuals are only located in correspondence with the inner disk and the B24 ring. They are the result of the observed inner disk and B24 ring being slightly shifted by ${\sim}25$~mas toward the northwest compared to the center of the extended emission. This offset becomes particularly evident in the polar plots. The center of our axisymmetric model coincides with the center of the extended emission (as proven by the absence of significant residuals beyond the B24 ring), constituting the bulk of the total emission and hence dominating the \galario fit. A possible interpretation of the observed residuals involves eccentricity effects, such as a companion on an eccentric orbit carving the gap D14. \cite{Hashimoto2021} and \cite{Francis2022_DMTau} studied DM~Tau with $0.035''$ resolution 1.3~mm ALMA data. Interestingly, the outer gap-ring pairs (D72-B90 and D102-B111), recovered by \frank in our dataset, become more evident in their higher angular resolution continuum image. Moreover, we confirm the asymmetry on the west side of the B24 ring, interpreted by \cite{Ribas2024} as a dust wall.

\hyperref[fig:gallery_single_sources_appendix1]{\textit{HD~135344B}}. Within our sample, HD~135344B exhibits the second-highest NAI (0.401, see Fig.~\ref{fig:gallery_residuals}). This is caused by the bright arc in the southern region of the disk contrasting with faint emission on the northern side. The \frank fit models this structure as full ring (B78), generating strong positive residuals in the southern region and negative residuals in the northern part. The data polar plot indicates a B51 ring that is not precisely horizontal, suggesting the possibility of either an eccentric ring or an imperfectly retrieved inclination. HD~135344B has been extensively explored with ALMA multiwavelength observations by \cite{Cazzoletti2018}, who found that the asymmetry is consistent with a dust trap where dust growth has occurred. \cite{Casassus2021} presented 1.3~mm observation at a high resolution of $0.03\arcsec$, but with lower surface brightness sensitivity (0.72~K) compared to our data (0.05~K). Their work revealed a tentative detection of a filament connecting the B51 and B78 rings. We identify strong residuals (SNR${>}35$) at the same location, specifically, the red residual aligning with the D66 gap in the southwestern region of the image and at an azimuthal angle of approximately $-15^\circ$ in the residual polar plot.

\hyperref[fig:gallery_single_sources_appendix2]{\textit{HD~143006}}. The continuum emission from this source has been extensively studied as part of the DSHARP large program by \cite{Perez2018}, utilizing $0.046''$ resolution data at 1.3~mm. Even though unresolved in the image and the intensity radial profile, our \frank model manages to retrieve the first ring at 7~au, consistent with the radial location of $6\pm1$~au found by \cite{Huang2018_DSHARPII}. \cite{Perez2018} derived an inclination of $24.1^\circ$ for the inner disk and $17.0^\circ$ for the outer disk, while our 2D \galario model includes a single disk inclination  retrieved at $18.7^\circ$. Our residuals around the inner ring might be an effect of the inner ring misalignment proposed by \cite{Benisty2018} and \cite{Perez2018}.  In addition, we observe a general pattern in our residuals where the eastern side is brighter than the western side, a feature that is also evident in the data image. This closely resembles the pattern observed in the Very Large Telescope (VLT) SPHERE images from \cite{Benisty2018}, which revealed a large-scale shadow on the western side, presumably caused by the warped inner disk. Our observations confirm these findings and further support the hypothesis of a misaligned inner disk. However, our residuals do not fully recover the spiral pattern indicated in the work of \cite{Andrews2021}, possibly due to their different approach, where they excised the large-scale asymmetry before fitting with \texttt{frank} and did not use a parametric fit with \galario to estimate the offsets in R.A. and decl. between the disk center and the observation phase center. Additionally, \cite{Ballabio2021} propose the presence of a strongly inclined binary and an outer planetary companion by comparing their simulations to the morphologies observed in the dust continuum and gas channel maps with ALMA, as well as NIR scattered light with VLT/SPHERE.

\hyperref[fig:gallery_single_sources_appendix2]{\textit{HD~34282}}. Our \frank model identifies a faint inner disk and three gap-ring pairs, which were not resolved in the lower-resolution (${\sim}0.14\arcsec$) Band~7 ALMA observations presented by \cite{vanderPlas2017}. The relevant nonaxisymmetric feature to the southeast of the disk generates the negative residuals as a counterpart due to the axisymmetric nature of the \frank fit. Red residuals along the minor axis may indicate an elevated dust surface, with a morphology consistent with the disk inclination derived from gas data, where the northeast side is the far side \citep{Galloway_exoALMA}.

\hyperref[fig:gallery_single_sources_appendix3]{\textit{J1604}}. We detect the presence of the shadows on the east and west sides of the B82 ring that were previously identified by \cite{Mayama2018} and \cite{Stadler2023} with angular resolutions of ${\sim}0.2''$ at 0.9~mm and ${\sim}0.05''$ at 1.3~mm, respectively, using ALMA observations. The high sensitivity of our data also allows us to reveal, in both the data and the \frank model intensity radial profiles, the presence of a potential new external pair of gap and ring, situated beyond $R_{90}$ and therefore not included in our classification of annular substructures. The gap is located at 139.5~au ($0.965''$) and the ring at 148.1~au ($1.024''$). The external ring in the \frank profile has a peak SNR of ${\sim}14$ with respect to the observed azimuthally averaged noise level at the same radius. We refer to \cite{Yoshida_exoALMA} for a detailed analysis of J1604, including a multiwavelength continuum study and a comparison with the retrieved gas surface density.

\hyperref[fig:gallery_single_sources_appendix3]{\textit{J1615}}. Similarly to DM~Tau, this source presents the inner disk and the first B26 ring slightly shifted by ${\sim}20$~mas to the southeast from the center of the outer disk, producing the visible residuals. Lower-resolution data of J1615 were presented in \cite{vanderMarel2015}, but with our observation, we can resolve a total of three pairs of gaps and rings.

\hyperref[fig:gallery_single_sources_appendix4]{\textit{J1842}}. This disk exhibits two shadows within the emission of the B36 ring (particularly evident in the image with the linear stretch in Fig.~\ref{fig:gallery_images_linear}) and shows clear signs of an elevated dust emission surface. In particular, the emission just inside the B36 ring on the west side of the cavity appears to originate from the inner edge of a vertically extended cavity wall. Gas kinematics data \citep{Galloway_exoALMA} confirm that the west side of the disk corresponds to the far side.  Moreover, the residual pattern, with alternating red and blue residuals along the minor axis, is consistent with the expected residuals obtained by applying a flat model (as \frank does) to an elevated emission surface, as illustrated in Appendix A of \cite{Andrews2021}. However, we note that this interpretation does not align with the pattern proposed by \cite{Ribas2024}. According to their work, an exposed inner cavity would result in a locally brighter emission, which is not observed in J1842. 
A possible explanation for this disagreement could be the presence of an inner disk (not detected in the continuum emission), which might prevent the cavity wall from receiving direct illumination from the central star. Additionally, this inner disk could also be responsible for the observed shadows.
In addition to the gap-ring pair D63-B70, the \frank model retrieves another pair beyond  $R_{90}$. The gap is estimated to be at 98.1~au ($0.650''$) and the ring at 105.6~au ($0.700''$). Differently from J1604, in this case, the substructures are visible only in the \frank profile and not in the azimuthally averaged profile of the CLEAN image. Therefore, we exercise caution regarding the presence of this particular gap-ring pair.

\begin{deluxetable*}{lp{0.85\textwidth}}
\tabletypesize{\footnotesize}
\tablewidth{1\textwidth} 
\tablecaption{Summary of the Observed Substructures in Each Disk.}
\tablehead{
\colhead{Source} & \colhead{Continuum substructures}
}
\startdata 
\hyperref[fig:gallery_single_sources_maintext]{AA Tau} & Four gap-ring pairs and an inner disk. Two shadows in the B42 ring with possible elevated surface in the brighter spots. Possible warped inner disk. \\ 
\hyperref[fig:gallery_single_sources_maintext]{CQ Tau} & Inner cavity and one ring with superimposed nonaxisymmetric spiral-like substructures, two on the northeast side and one on the southwest side.\\
\hyperref[fig:gallery_single_sources_appendix1]{DM Tau} & Inner disk and three pairs of rings and gaps. Offset between the inner disk and the center of the outer disk. Asymmetry on the west side of the B24 ring.  \\
\hyperref[fig:gallery_single_sources_appendix1]{HD 135344B} & Inner cavity with one ring and a bright arc on the southern side. Possible confirmation of the filament connecting the arc with the B51 ring observed by \cite{Casassus2021}. \\
\hyperref[fig:gallery_single_sources_appendix2]{HD 143006} & Inner ring with two pairs of rings and gaps and a bright asymmetry on the southeast side. Eastern side brighter than the western side, consistent with VLT/SPHERE images by \cite{Benisty2018}. Residuals indicate inner disk misalignment, supporting the warped inner disk interpretation by \cite{Benisty2018} and \cite{Perez2018}. \\
\hyperref[fig:gallery_single_sources_appendix2]{HD 34282} & Four gap-ring pairs surrounding an inner cavity and a bright asymmetry on the southeast side. Possible elevated dust surface. \\
\hyperref[fig:gallery_single_sources_appendix3]{J1604} & Single ring with an inner cavity. Shadows on the east and west sides of the ring. Possible external gap-ring pair. \\
\hyperref[fig:gallery_single_sources_appendix3]{J1615} & Three pairs of rings and gaps with a faint inner disk. Offset between the inner disk and the center of the outer disk. \\
\hyperref[fig:gallery_single_sources_appendix4]{J1842} & Inner cavity surrounded by a ring and an additional gap-ring pair. Two shadows in the B36 ring. Signs of an elevated dust surface. Possible external gap-ring pair. \\
\hyperref[fig:gallery_single_sources_appendix4]{J1852} & Bright annular ring surrounding an inner cavity hosting a faint ring. Possible point-source feature within the D31 gap.\\
\hyperref[fig:gallery_single_sources_appendix5]{LkCa 15} &  Two rings surrounding an inner cavity with indications of a third inner ring evident in higher-resolution observations (\citealt{Long2022}, \citealt{Gardner_exoALMA}). Confirmation of the residuals around the Lagrangian points presented by \cite{Long2022}. Residuals along the minor axis indicating an elevated dust surface. Presence of a shoulder in the faint outer emission around 170~au. \\
\hyperref[fig:gallery_single_sources_appendix5]{MWC 758} & Two rings each with a superimposed bright asymmetry. Eccentric inner cavity with a faint inner disk that is offset from the center of the outer disk. \\
\hyperref[fig:gallery_single_sources_appendix6]{PDS 66} &  No clear substructures, only a subtle change in the intensity radial profile slope at 45~au.\\
\hyperref[fig:gallery_single_sources_appendix6]{SY Cha} & Inner disk and one ring with an extended outer shoulder. Bright asymmetry on the northern side of the ring.\\
\hyperref[fig:gallery_single_sources_appendix7]{V4046 Sgr} & Inner disk and two rings, with the outer one having extended external emission. Offset between the inner disk and the center of the outer disk.
\label{tab:summary_substructures}
\enddata
\end{deluxetable*}

\hyperref[fig:gallery_single_sources_appendix4]{\textit{J1852}}. The source is composed by the ring B50 and then both the azimuthally averaged CLEAN intensity radial profile and the \frank model resolve the faint inner ring B19 inside the cavity. Notably, this inner ring was predicted by  \cite{Villenave2019}, who performed a radiative transfer model to match the SPHERE data, spectral energy distribution, and low-resolution ALMA data for this disk. Their model produced a prediction for an ALMA image before convolution presenting a faint inner ring, also suggesting a possible composition of small dust grains with low millimeter opacity. In addition to the faint inner ring, an intriguing feature emerges both in the image data and in the residuals, located at gap D31 on the southern side of the disk. The feature has a significance of ${\sim}5\sigma$ (with $\sigma$ being the rms noise in the image) and is situated adjacent to a negative residual with the same significance, tracing a small region where the observed B50 ring emission contracts compared to the \frank model. Future higher-resolution observations are required to inspect the nature of this feature and determine whether it is genuine or an artifact.

\hyperref[fig:gallery_single_sources_appendix5]{\textit{LkCa~15}}. We recover with a significance of 3$\sigma$-4$\sigma$ the residuals around the Lagrangian points previously studied by \cite{Long2022} using ${\sim}0.05''$ resolution images at 0.9 and 1.3~mm. We also identify pronounced residuals along the minor axis, resembling the residuals presented in \cite{Facchini2020}. This could be indicative of emission coming from a geometrically thick ring (see also \citealt{Huang2020}). Moreover, both the azimuthally averaged CLEAN intensity radial profile and the \frank model present a shoulder in the extended emission at ${\sim}170$~au. For a comprehensive study on the origins of the observed dust and gas substructures in LkCa~15, combining higher-resolution observations and comparing with numerical simulations, see \cite{Gardner_exoALMA}. Moreover, note that at higher resolution, an inner ring (B43) becomes visible, while with our resolution, it does not meet the criteria defining annular substructures (Sect.~\ref{sect:axisymm_subs}).

\hyperref[fig:gallery_single_sources_appendix5]{\textit{MWC~758}}. The disk of MWC~758 has the highest NAI value in our sample (0.429, see Fig.~\ref{fig:gallery_residuals}). The \frank model identifies an inner disk and then two gap-ring pairs. This disk has been extensively studied, e.g., by \cite{Dong2018} with $0.04''$ resolution ALMA observations at 0.9~mm. Their work revealed the eccentricity of the central cavity and indicated that the spirals observed in NIR scattered light \citep{Benisty2015} align with the continuum asymmetries.

\hyperref[fig:gallery_single_sources_appendix6]{\textit{PDS~66}}. This source stands out in the exoALMA sample as the only one that does not exhibit substructures in the continuum emission. All residuals show a significance of less than $4\sigma$. \frank, however, only identifies a subtle change in slope in the intensity radial profile at 45~au. Recently, PDS~66 was analyzed by \cite{Ribas2023} with multiwavelength ALMA observations. Their 1.3~mm observations at $0.05''$ resolution also reveal a smooth disk. Our measured $R_{68}$ and $R_{90}$ align perfectly with the estimates made by \cite{Ribas2023} using data at 1.3 and 2.2mm, indicating a consistent dust continuum extent between 0.9 and 2.2~mm observing wavelengths. This provides additional evidence for optically thick emission at these wavelengths, where gaps, rings, and other substructures would be challenging to detect unless they involve a very large depletion or concentration of material, as the emission would primarily trace the uniform surface of the disk.

\hyperref[fig:gallery_single_sources_appendix6]{\textit{SY~Cha}}. Our data identify an inner disk in the middle of a cavity surrounded by the B101 ring and an extended fainter emission reaching an $R_{95}$ of 228.1~au. The B101 ring shows a nonaxisymmetric feature on its northern side. This structure reflects what was observed by \cite{Orihara2023} using ALMA observations at 1.3~mm at $0.04''$ resolution.

\hyperref[fig:gallery_single_sources_appendix7]{\textit{V4046~Sgr}}. This disk exhibits both an inner disk and the first B13 ring shifted by ${\sim}27$~mas to the north relative to the center of the outer disk, akin to the cases of DM~Tau and J1615. This causes the alternating red-blue residuals, evident also in the residual polar plot. This system hosts a tight binary system \citep{Quast2000} and the gas emission is very smooth \citep{Pinte_exoALMA}. A possible explanation of the observed morphology might be a misalignment of the inner binary, causing the formation of two dust rings as proposed by \cite{Aly2020, Longarini2021}.  However, this is challenging given the system's tight binary configuration, presenting a semimajor axis of ${\approx}0.041$~au (corresponding to an orbital period of 2.42 days) in a circular orbit ($e\le0.01$) and stars with very similar masses ($0.90\pm0.05\,M_\odot$ and $0.85\pm0.05\,M_\odot$, \citealt{Quast2000, Rosenfeld2012}). Additionally, if there were a misalignment, the dynamical mass derived from the disk would be inconsistent with the one derived from the binary orbit, as noted by \cite{Rosenfeld2012}. Another hypothesis could involve an eccentric planet within the D8 gap. No local perturbations are identified in the gas channel maps \citep{Pinte_exoALMA}, but, interestingly, \cite{Stadler_exoALMA} detect a negative gradient in the $^{12}$CO azimuthal velocity deviation from Keplerian rotation colocated with the B13 ring, indicating that the ring just outside the D8 gap is consistent with a dust trap. Higher-resolution continuum observations of V4046~Sgr at 1.3mm, as presented by \cite{MartinezBrunner2022} and \cite{Weber2022}, reveal similar structures, although the offset of the inner disk is less pronounced.

\subsection{Inner-Outer Disk Connection} \label{subsect:correlations_dust_star}

\begin{figure*}[t]
\centering
\includegraphics[width=1\hsize]{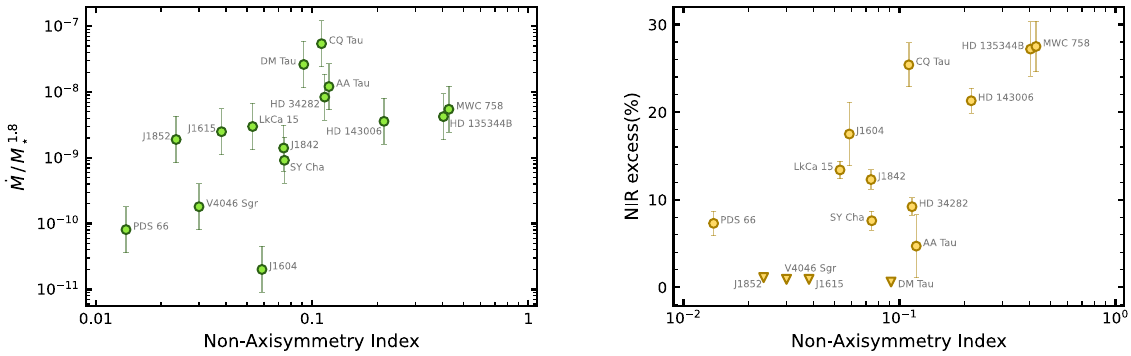}
\caption{Mass accretion rate and NIR excess as function of the NAI (higher values indicating more asymmetric disks). Left plot: log-log plot of the mass accretion rate, normalized for the correlation with the stellar mass assuming $\dot{M} \propto M_\star^{1.8}$, \citep{Manara2023}, as a function of the NAI (defined in Sect.~\ref{subsect:non-axisymm_subs}). Right panel: lin-log plot of NIR excess vs. the NAI. Downward-pointing triangles represent upper limits. Values of the NIR excess are reported in Table~\ref{tab:Macc_NIR}.
}
\label{fig:Macc_NIR_vs_NAI}
\end{figure*}


The definition of the NAI (see Sect.~\ref{subsect:non-axisymm_subs}) is valuable for quantifying the morphological characteristics of each disk and investigating potential explanations by identifying patterns with other properties of the disk and its central star. The left plot of Fig~\ref{fig:Macc_NIR_vs_NAI} presents the mass accretion rate versus the NAI. The mass accretion rate $\dot{M}$ scales with the stellar mass $M_\star$ following a steeper-than-linear relation $\dot{M}\propto M_\star^\gamma$ with $\gamma \sim 1.6-2$ (compilation by \citealt{Manara2023}). To homogeneously compare the mass accretion rates across the 15 disks in our sample, we normalized the mass accretion rate to account for its dependence on the stellar mass by considering $\dot{M} / M_\star^{1.8}$, assuming $\gamma = 1.8$. This value is plotted on the y-axis of the left panel in Fig~\ref{fig:Macc_NIR_vs_NAI}. 

The relation between the measured NIR excess of each disk and the corresponding NAI is presented in the right plot of Fig.~\ref{fig:Macc_NIR_vs_NAI}, where the NIR excess quantifies the excess flux in the NIR above the stellar photosphere, typically tracing hot dust in the inner disk. For some disks, we report the NIR excess from \cite{Garufi2018}. For sources not included in that work, the NIR excess was calculated following the same procedure, namely, integrating the dereddened flux measured by the Two Micron All-Sky Survey (2MASS) and Wide-field Infrared Survey Explorer (WISE) photometry from $1.2\,\mu\mathrm{m}$ to $4.5\,\mu\mathrm{m}$ in excess over a Phoenix model of the stellar photosphere \citep{Hauschildt1999} with the $T_\mathrm{eff}$ of the specific source. The final value is then normalized to the total stellar flux. Values of the mass accretion rate and NIR excess are reported in Appendix~\ref{sect:Macc_NIR} and Table~\ref{tab:Macc_NIR}. 

The Kendall’s tau coefficient for the relation between the mass accretion rate (normalized for stellar mass dependence) and the NAI is 0.45 with a p-value of 0.02, while for the relation between NIR excess and the NAI, it is 0.48 with a p-value of 0.01. Both indicate moderate, statistically significant positive correlations. We verified that assuming an exponent of 1.6 or 2 for the normalization of the mass accretion rate to stellar mass yields minimal differences, as well as completely omitting the normalization to stellar mass (see Fig.~\ref{fig:Macc_NOnorm_vs_NAI}). Figure~\ref{fig:StellaMass_and_dust_mass_vs_NAI} in the Appendix shows the correlations between stellar mass from \discminer \citep{Izquierdo_exoALMA} and dust disk mass, calculated as explained in Sect.~\ref{sect:data}, with  NAI . A weak correlation between dust disk mass and  NAI  is observed (Kendall’s tau coefficient of 0.31 with a p-value of 0.11). In contrast, for stellar mass and  NAI, while the Kendall’s tau test suggests no significant correlation (0.10 with a p-value of 0.62), the plot interestingly shows that the most asymmetric sources are also the ones with higher stellar masses. While our findings are robust within the exoALMA sample, it is important to note that the sample selection may introduce biases that influence these results, as it primarily targets bright, extended disks with significant substructures. Future studies with a more diverse sample could help confirm these trends.

Each plot clearly shows that the most asymmetric sources also exhibit the highest values of accretion rate and NIR excess. Interestingly, of the six most nonaxisymmetric disks in our sample (CQ~Tau, HD~34282, AA~Tau, HD~143006, HD~135344B, MWC~758, all with $NAI>0.1$), five exhibit inner cavities. \cite{Garufi2018} analyzed a substantial NIR dataset on protoplanetary disks, concluding that the presence of spirals and shadows is associated with a high NIR excess. Our results align with their findings, as NIR spirals often coincide with strong nonaxisymmetric features in the millimeter dust continuum emission. A plausible explanation for this involves a massive perturber generating the NIR spirals, such as a stellar or planetary companion within the observed cavities in our most asymmetric sources, potentially triggering higher mass accretion onto the central star. Future theoretical and numerical work should explore this possibility in greater detail.

\subsection{Presence of Companions}
\label{subsect:companion}

\begin{figure*}[t]
\centering
\includegraphics[width=0.86\hsize]{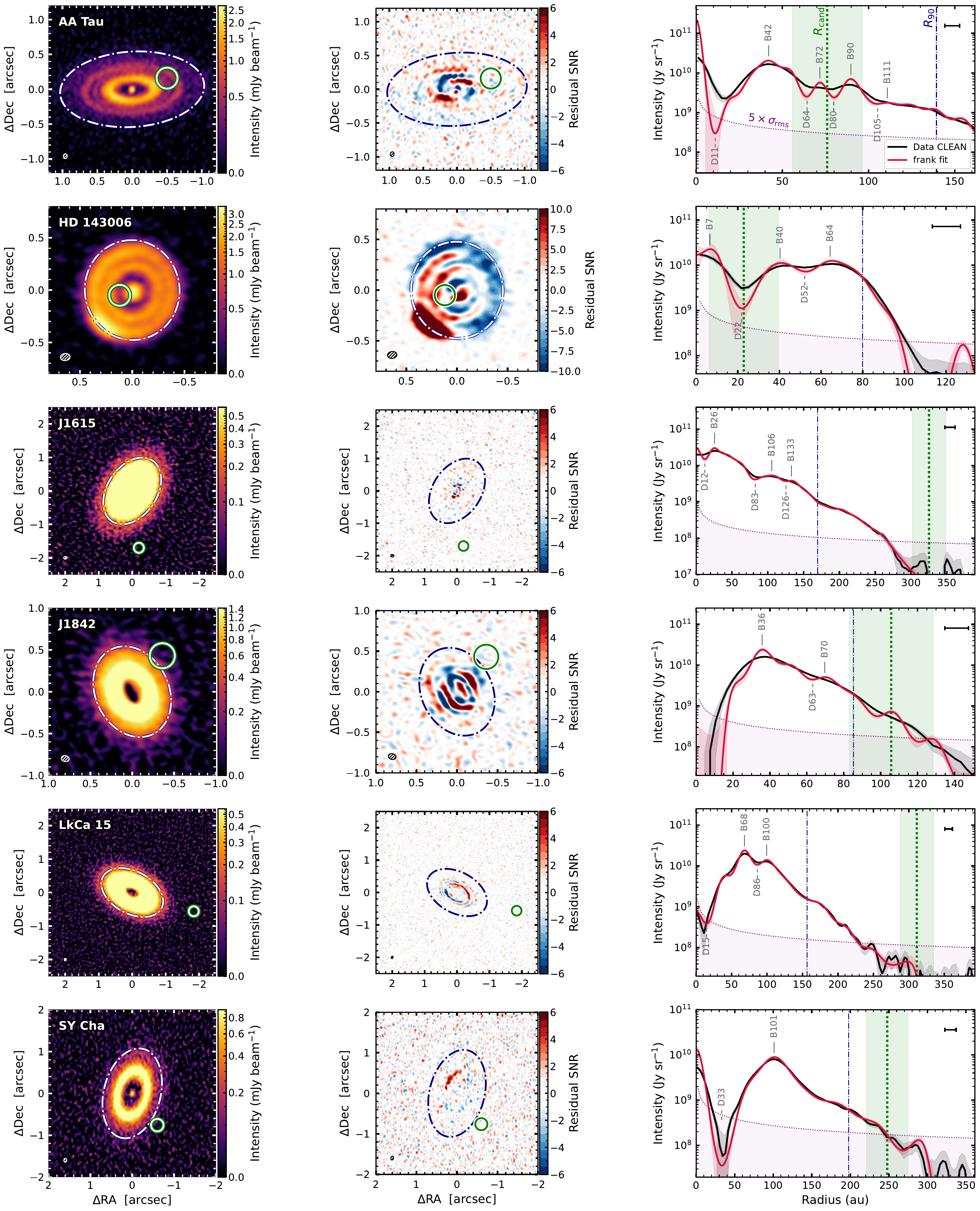}
\caption{Comparison between the continuum emission and the $^{12}$CO velocity kinks identified by \cite{Pinte_exoALMA}. The left panels show the continuum emission from the fiducial CLEAN data images, and the middle panels display the \frank residuals. Dashed-dotted ellipses represent $R_{90}$, while solid circles indicate the positions of the velocity kinks, deprojected onto the midplane. Ellipses in the lower left corner indicate the synthesized beam. The right panels show the intensity radial profile of the azimuthally averaged fiducial CLEAN images (black) and the \frank model (red). Vertical blue dashed-dotted lines represent $R_{90}$, while the vertical green dotted lines indicate the radial location deprojected onto the midplane of the planet candidates generating the kinks ($R_\mathrm{cand}$). The green shaded area in the intensity profile and the size of the green circles in the 2D images represent the uncertainty due to the gas image beam size \citep{Pinte_exoALMA}.
The purple shaded area represents intensity values below 5 times the rms noise ($\sigma_\mathrm{rms}$) measured in the CLEAN image.}
\label{fig:companions}
\end{figure*}

In this section, we explore the possibility of companions, either stellar or planetary, in the disks of our sample. This analysis is first carried out by comparing the observed continuum substructures with numerical simulations from the literature, followed by assessing correspondences with gas kinematic signatures from exoALMA $^{12}$CO observations presented by \cite{Pinte_exoALMA}. Our findings suggest that while massive companions could explain some observed substructures, such as central cavities and major asymmetries,  no clear evidence of direct companion emission is found in the continuum data. The comparison with gas kinematic data yields mixed conclusions, with some continuum substructures aligning with kinematic signatures, while others do not provide conclusive evidence.

In Sect.~\ref{subsect:correlations_dust_star}, we propose that massive companions (either stellar or planetary) may plausibly explain why the most asymmetric disks with higher NAI values tend to have a central inner cavity and higher mass accretion rates and NIR excess. This interpretation is supported by the work of \cite{Calcino2023}, who define criteria linking gas kinematic asymmetries and central cavities to the presence of inner binaries.

Regarding planetary companions, \cite{Speedie2022} used synthetic ALMA Band~7 observations from hydrodynamic and radiative transfer simulations to show how thermal mass planets at tens of astronomical units can drive spirals in the dust continuum emission, which are effectively highlighted in residual maps. However, they note that gaps and rings can obscure spirals by limiting the disk area where spirals are visible. Furthermore, \cite{Sturm2020} demonstrated that planet-induced spirals in the dust are significantly weaker than those in the gas, with the amplitude of the dust spirals decreasing with higher Stokes numbers. 

In our sample, apart from the spiral-like asymmetries in CQ~Tau, no disks exhibit clear full spirals in the dust continuum. While crescent-shaped features are observed and known to sometimes coincide with spirals in the NIR (see Fig.~1 in \citealt{Wölfer_exoALMA}), our residuals show no unambiguous spiral structures. The only spiral-like feature in the \frank residuals (Fig.~\ref{fig:gallery_residuals}) is seen in DM~Tau, but as detailed in Sect.\ref{subsect:each_source}, we interpret this as an artifact caused by the offset between the inner disk and the first bright ring relative to the outer disk center. This lack of clear spiral structures prevents us from inferring planetary companions solely from continuum morphology.

Additional insights come from comparing the continuum substructures to the work of \cite{Pinte_exoALMA}, who analyzed $^{12}$CO data cubes from exoALMA and identified six disks with kinematic signatures consistent with planet wakes: AA~Tau, HD~143006, J1615, J1842, LkCa15, and SY~Cha. The kink locations deprojected to the midplane is compared with our continuum morphologies in Fig.~\ref{fig:companions}. Given the difficulties in assigning a robust uncertainty to the kink location, we estimate the uncertainty using the gas image beam size \citep{Pinte_exoALMA}. Specifically, in the data images and residuals, the kink locations are marked with circles centered at the deprojected kink positions, with a radius equal to the gas beam size. In the intensity radial profiles, the radial locations of the planet candidates generating the kinks ($R_{\mathrm{cand}}$) are indicated with green dashed lines, while the associated uncertainties are represented by green shaded areas spanning  $R_{\mathrm{cand}} \pm$  one gas beam size. We found no evidence of direct emission that could be interpreted as coming from a companion, either in the fiducial images or in the \frank residuals. However, valuable observations can be made by comparing the candidate locations to the substructures in the intensity radial profiles. 

For AA~Tau, the kink aligns with the D80 gap, further supporting the hypothesis of a planetary companion carving this gap. In HD~143006, they detected hints of the same kink observed in DSHARP data, potentially explained by a giant planet located within the continuum D22 gap \citep{Perez2018,Pinte2020, Ballabio2021}. For J1615 and LkCa15, the kinematic candidates are situated at distances where both the azimuthally averaged CLEAN image and the \frank profile fall below the noise level. However, as noted by \cite{Pinte_exoALMA}, it is interesting to observe that the locations of these candidates lie just outside the region where the dust emission drops, potentially hinting that these candidates could be truncating the disk. In J1842, the proposed kink is located beyond $R_{90}$ but still has SNR${>}5$ in the azimuthally averaged intensity radial profile. In this region, the \frank profile reveals substructures not visible in the azimuthally averaged intensity radial profile from the CLEAN image. In SY~Cha, the candidate location corresponds to a region beyond $R_{90}$ where both the azimuthally averaged CLEAN profile and the \frank\ model identify a gap at ${\sim}270$~au, followed by a slight increase in intensity and a sharp drop at ${\sim}300$~au. This outer gap-ring pair has an SNR ranging from 2 to 5. Despite the low SNR, the correspondence between the candidate location and these outer substructures could suggest a candidate carving a gap and creating a faint ring in the disk outer region.

Finally, we aim to provide a flux density upper limit for the undetected circumplanetary disks (CPDs). For AA~Tau and HD~143006, where the kink location corresponds to a dust gap, a statistical approach would be necessary, with an injection-recovery test to characterize CPDs in residual images, as done in \cite{Andrews2021}. This is because pixels in the gaps are highly correlated, and nonaxisymmetric residuals can still influence the estimate. This would be best approached with a dedicated study that can invest more effort into asymmetric models of the circumstellar material. However, we can straightforwardly provide a flux density upper limit for the remaining four disks, where the kink location is beyond  $R_{90}$. The ($3\sigma$) upper limit on the emission is derived as 3 times the rms measured in the radial range $[R_{\mathrm{cand}} - \mathrm{gas\,beam\,size}, R_{\mathrm{cand}} + \mathrm{gas\,beam\,size}]$, that is, from an aperture centered on the putative companion location, with a width accounting for the gas beam size. The computed flux density upper limits are 111~$\mu$Jy for J1615, 312~$\mu$Jy for J1842, 105~$\mu$Jy for LkCa~15, and 165~$\mu$Jy for SY~Cha.


\subsection{Analysis of the Extended Emission}

\begin{figure*}[t]
\centering
\includegraphics[width=0.97\hsize]{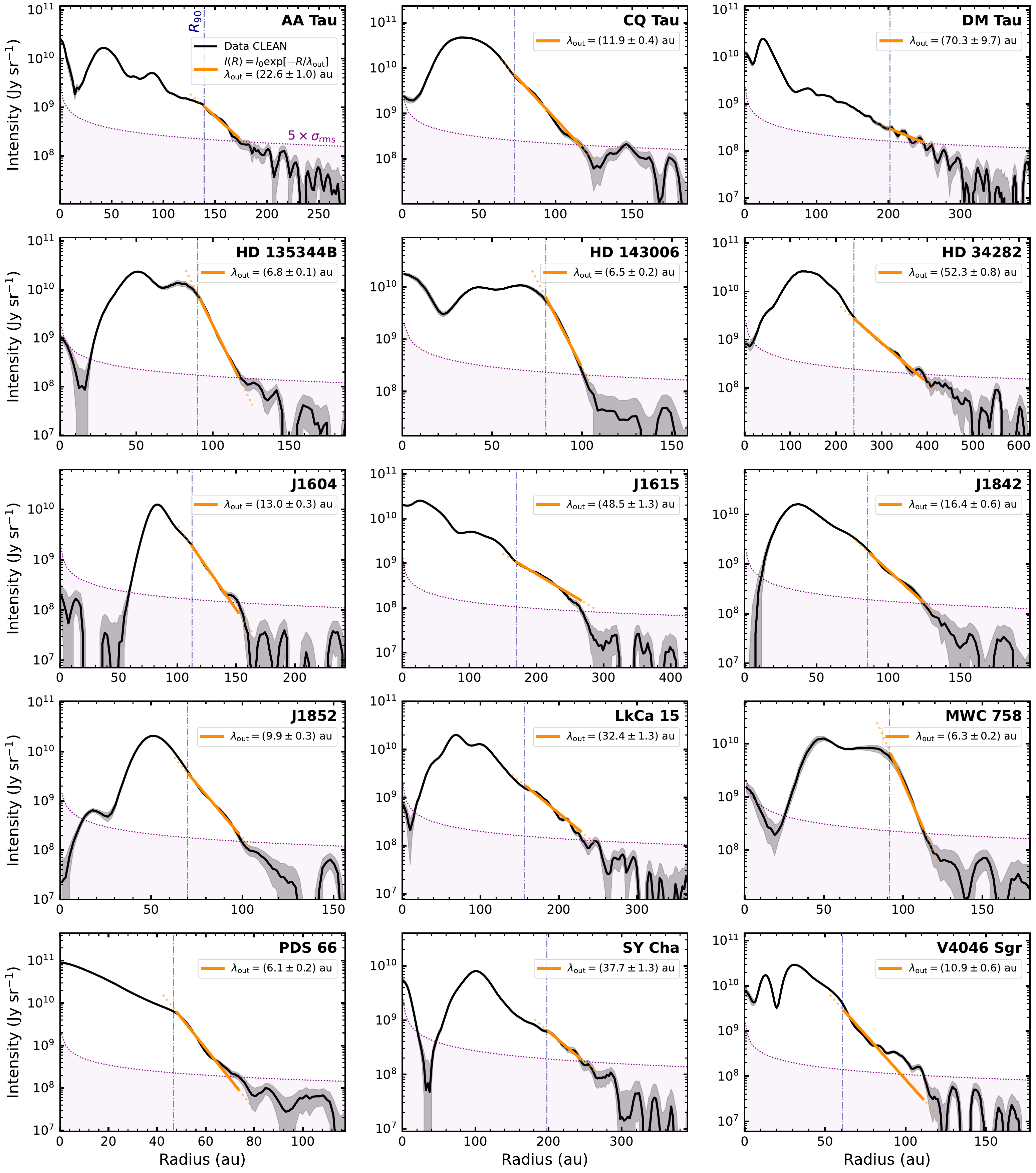}
\caption{Gallery showing the fits of the continuum extended emission with an exponential function in a log-lin scale. The intensity radial profiles are from the azimuthally averaged CLEAN images with robust -0.5. The blue vertical dashed-dotted line indicates $R_{90}$, while the purple shaded area represents intensity values below 5 times the rms noise $\sigma_\mathrm{rms}$ measured in the CLEAN image. The rms noise is properly scaled accounting for the radial dependence of the azimuthal average, that is, dividing the $\sigma_\mathrm{rms}$ by the square root of the number of beams within the corresponding radial annulus. The orange line shows the best fit using the exponential model $I(R) = I_0 \exp[- R/\lambda_\mathrm{out}]$ of the region between $R_{90}$ and the radius where the intensity intersects the $5\times\sigma_\mathrm{rms}$ line.}
\label{fig:gallery_extended_emission}
\end{figure*}

\begin{figure*}[t]
\centering
\includegraphics[width=1\textwidth]{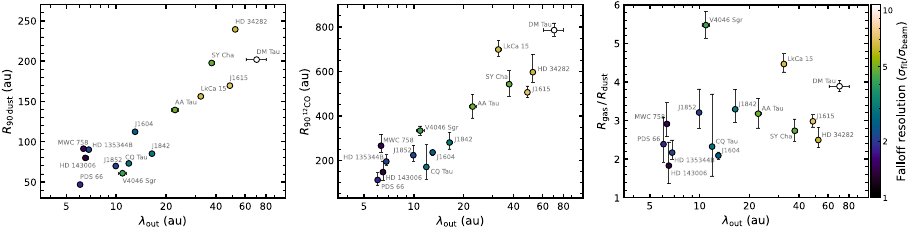}
\caption{From left to right: radius enclosing $90\%$ of the continuum emission ($R_{90\,\mathrm{dust}}$), radius enclosing $90\%$ of the $^{12}$CO emission ($R_{90\,^{12}\mathrm{CO}}$, \citealt{Galloway_exoALMA}), and their ratio ($R_\mathrm{gas}/R_\mathrm{dust}$)  as a function of the parameter $\lambda_\mathrm{out}$ from the exponential model of the continuum extended emission $I(R) = I_0 \exp[-R/\lambda_\mathrm{out}]$. Uncertainties on $R_{90\,\mathrm{dust}}$ are present but smaller than the data points. The color scale indicates the number of beams resolving the outer disk falloff, with values of $\sigma_\mathrm{fit}/\sigma_\mathrm{beam}$ (Table~\ref{tab:ext_emission}) displayed on a logarithmic scale.}
\label{fig:radii_vs_lambdaout}
\end{figure*}

So far, we characterized the dust emission within $R_{90}$ (see the criteria for defining axisymmetric substructures in Sect.~\ref{sect:axisymm_subs}). However, the high surface brightness sensitivity of our dataset also allows us to inspect the faint outer disk continuum emission.  As seen in the intensity radial profiles in Fig.\ref{fig:gallery_intensity_profiles},  there is a region beyond $R_{90}$ with clear signal before noise becomes dominant at even larger radii. In this region, the azimuthally averaged profile from the CLEAN images and the frank fit are usually in good agreement. These are areas where we detect a reliable signal that is not visible in the CLEAN images but is revealed in the profiles due to the azimuthal average boosting the local SNR. 

We also note that in our data, the continuum flux density in these outer regions is robustly recovered without the maximum recoverable scale being a limiting factor. This is because exoALMA was primarily designed to study gas emission, which extends beyond the dust continuum emission (see column (4) in Table~\ref{tab:ext_emission}). To capture large-scale structures, the observations combine a compact ALMA configuration and, for the most extended sources, also include the ACA.

To quantitatively characterize this continuum's outer regions, considering only the azimuthally averaged CLEAN profile, we focus on the radius range beyond $R_{90}$ and out to where the intensity is above 5 times the rms noise. Within this radial range, the intensity profiles appear generally linear in a log-lin plot. Therefore, we fitted these regions with an exponential function:
\begin{equation}
    I(R) = I_0 \,\exp{\left[-\frac{R}{\lambda_\mathrm{out}}\right]}.
\end{equation}
The parameter $\lambda_\mathrm{out}$ represents the scale length of the outer disk emission taper, with higher values indicating a flatter outer disk. Figure~\ref{fig:gallery_extended_emission} presents a gallery with the results of these fits for each exoALMA source. We observe that the exponential function well reproduces the overall intensity profile trend in the outer regions for most of the disks. It only partially fails in the case of V4046~Sgr, which does not exhibit a single slope.

To determine whether we have sufficient angular resolution to accurately resolve the steepness of the extended emission, we first fitted the same radial range with a Gaussian model centered on $R_{90}$, i.e.,  
\begin{equation}
    I(R) = A \,\exp{\left[-\frac{(x-R_{90})^2}{2\sigma_\mathrm{fit}^2}\right]}.
\end{equation}
Next, we divided $\sigma_\mathrm{fit}$ by  $\sigma_\mathrm{beam}$, which is the average of the $\sigma$ values of the major and minor axes of the synthesized beam. Values of $\gamma$ and $\sigma_\mathrm{fit}/\sigma_\mathrm{beam}$ are listed in Table~\ref{tab:ext_emission}. We consider the descent to be resolved if $\sigma_\mathrm{fit}/\sigma_\mathrm{beam}>2$. Thus, the steepness of HD~143006, MWC~758, and PDS~66 is not resolved.

\begin{deluxetable}{lccc}
	\tabletypesize{\footnotesize}
	\tablewidth{1\textwidth} 
	\tablecaption{Continuum Outer Disk Fit}
	\tablehead{
		\colhead{Source} & 
		\colhead{$\lambda_\mathrm{out}$} &  \colhead{$\sigma_\mathrm{fit}\,/\,\sigma_\mathrm{beam}$} &
        \colhead{$R_\mathrm{gas}\,/\,R_{dust}$} \\
        \colhead{} & 
		\colhead{(au)} &  \colhead{} &
        \colhead{}
		} 
	\colnumbers
	\startdata 
AA Tau     &    $22.6\pm1.0$   &    4.9  &  $3.18_{-0.37}^{+0.40}$ \\
CQ Tau     &    $11.9\pm0.4$   &   2.9 &  $2.33_{-0.76}^{+1.37}$\\
DM Tau     &    $70.3\pm9.7$   &  10.7  &  $3.89_{-0.14}^{+0.16}$ \\
HD 135344B     &    $6.8\pm0.1$   &  2.1  &  $2.17_{-0.18}^{+0.33}$\\
HD 143006     &    $6.5\pm0.2$   &   1.4  & $1.83_{-0.46}^{+0.62}$\\
HD 34282    &    $52.3\pm0.8$   &   6.3    & $2.50_{-0.20}^{+0.34}$\\
J1604    &    $13.0\pm0.3$   &   2.5   &  $2.09_{-0.09}^{+0.10}$\\
J1615     &    $48.5\pm1.3$   &   7.1   & $2.98_{-0.14}^{+0.18}$\\
J1842   &   $16.4\pm0.6$   &   3.0    &  $3.29_{-0.34}^{+0.52}$\\
J1852   &    $9.9\pm0.3$   &    2.1  & $3.21_{-0.40}^{+0.61}$\\
LkCa 15    &    $32.4\pm1.3$   &   6.6   &  $4.47_{-0.20}^{+0.27}$ \\
MWC 758    &    $6.3\pm0.2$   &   1.5   &  $2.91_{-0.33}^{+0.55}$\\
PDS 66     &    $6.1\pm0.2$   &   1.9  &  $2.39_{-0.47}^{+0.70}$  \\
SY Cha     &    $37.7\pm1.3$    &   4.8  & $2.74_{-0.28}^{+0.31}$  \\
V4046 Sgr     &    $10.9\pm0.6$   &  4.2  &  $5.47_{-0.35}^{+0.36}$  \\
	\enddata
	\tablecomments{Column (1): target name. Column (2): scale length of the outer disk taper $\lambda_\mathrm{out}$ from the exponential model $I(R) = I_0 \exp[-R/\lambda_\mathrm{out}]$. Column (3): ratio between the $\sigma_\mathrm{fit}$ from the Gaussian model $I(R)=A\exp[-(x-R_{90})^2/2\sigma_\mathrm{fit}^2]$ and the $\sigma_\mathrm{beam}$ obtained by averaging the major and minor axis $\sigma$ values of the synthesized beam. Column 4: ratio between $R_{90}$ from $^{12}$CO \citep{Galloway_exoALMA} and from the dust continuum (Table~\ref{tab:disk_geometrical_params}).}
 \label{tab:ext_emission}
\end{deluxetable}

In Fig.~\ref{fig:radii_vs_lambdaout}, we present $R_{90}$ from the continuum emission, the $^{12}$CO emission, and their ratio as a function of the parameter $\lambda_\mathrm{out}$ from the exponential model. The color scale represents the number of beams resolving the outer disk falloff, showing values of $\sigma_\mathrm{fit}/\sigma_\mathrm{beam}$ on a logarithmic scale. These plots illustrate that larger disks (both in dust and in gas) within our sample systematically have a shallower slope in the falloff of their outer regions, whereas more compact disks exhibit a steeper outer edge. The steepness may be even greater for sources whose outer descent we are not fully resolving. On the other hand, there is no correlation between the ratio of the gas and dust radii and the $\lambda_\mathrm{out}$ parameter. Therefore, even though dust radial drift could play an important role in shaping the outer disk continuum falloff, it is not straightforward to establish this connection with the current findings. Other phenomena that might directly influence the appearance of the faint outer disk emission include late infall events or disk truncation caused by flybys or outer companions. Additionally, it is worth noting that a metric like the one presented here has not been included in theoretical studies examining disk size in dust with and without substructures (e.g., \citealt{Rosotti2019, Zormpas2022, Delussu2024}). Future modeling efforts could benefit from exploring this metric in more detail.




\section{Conclusions} \label{sect:conlusion}

In this paper, we analyzed the continuum emission from the ALMA Band~7 data of the 15 disks in the exoALMA Large Program. In the quest to understand the origin of the observed dust morphologies,  we characterized both the axisymmetric and nonaxisymmetric substructures, as well as the bright inner regions and faint outer regions of each disk. 

We developed a pipeline focused on visibility fitting to characterize axisymmetric substructures (rings and gaps) and nonaxisymmetric residuals obtained by subtracting an axisymmetric model from the data.

\begin{enumerate}
    \item Our procedure begins with a parametric fit using the code \galario \citep{Tazzari2018_galario} to retrieve solid estimates of the geometrical parameters (inclination, PA, offsets in R.A. and decl.). These parameters are then employed in a nonparametric fit with the package \frank \citep{Jennings2020_frank}, resulting in a superresolution 1D best-fit model of the radial intensity profile.
    \item We use the \frank model intensity profile to define the radial location, width, and depth of rings and gaps, limiting this characterization within $R_{90}$. Next, we use the same axisymmetric model to extract nonaxisymmetric residuals from the data. We define the NAI as a measure to quantify the level of asymmetry for each disk.
\end{enumerate}

Our main findings are summarized below. It should be noted that these results have been obtained from a biased sample of large and bright disks. Future, more complete surveys will be essential to determine whether these findings hold for the broader population of protoplanetary disks.
\begin{enumerate}
    \item The data angular resolution and sensitivity allowed us to retrieve specific features for the various disks. These include prominent shadows (AA~Tau, HD 143006, J1604, J1842), inner disks offset with respect to the outer disk center (DM~Tau, J1615, V4046~Sgr), possible warped inner disks (AA~Tau, HD~143006) indications of a dust wall (J1842) and of a geometrically thick disk (AA~Tau, HD~34282, LkCa~15), potential external substructures (an outer ring in J1604 and J1842, and an outer shoulder in LkCa~15), and a seemingly structureless disk (PDS~66). 
    \item Except for PDS~66, all other disks in our sample exhibit some form of nonaxisymmetric features. Only CQ~Tau hosts clear spiral-like structures, while five disks (HD~135344B, HD~143006, HD~34282, MWC~758, and SY~Cha) show crescent-shaped asymmetries. The remaining eight disks present other types of irregularities. This suggests that, given sufficient angular resolution and sensitivity, nonaxisymmetries may be a common characteristic of protoplanetary disks.
    \item In our attempt to gain a deeper understanding of the origin of the observed strong asymmetries, we found tentative correlations between the NAI and mass accretion rate and NIR excess. Notably, the more asymmetric disks almost all feature inner cavities and consistently exhibit higher values of these parameters. This finding suggests a connection between the outer disk structures and the inner disk properties. 
    \item Capitalizing on the high surface brightness sensitivity of our data, we provided a preliminary characterization of the continuum extended emission. This outer emission can generally be reproduced with an exponential fit. We found that larger disks exhibit a shallower falloff in the outer regions, while more compact disks present a sharper outer edge.
\end{enumerate}

The data and disk parameters presented in this paper are provided as a publicly available value-added data product. These include CLEAN images of the continuum data and the residuals from the \frank fit at different robust values, intensity radial profiles from the fiducial CLEAN images and the \frank fits, radial locations of gaps and rings, geometrical parameters from \galario ($i$, PA, $\Delta$R.A., $\Delta$decl.), and values of the continuum radii ($R_{68}$, $R_{90}$, $R_{95}$).

\section*{Acknowledgments}
The authors thank the anonymous referee for the thorough and detailed review, which greatly helped improve the manuscript. This paper makes use of the following ALMA data: ADS/JAO.ALMA\#2021.1.01123.L. ALMA is a partnership of ESO (representing its member states), NSF (USA) and NINS (Japan), together with NRC (Canada), MOST and ASIAA (Taiwan), and KASI (Republic of Korea), in cooperation with the Republic of Chile. The Joint ALMA Observatory is operated by ESO, AUI/NRAO and NAOJ. The National Radio Astronomy Observatory is a facility of the National Science Foundation operated under cooperative agreement by Associated Universities, Inc. We thank the North American ALMA Science Center (NAASC) for their generous support including providing computing facilities and financial support for student attendance at workshops and publications. 

P.C. thanks Antonio Garufi for providing the NIR excess values for the exoALMA sources and Laura Pérez, Anibal Sierra, Enrique Macías, Francesco Zagaria, and María Jesús Mellado for helpful discussions. 

P.C. acknowledges support by the Italian Ministero dell'Istruzione, Universit\`a e Ricerca through the grant Progetti Premiali 2012 – iALMA (CUP C52I13000140001) and by the ANID BASAL project FB210003. S.F. is funded by the European Union (ERC, UNVEIL, 101076613), and acknowledges financial contribution from PRIN-MUR 2022YP5ACE. J.B. acknowledges support from NASA XRP grant No. 80NSSC23K1312. M.B., D.F., J.S., and A.W. have received funding from the European Research Council (ERC) under the European Union’s Horizon 2020 research and innovation programme (PROTOPLANETS, grant agreement No. 101002188). Computations by J.S. have been performed on the `Mesocentre SIGAMM' machine, hosted by Observatoire de la Cote d’Azur. M.F. has received funding from the European Research Council (ERC) under the European Unions Horizon 2020 research and innovation program (grant agreement No. 757957). M.F. is supported by a Grant-in-Aid from the Japan Society for the Promotion of Science (KAKENHI: No. JP22H01274). C.H. acknowledges support from NSF AAG grant No. 2407679. J.D.I. acknowledges support from an STFC Ernest Rutherford Fellowship (ST/W004119/1) and a University Academic Fellowship from the University of Leeds. A.I. acknowledges support from the National Aeronautics and Space Administration under grant No. 80NSSC18K0828. Support for A.F.I. was provided by NASA through the NASA Hubble Fellowship grant No. HST-HF2-51532.001-A awarded by the Space Telescope Science Institute, which is operated by the Association of Universities for Research in Astronomy, Inc., for NASA, under contract NAS5-26555. G.L. has received funding from the European Research Council (ERC) under the European Union Horizon 2020 research and innovation program (Grant agreement no. 815559 (MHDiscs)). G.L. and C.L. have received funding from the European Union's Horizon 2020 research and innovation program under the Marie Sklodowska-Curie grant agreement No. 823823 (DUSTBUSTERS). C.L. acknowledges support from the UK Science and Technology research Council (STFC) via the consolidated grant ST/W000997/1. C.P. acknowledges Australian Research Council funding via FT170100040, DP18010423, DP220103767, and DP240103290. D.P. acknowledges Australian Research Council funding via DP18010423, DP220103767, and DP240103290. G.R. acknowledges funding from the Fondazione Cariplo, grant no. 2022-1217, and the European Research Council (ERC) under the European Union’s Horizon Europe Research \& Innovation Programme under grant agreement no. 101039651 (DiscEvol). F.M. received funding from the European Research Council (ERC) under the European Union’s Horizon Europe research and innovation program (grant agreement No. 101053020, project Dust2Planets). N.C. has received funding from the European Research Council (ERC) under the European Union Horizon Europe research and innovation program (grant agreement No. 101042275, project Stellar-MADE). L.T. acknowledges funding from Progetti Premiali 2012 iALMA (CUP C52I13000140001), Deutsche Forschungs-gemeinschaft (German Research Foundation) ref no. 325594231 FOR 2634/1 TE 1024/1-1, European Union’s Horizon 2020 research and innovation programme under the Marie Sklodowska-Curie grant no. 823823 (DUSTBUSTERS) and the ERC via the ERC Synergy Grant ECOGAL (grant no. 855130). T.C.Y. acknowledges support by Grant-in-Aid for JSPS Fellows JP23KJ1008. H.-W.Y. acknowledges support from National Science and Technology Council (NSTC) in Taiwan through grant NSTC 113-2112-M-001-035- and from the Academia Sinica Career Development Award (AS-CDA-111-M03).  G.W.F. acknowledges support from the European Research Council (ERC) under the European Union Horizon 2020 research and innovation program (Grant agreement no. 815559 (MHDiscs)). G.W.F. was granted access to the HPC resources of IDRIS under the allocation A0120402231 made by GENCI. Support for B.Z. was provided by The Brinson Foundation.  Views and opinions expressed by ERC-funded scientists are however those of the author(s) only and do not necessarily reflect those of the European Union or the European Research Council. Neither the European Union nor the granting authority can be held responsible for them.

\bibliography{bibliography}{}
\bibliographystyle{aasjournal}

\appendix
\counterwithin{figure}{section}
\counterwithin{table}{section}

\section{Supplementary table and figures} \label{sect:Appendix_Galleries}

\begin{figure}[]
\centering
\includegraphics[width=1\hsize]{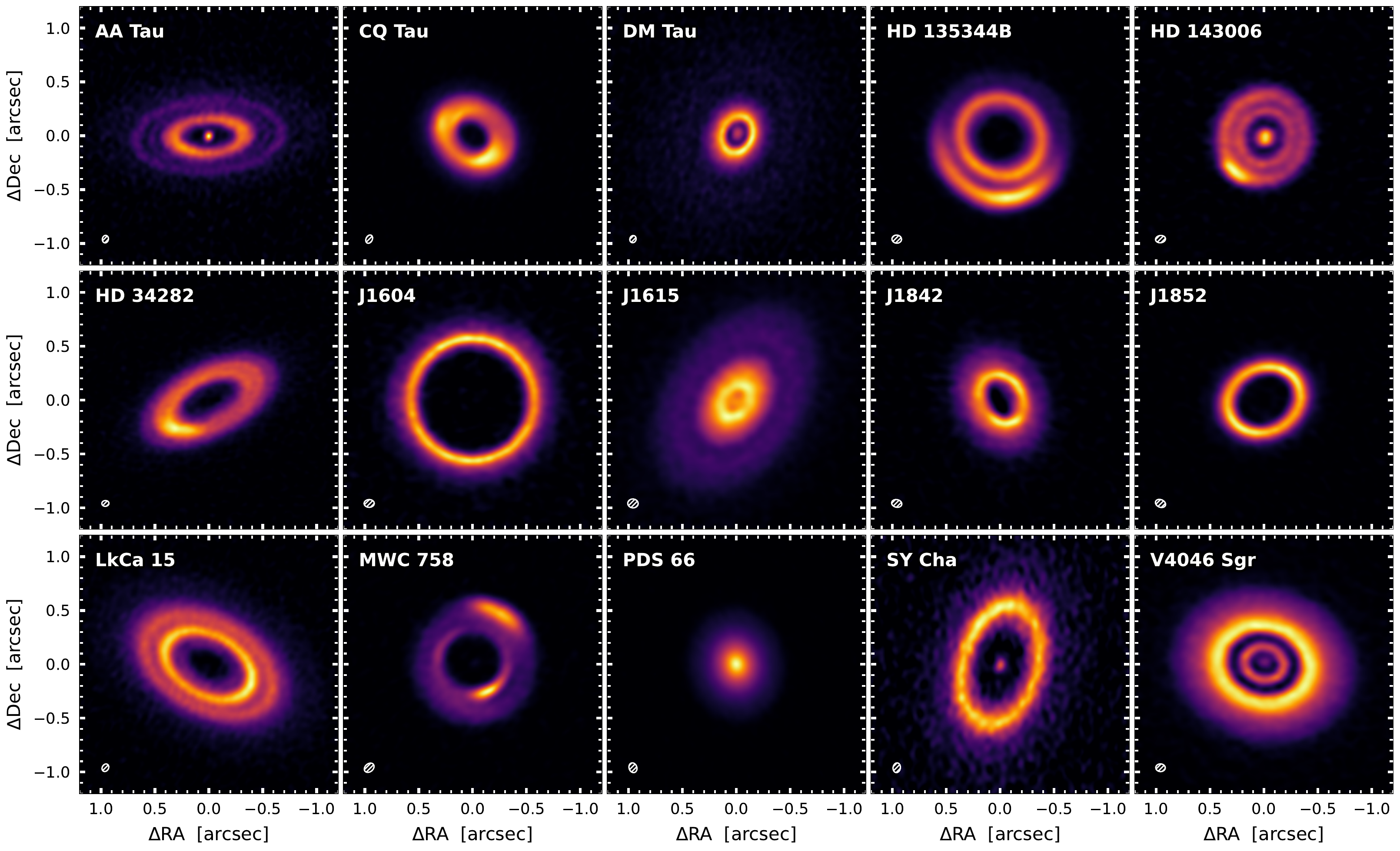}
\caption{Same as in Fig.~\ref{fig:gallery_images_asinh} but with a linear stretch in the color scale to highlight the changes in intensity within the brightest regions.}
\label{fig:gallery_images_linear}
\end{figure}

\begin{figure}[]
\centering
\includegraphics[width=0.97\hsize]{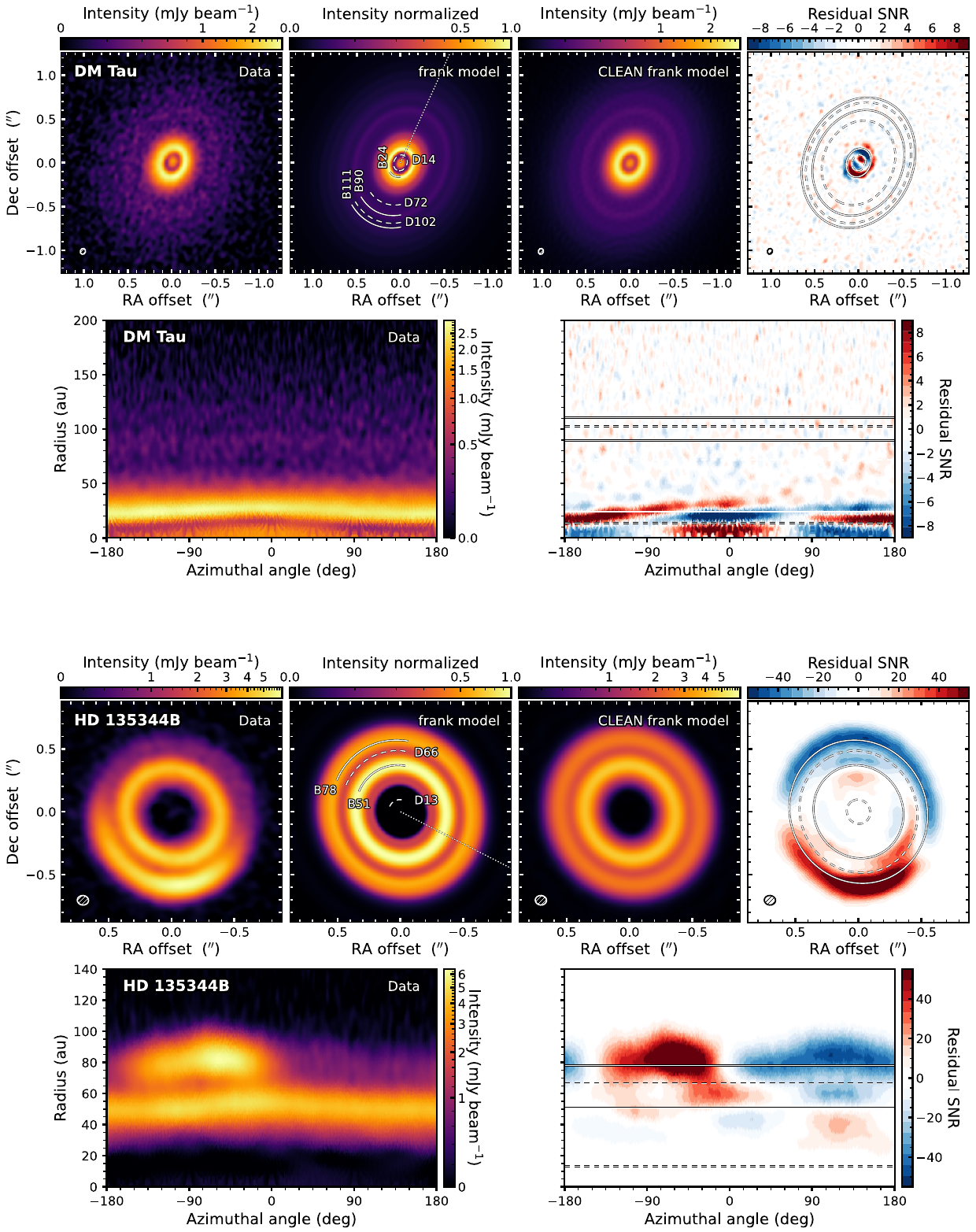}
\caption{Same as Fig.~\ref{fig:gallery_single_sources_maintext} but for DM~Tau and HD~135344B.}
\label{fig:gallery_single_sources_appendix1}
\end{figure}

\begin{figure}[]
\centering
\includegraphics[width=0.97\hsize]{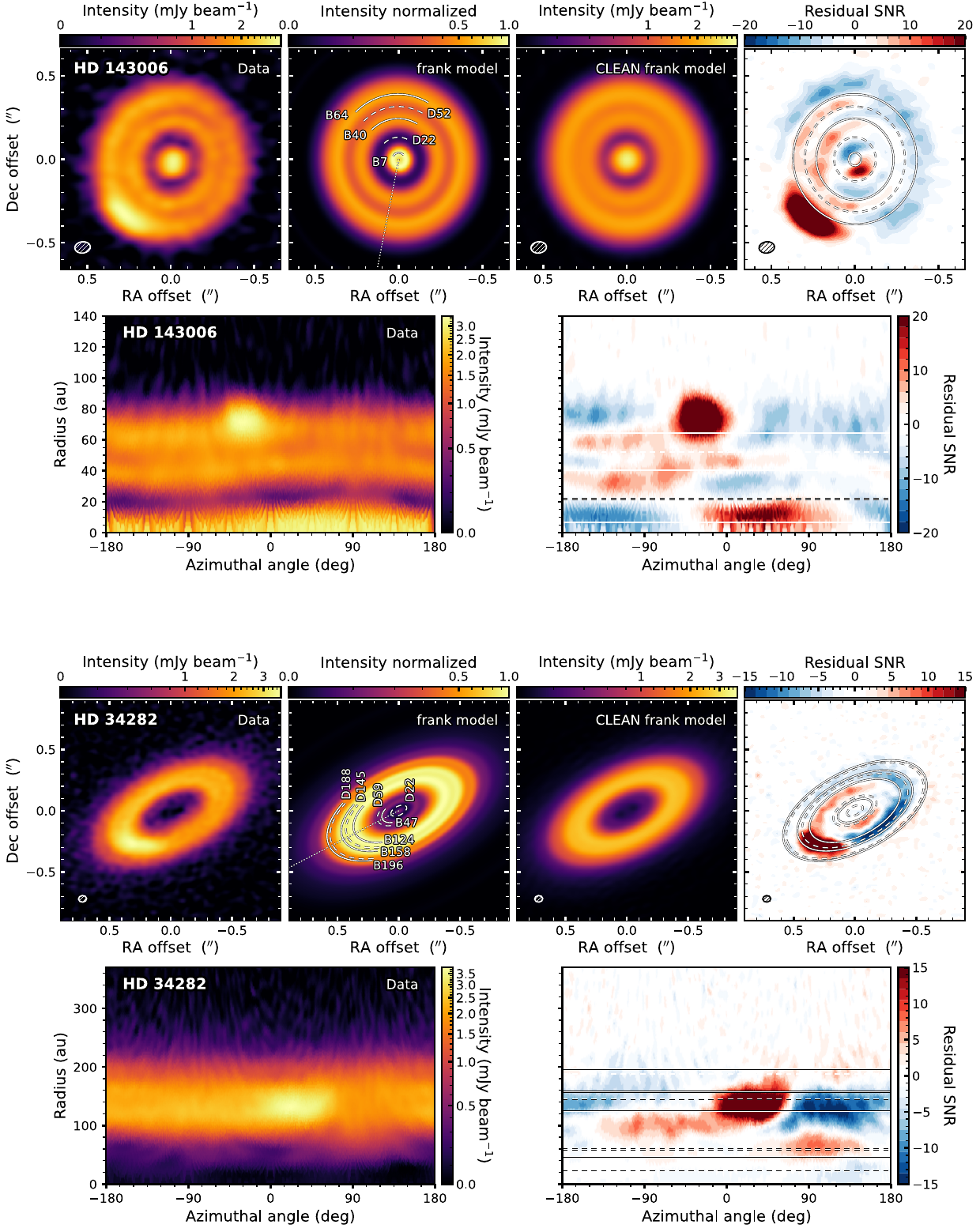}
\caption{Same as Fig.~\ref{fig:gallery_single_sources_maintext} but for HD~143006 and HD~34282.}
\label{fig:gallery_single_sources_appendix2}
\end{figure}

\begin{figure}[]
\centering
\includegraphics[width=0.97\hsize]{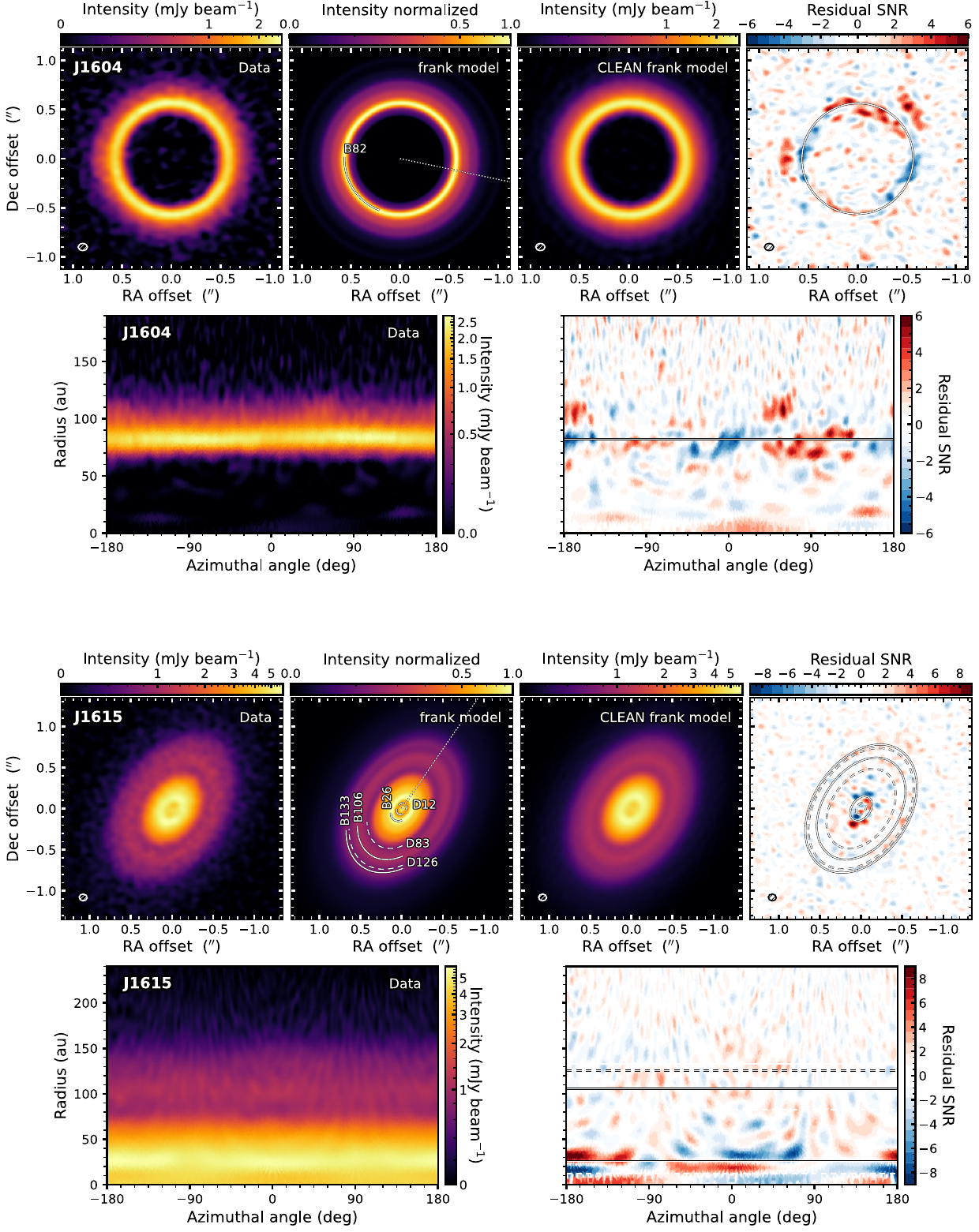}
\caption{Same as Fig.~\ref{fig:gallery_single_sources_maintext} but for J1604 and J1615.}
\label{fig:gallery_single_sources_appendix3}
\end{figure}

\begin{figure}[]
\centering
\includegraphics[width=0.97\hsize]{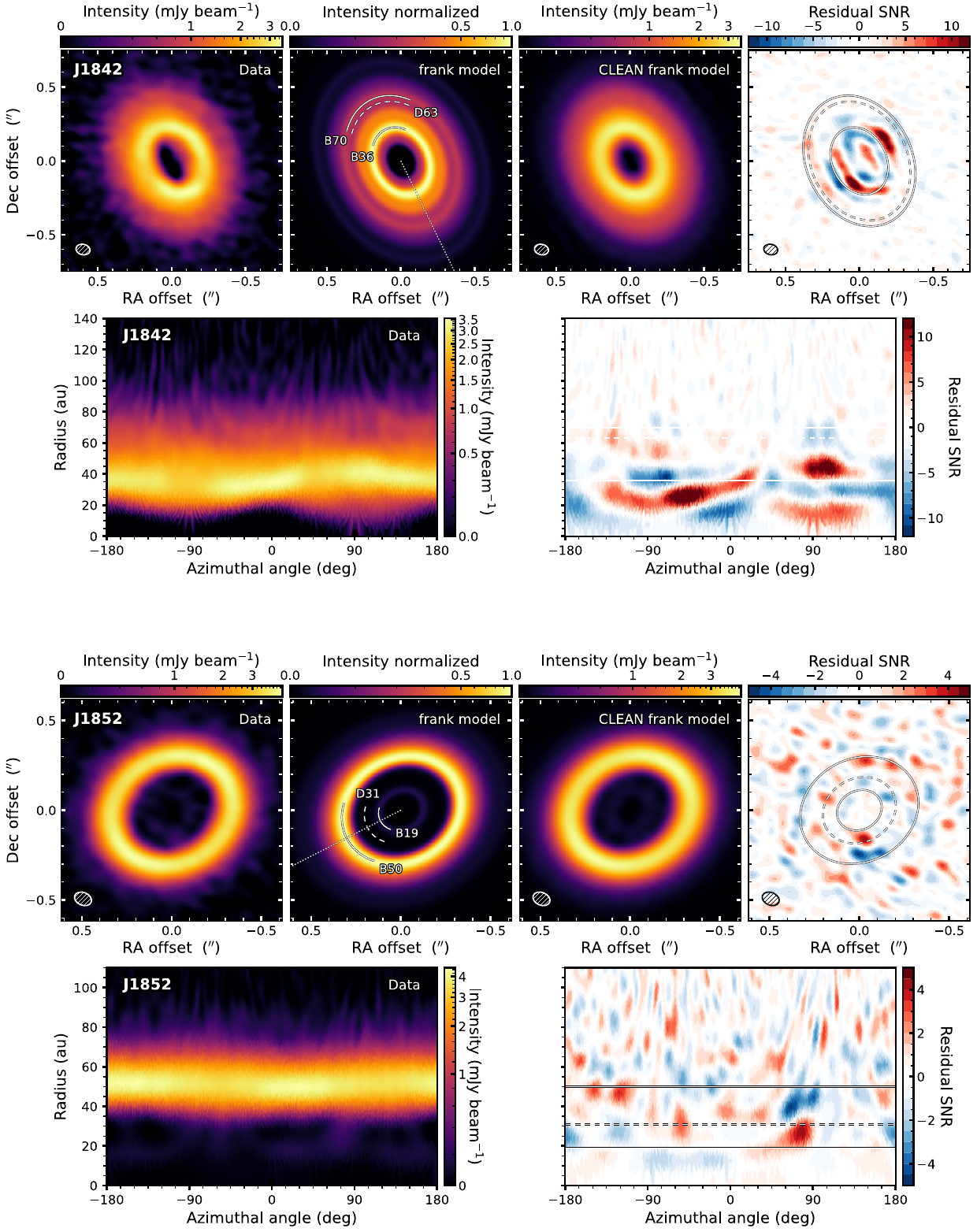}
\caption{Same as Fig.~\ref{fig:gallery_single_sources_maintext} but for J1842 and J1852.}
\label{fig:gallery_single_sources_appendix4}
\end{figure}

\begin{figure}[]
\centering
\includegraphics[width=0.97\hsize]{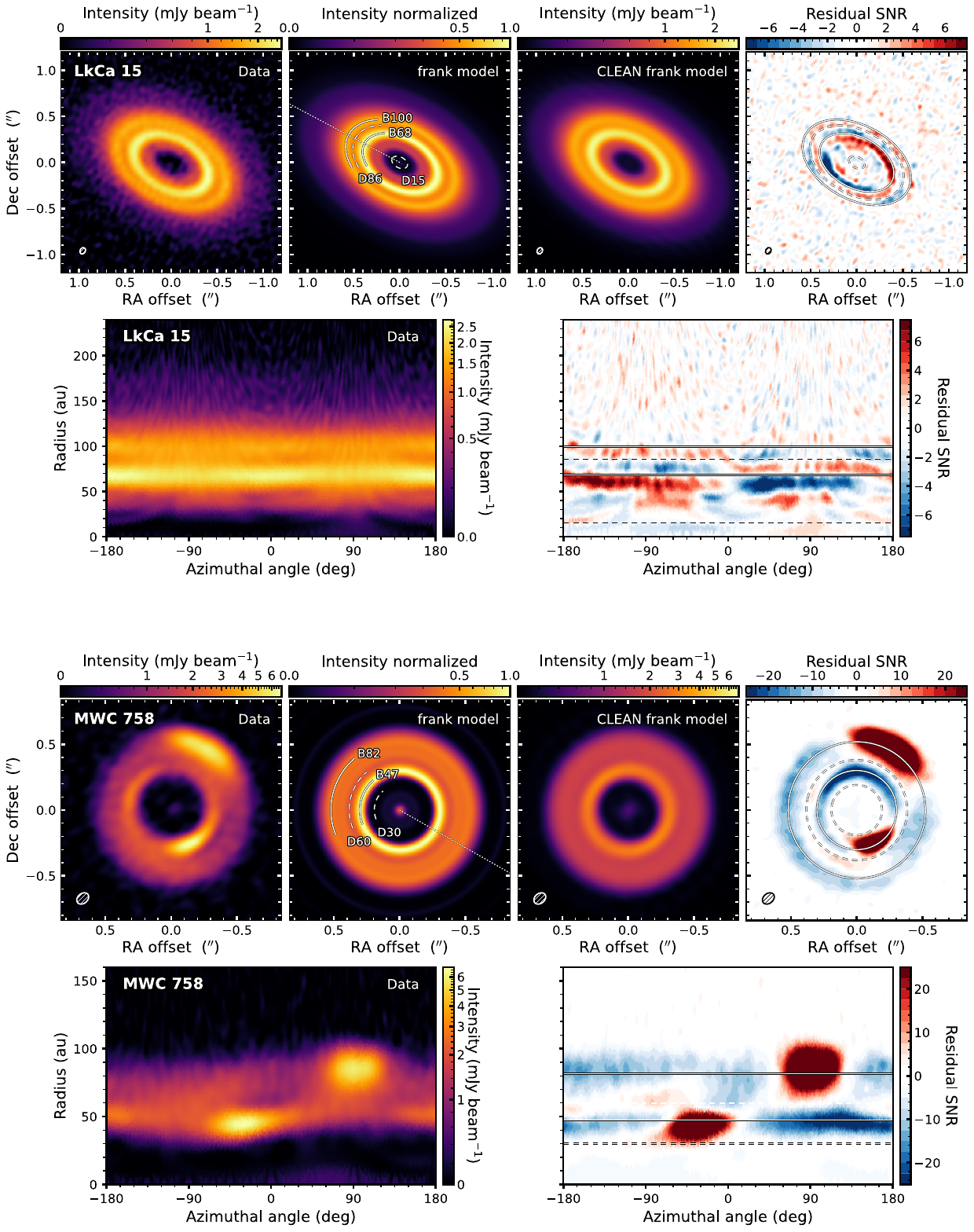}
\caption{Same as Fig.~\ref{fig:gallery_single_sources_maintext} but for LkCa~15 and MWC~758.}
\label{fig:gallery_single_sources_appendix5}
\end{figure}

\begin{figure}[]
\centering
\includegraphics[width=0.97\hsize]{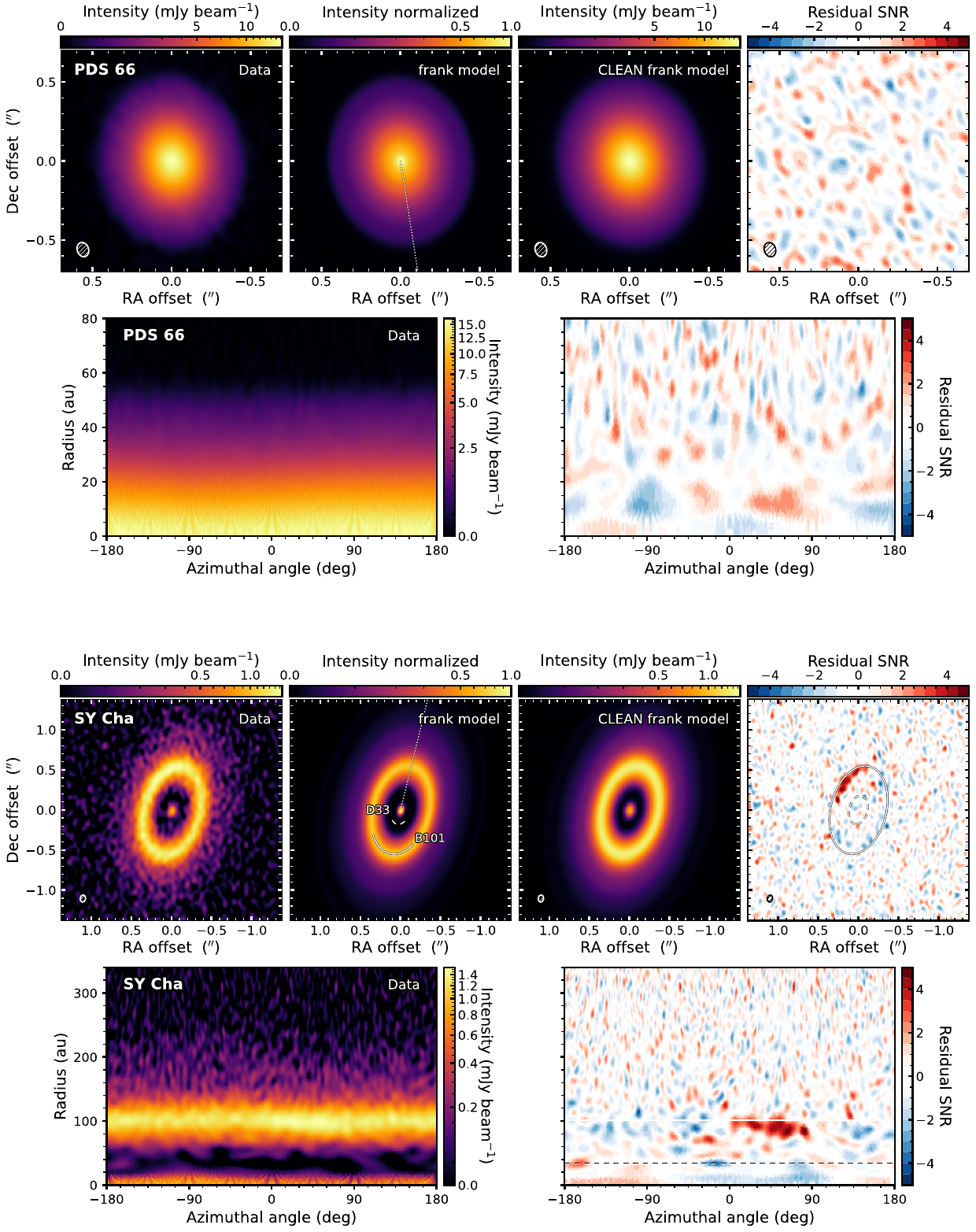}
\caption{Same as Fig.~\ref{fig:gallery_single_sources_maintext} but for PDS~66 and SY~Cha.}
\label{fig:gallery_single_sources_appendix6}
\end{figure}

\begin{figure}[]
\centering
\includegraphics[width=0.97\hsize]{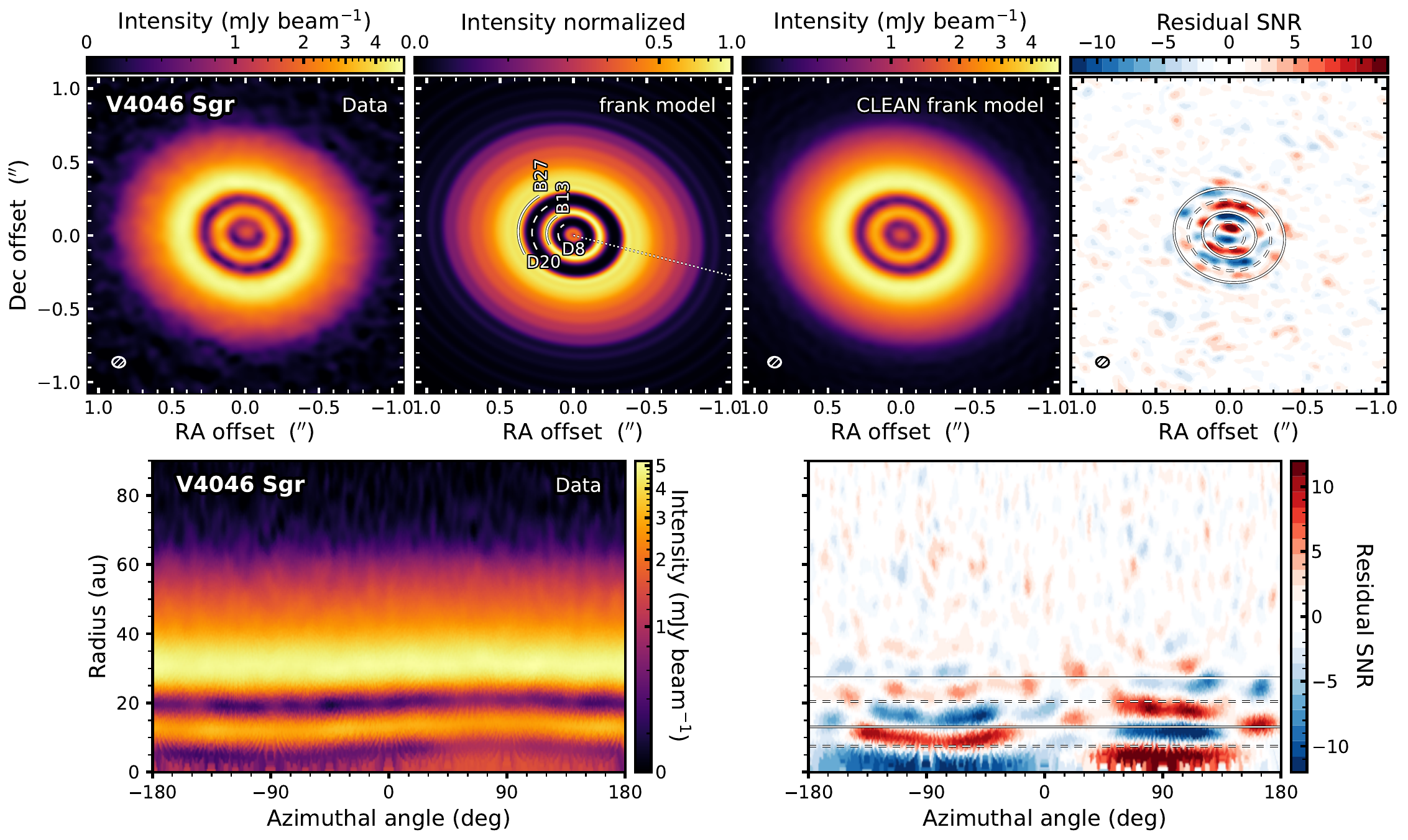}
\caption{Same as Fig.~\ref{fig:gallery_single_sources_maintext} but for V4046~Sgr.}
\label{fig:gallery_single_sources_appendix7}
\end{figure}

Table~\ref{tab:substructures} presents all the substructure properties for each disk.
Figure~\ref{fig:gallery_images_linear} shows a gallery of the continuum emission from the exoALMA sample with a linear stretch in the color scale. 
Figures~\ref{fig:gallery_single_sources_appendix1}, \ref{fig:gallery_single_sources_appendix2}, \ref{fig:gallery_single_sources_appendix3}, \ref{fig:gallery_single_sources_appendix4}, \ref{fig:gallery_single_sources_appendix5}, \ref{fig:gallery_single_sources_appendix6}, \ref{fig:gallery_single_sources_appendix7} complete the disk-specific results gallery introduced in Sect.~\ref{subsect:non-axisymm_subs}.

\startlongtable
\begin{deluxetable*}{llrrrrc|c}
\tabletypesize{\footnotesize}
\tablewidth{1\textwidth} 
\tablecaption{Properties of the Continuum Substructures}
\tablehead{
\colhead{Source} & \colhead{Feature}    &   \colhead{Radial Location}   &   \colhead{Width}   &  \colhead{$R_\mathrm{in}$}   &   \colhead{$R_\mathrm{out}$}   &   \colhead{Depth} & \colhead{NAI}\\
\colhead{}  &  \colhead{}    &  \colhead{(au, arcsec)}   &    \colhead{(au, arcsec)}  &  \colhead{(au, arcsec)}  &  \colhead{(au, arcsec)}  & \colhead{} & \colhead{}
} 
\colnumbers
\startdata 
AA Tau  &  D11  &  11.0, 0.082  &  28.1, 0.209  &  4.9, 0.037  &  33.0, 0.245  &  0.01  &  0.120  \\
 &  B42  &  42.0, 0.312  &  22.8, 0.169  &  33.0, 0.245  &  55.8, 0.414  &  \nodata  &  \\
 &  D64  &  64.3, 0.478  &  8.2, 0.061  &  60.3, 0.448  &  68.5, 0.508  &  0.44  &  \\
 &  B72  &  71.8, 0.533  &  6.8, 0.051  &  68.5, 0.508  &  75.3, 0.559  &  \nodata  &  \\
 &  D80  &  79.8, 0.593  &  10.2, 0.076  &  75.3, 0.559  &  85.5, 0.635  &  0.34  &  \\
 &  B90  &  89.8, 0.666  &  8.7, 0.065  &  85.5, 0.635  &  94.2, 0.699  &  \nodata  &  \\
 &  D105  &  105.3, 0.782  &  4.9, 0.036  &  103.1, 0.766  &  108.0, 0.802  &  0.94  &  \\
 &  B111  &  110.9, 0.823  &  6.0, 0.044  &  108.0, 0.802  &  114.0, 0.846  &  \nodata  &  \\
\hline
CQ Tau  &  B41  &  41.2, 0.276  &  33.4, 0.223  &  26.1, 0.175  &  59.5, 0.398  &  \nodata  &  0.111  \\
\hline
DM Tau  &  D14  &  13.5, 0.094  &  12.7, 0.088  &  7.7, 0.053  &  20.4, 0.142  &  0.08  &  0.092  \\
 &  B24  &  24.1, 0.167  &  10.7, 0.074  &  20.4, 0.142  &  31.1, 0.216  &  \nodata  &  \\
 &  D72  &  71.8, 0.498  &  18.7, 0.130  &  64.4, 0.447  &  83.1, 0.577  &  0.78  &  \\
 &  B90  &  89.5, 0.622  &  10.6, 0.074  &  83.1, 0.577  &  93.7, 0.651  &  \nodata  &  \\
 &  D102  &  102.4, 0.712  &  6.6, 0.046  &  99.2, 0.689  &  105.8, 0.735  &  0.92  &  \\
 &  B111  &  110.6, 0.768  &  9.4, 0.065  &  105.8, 0.735  &  115.2, 0.800  &  \nodata  &  \\
\hline
HD 135344B  &  D13  &  13.2, 0.098  &  40.8, 0.302  &  1.5, 0.011  &  42.3, 0.313  &  0.0  &  0.405  \\
 &  B51  &  50.8, 0.376  &  17.3, 0.128  &  42.3, 0.313  &  59.6, 0.441  &  \nodata  &  \\
 &  D66  &  66.4, 0.493  &  11.4, 0.084  &  60.9, 0.451  &  72.3, 0.535  &  0.47  &  \\
 &  B78  &  78.1, 0.578  &  13.6, 0.101  &  72.3, 0.535  &  85.9, 0.636  &  \nodata  &  \\
\hline
HD 143006  &  B7  &  6.6, 0.040  &  5.4, 0.033  &  3.6, 0.022  &  9.0, 0.055  &  \nodata  &  0.215  \\
 &  D22  &  21.8, 0.132  &  18.3, 0.111  &  13.8, 0.084  &  32.1, 0.195  &  0.1  &  \\
 &  B40  &  40.3, 0.244  &  12.6, 0.076  &  32.1, 0.195  &  44.7, 0.271  &  \nodata  &  \\
 &  D52  &  52.2, 0.316  &  13.8, 0.084  &  44.7, 0.271  &  58.5, 0.354  &  0.58  &  \\
 &  B64  &  64.4, 0.390  &  12.9, 0.078  &  58.5, 0.354  &  71.4, 0.433  &  \nodata  &  \\
\hline
HD 34282  &  D22  &  21.8, 0.071  &  26.8, 0.087  &  8.8, 0.029  &  35.6, 0.115  &  0.2  &  0.114  \\
 &  B47  &  46.8, 0.152  &  16.9, 0.055  &  35.6, 0.115  &  52.5, 0.170  &  \nodata  &  \\
 &  D59  &  59.3, 0.192  &  44.4, 0.144  &  52.5, 0.170  &  96.9, 0.314  &  0.1  &  \\
 &  B124  &  124.4, 0.403  &  42.7, 0.138  &  96.9, 0.314  &  139.7, 0.453  &  \nodata  &  \\
 &  D145  &  145.2, 0.470  &  12.0, 0.039  &  139.7, 0.453  &  151.7, 0.492  &  0.96  &  \\
 &  B158  &  157.7, 0.511  &  9.6, 0.031  &  151.7, 0.492  &  161.3, 0.523  &  \nodata  &  \\
 &  D188  &  188.2, 0.610  &  5.7, 0.018  &  186.2, 0.603  &  191.9, 0.622  &  0.96  &  \\
 &  B196  &  196.4, 0.637  &  7.0, 0.023  &  191.9, 0.622  &  198.9, 0.645  &  \nodata  &  \\
\hline
J1604  &  B82  &  82.1, 0.568  &  9.5, 0.066  &  78.0, 0.539  &  87.6, 0.606  &  \nodata  &  0.059  \\
\hline
J1615  &  D12  &  12.3, 0.079  &  14.0, 0.090  &  5.4, 0.035  &  19.3, 0.124  &  0.49  &  0.038  \\
 &  B26  &  25.9, 0.167  &  15.7, 0.101  &  19.3, 0.124  &  35.0, 0.225  &  \nodata  &  \\
 &  D83  &  82.6, 0.531  &  14.1, 0.090  &  76.7, 0.493  &  90.8, 0.583  &  0.77  &  \\
 &  B106  &  105.6, 0.679  &  23.3, 0.150  &  90.8, 0.583  &  114.1, 0.733  &  \nodata  &  \\
 &  D126  &  125.5, 0.807  &  6.3, 0.040  &  122.8, 0.789  &  129.0, 0.829  &  0.97  &  \\
 &  B133  &  132.9, 0.854  &  6.2, 0.040  &  129.0, 0.829  &  135.3, 0.869  &  \nodata  &  \\
\hline
J1842  &  B36  &  35.8, 0.237  &  14.0, 0.092  &  30.2, 0.200  &  44.2, 0.293  &  \nodata  &  0.074  \\
 &  D63  &  63.2, 0.419  &  5.9, 0.039  &  60.6, 0.402  &  66.5, 0.440  &  0.87  &  \\
 &  B70  &  69.7, 0.461  &  5.9, 0.039  &  66.5, 0.440  &  72.4, 0.480  &  \nodata  &  \\
\hline
J1852  &  B19  &  19.0, 0.129  &  7.2, 0.049  &  15.6, 0.106  &  22.8, 0.155  &  \nodata  &  0.024  \\
 &  D31  &  30.9, 0.210  &  22.3, 0.152  &  22.8, 0.155  &  45.2, 0.307  &  0.01  &  \\
 &  B50  &  50.0, 0.340  &  12.3, 0.084  &  45.2, 0.307  &  57.5, 0.391  &  \nodata  &  \\
\hline
LkCa 15  &  D15  &  14.6, 0.093  &  51.4, 0.327  &  6.8, 0.043  &  58.2, 0.370  &  0.02  &  0.053  \\
 &  B68  &  68.2, 0.434  &  22.9, 0.146  &  58.2, 0.370  &  81.1, 0.516  &  \nodata  &  \\
 &  D86  &  86.3, 0.549  &  12.0, 0.077  &  81.1, 0.516  &  93.2, 0.593  &  0.76  &  \\
 &  B100  &  99.5, 0.633  &  12.7, 0.081  &  93.2, 0.593  &  105.9, 0.673  &  \nodata  &  \\
\hline
MWC 758  &  D30  &  30.1, 0.193  &  38.5, 0.247  &  4.5, 0.029  &  43.0, 0.276  &  0.01  &  0.429  \\
 &  B47  &  47.3, 0.303  &  10.4, 0.067  &  43.0, 0.276  &  53.4, 0.342  &  \nodata  &  \\
 &  D60  &  59.9, 0.384  &  8.6, 0.055  &  56.4, 0.362  &  65.0, 0.417  &  0.77  &  \\
 &  B82  &  81.6, 0.523  &  21.1, 0.135  &  65.0, 0.417  &  86.1, 0.552  &  \nodata  &  \\
\hline
PDS 66      &    \nodata     &       \nodata         &    \nodata      &    \nodata &   \nodata &     \nodata &   0.014\\
\hline
SY Cha  &  D33  &  33.3, 0.184  &  74.0, 0.409  &  7.2, 0.040  &  81.1, 0.449  &  0.04  &  0.075  \\
 &  B101  &  101.2, 0.560  &  39.9, 0.221  &  81.1, 0.449  &  121.1, 0.670  &  \nodata  &  \\
\hline
V4046 Sgr  &  D8  &  7.5, 0.105  &  7.3, 0.101  &  4.2, 0.059  &  11.5, 0.161  &  0.01  &  0.030  \\
 &  B13  &  13.1, 0.184  &  3.4, 0.047  &  11.5, 0.161  &  14.9, 0.208  &  \nodata  &  \\
 &  D20  &  20.4, 0.285  &  10.1, 0.141  &  15.1, 0.211  &  25.2, 0.352  &  0.01  &  \\
 &  B27  &  27.2, 0.380  &  15.4, 0.216  &  25.2, 0.352  &  40.6, 0.568  &  \nodata  &  \\
\enddata
\tablecomments{Column~(1): target name. Column~(2): annular substructure label. “B" (for bright) indicates a ring, while “D" (for dark) indicates a gap. The number in the label is the feature distance from the central star measured in au. Column~(3): substructure radial location, extracted as explained in Sect.~\ref{sect:axisymm_subs}. Column~(4): annular substructure width. Columns~(5) and (6): inner and outer edge of the substructure. Column~(7): gap depth. Substructure width, edges, and gap depth are derived following the criteria of \cite{Huang2018_DSHARPII}. Column~(8): NAI, computed as described in Sect.~\ref{subsect:non-axisymm_subs}.}
\label{tab:substructures}
\end{deluxetable*}

\section{Visibility modeling} \label{sect:Appendix_visibilities_profiles}

\begin{figure}[h]
\centering
\includegraphics[width=0.97\hsize]{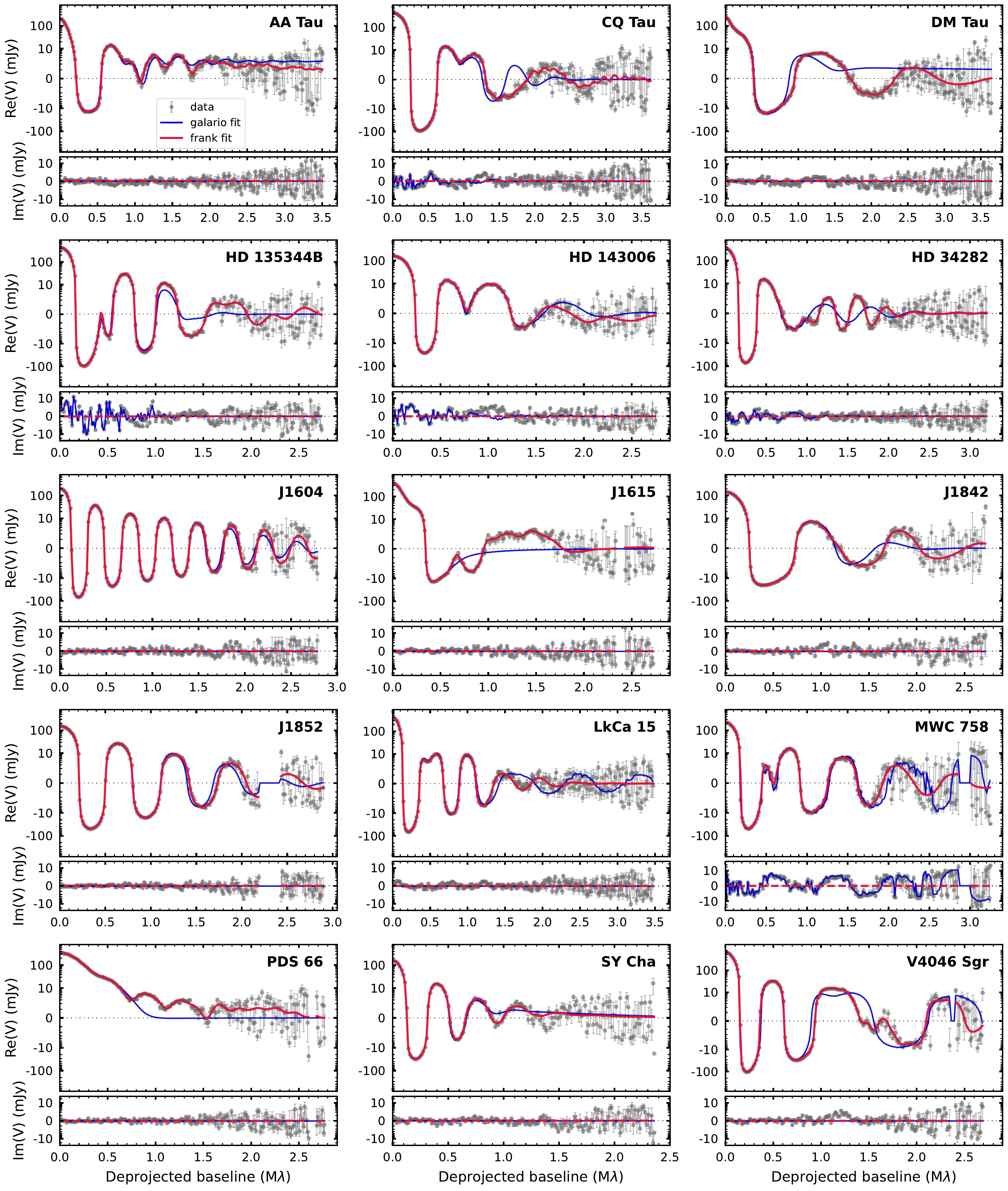}
\caption{Real and imaginary parts of the recentered and deprojected visibilities azimuthally averaged into 15~k$\lambda$ wide bins on an asinh scale as a function of the deprojected baseline length for the data (gray points) and the best-fit models from \galario (blue line) and \frank (red line). Note that the imaginary part is fitted only when employing a 2D nonaxisymmetric \galario model (i.e., for CQ~Tau, HD~135344B, HD~143006, HD~34282, and MWC~758), while the imaginary components of the \frank and 1D \galario axisymmetric models are null for all spatial frequencies by definition.}
\label{fig:uv_profiles}
\end{figure}

Tables~\ref{tab:galario_results_1D} and~\ref{tab:galario_results_2D} present the \galario best-fit results for each parameter in the 1D and 2D parametric models, respectively. Figure~\ref{fig:uv_profiles} displays a gallery of the visibility profiles as a function of deprojected baseline for each source, along with the best-fit profiles from \galario and \frank.

\begin{deluxetable}{lllllll} 
	\tabletypesize{\footnotesize}
	\tablewidth{1\textwidth} 
	\tablecaption{\galario Best-Fit Results for 1D Parametric Models}
	\tablehead{
		\colhead{Source} & 
		\colhead{Model} &  
		\colhead{Inner disk} & 
		\colhead{} &  
		\colhead{Ring} &   
		\colhead{}& 
		\colhead{} \\
		\colhead{}& 
		\colhead{} &  
		\colhead{$\log_{10}(f)$} & 
		\colhead{$\sigma$}   & 
		\colhead{$\log_{10}(f)$} & 
		\colhead{$r$} & 
		\colhead{$\sigma$} \\
		\colhead{} &
		\colhead{} & 
		\colhead{(Jy sr$^{-1}$)} & 
		\colhead{(mas)} &   
		\colhead{(Jy sr$^{-1}$)}  & 
		\colhead{(mas)} & 
		\colhead{(mas)}
		} 
	\colnumbers
	\startdata 
AA Tau     &   Central point source + four rings   &    $15.444^{+0.003}_{-0.003}$    &    \nodata    &    $10.11^{+0.01}_{-0.01}$   &    $344^{+2}_{-2}$    &     $55^{+1}_{-1}$  \\
 & & & &  $9.90^{+0.03}_{-0.03}$    &    $281^{+1}_{-1}$    &       $33^{+1}_{-1}$  \\  
 & & & &  $9.66^{+0.01}_{-0.01}$    &     $663.4^{+0.5}_{-0.5}$    &     $23.0^{+0.9}_{-0.9}$  \\  
 & & & &  $9.524^{+0.003}_{-0.003}$  &    $423^{+4}_{-4}$    & $387^{+2}_{-2}$   \\  
 \hline
 DM Tau     &   Central Gaussian + three rings   &    $11.5^{+0.1}_{-0.2}$    &     $6^{+1}_{-1}$   &    $10.357^{+0.001}_{-0.001}$   &    $192.9^{+0.1}_{-0.1}$    &     $57.2^{+0.2}_{-0.2}$   \\
 & & & &  $9.40^{+0.02}_{-0.02}$    &    $124^{+14}_{-13}$    &       $485^{+8}_{-8}$   \\  
 & & & &  $8.71^{+0.06}_{-0.07}$    &     $458^{+111}_{-101}$    &     $771^{+25}_{-25}$   \\  
 \hline
 J1604     &   Two rings   &    \nodata    &     \nodata   &    $10.210^{+0.001}_{-0.001}$   &    $572.7^{+0.1}_{-0.1}$    &     $30.5^{+0.2}_{-0.2}$    \\
 & & & &  $9.546^{+0.002}_{-0.002}$    &    $670.3^{+0.6}_{-0.6}$    &       $108.4^{+0.3}_{-0.3}$  \\  
 \hline
 J1615     &   Central Gaussian + three rings   &    $8.15^{+0.01}_{-0.06}$  &    $0.08^{+0.02}_{-0.01}$  &    $10.3982^{+0.0003}_{-0.0002}$  &    $0.1795^{+0.0002}_{-0.0002}$     &   $0.1551^{+0.0003}_{-0.0003}$   \\
 & & & &  $9.665^{+0.001}_{-0.001}$   &    $0.7122^{+0.0006}_{-0.0006}$    &     $0.1882^{+0.0009}_{-0.0008}$   \\  
 & & & &  $8.707^{+0.005}_{-0.003}$   &    $1.249^{+0.005}_{-0.006}$    &     $0.227^{+0.004}_{-0.003}$  \\  
 \hline
 J1842     &   Three rings   &    \nodata    &     \nodata   &     $10.247^{+0.002}_{-0.002}$   &    $242.2^{+0.2}_{-0.2}$    &     $47.2^{+0.3}_{-0.3}$   \\
& & & &  $9.759^{+0.005}_{-0.005}$    &    $373^{+2}_{-2}$    &       $109.4^{+1}_{-1}$    \\  
& & & & $8.77^{+0.03}_{-0.03} $   &     $551^{+14}_{-13}$    &    $177^{+4}_{-5}$  \\  
\hline
 J1852     &   Three rings   &    \nodata    &     \nodata   &     $10.456^{+0.001}_{-0.001}$   &    $348.9^{+0.2}_{-0.1}$    &     $35.1^{+0.2}_{-0.1}$  \\
& & & & $9.5^{+0.1}_{-0.2}$    &    $134^{+2}_{-3}$    &       $6.5^{+3}_{-2}$    \\  
& & & & $9.562^{+0.006}_{-0.008} $   &     $430^{+1}_{-1}$    &    $82.2^{+0.6}_{-0.5}$  \\  
\hline
LkCa 15    &  Four rings   &    \nodata    &    \nodata    &    $10.295^{+0.001}_{-0.001}$  &    $429.0^{+0.1}_{-0.1}$    &     $53.7^{+0.1}_{-0.1}$  \\
& & & &  $10.000^{+0.001}_{-0.001}$    &    $635^{+0.3}_{-0.3}$    &       $88.2^{+0.4}_{-0.4}$  \\  
& & & &   $10.69^{+0.02}_{-0.03}$    &     $259.5^{+0.1}_{-0.1}$    &     $2.5^{+0.1}_{-0.1}$    \\  
& & & & $9.513^{+0.003}_{-0.004}$  &    $606^{+2}_{-2}$    & $341^{+1}_{-1}$    \\  
\hline
PDS 66     &   Central Gaussian + one ring   &    $10.9405^{+0.0004}_{-0.0004}$   &     $123.8^{+0.4}_{-0.4}$  &    $10.106^{+0.004}_{-0.003}$   &    $295^{+2}_{-2}$    &     $141.5^{+0.7}_{-0.7}$   \\
\hline
SY Cha     &   Central Gaussian + two rings   &    $10.2^{+0.04}_{-0.04}$    &     $27^{+2}_{-2}$   &    $9.879^{+0.001}_{-0.001}$   &    $55.6^{+0.4}_{-0.4}$    &     $95.1^{+0.4}_{-0.4}$    \\
& & & &  $9.062^{+0.004}_{-0.004}$    &    $747^{+4}_{-4}$    &       $326^{+2}_{-2}$  \\  
\hline
V4046 Sgr     &   Central Gaussian + three rings   &    $10.033^{+0.006}_{-0.007}$    &     $58.8^{+0.7}_{-0.7}$    &    $11.296^{+0.005}_{-0.005}$   &    $183.34^{+0.03}_{-0.03}$    &     $2.97^{+0.04}_{-0.03}$   \\
& & & &  $10.3812^{+0.0004}_{-0.0004}$    &    $441.41^{+0.06}_{-0.06}$    &       $65.7^{+0.1}_{-0.1}$  \\  
& & & & $10.0773^{+0.0005}_{-0.0005}$    &     $589.5^{+0.3}_{-0.3}$    &     $181.7^{+0.1}_{-0.1}$   \\  
	\enddata
	\tablecomments{Column (1): target name. Column (2): parametric model assumed for the \galario fit. Columns (3) and (4): best-fit parameters for inner disk emission. In the case of AA~Tau, $\sigma$ is undefined because the inner disk was modeled with an unresolved point source. Columns (5)-(7): best-fit parameters for ring emission. The median of the marginalized posterior distribution is shown, along with the associated statistical uncertainties from the 16th and 84th percentiles of the MCMC marginalized distribution.}
    \label{tab:galario_results_1D}
\end{deluxetable}

\begin{deluxetable}{llllllllll} 
	\tabletypesize{\footnotesize}
	\tablewidth{1\textwidth} 
	\tablecaption{\galario Best-Fit Results for 2D Parametric Models}
	\tablehead{
		\colhead{Source} & 
		\colhead{Model} &  
		\colhead{Ring} &  
		\colhead{} & 
		\colhead{} & 
		\colhead{Arc} &
		\colhead{} &  
		\colhead{} & 
		\colhead{} &
		\colhead{} \\
		\colhead{} &
		\colhead{} &  
		\colhead{$\log_{10}(f)$} & 
		\colhead{$r$}  & 
		\colhead{$\sigma$}  & 
		\colhead{$\log_{10}(f)$} & 
		\colhead{$r$} & 
		\colhead{$\sigma$}  & 
		\colhead{$\phi$} & 
		\colhead{$\sigma_{\phi}$} \\
		\colhead{} & 
		\colhead{} & 
		\colhead{(Jy sr$^{-1}$)} & 
		\colhead{(mas)} & 
		\colhead{(mas)}  &   
		\colhead{(Jy sr$^{-1}$)} & 
		\colhead{(mas)} & 
		\colhead{(mas)}  & 
		\colhead{(deg)} & 
		\colhead{(deg)}
		} 
	\colnumbers
	\startdata 
CQ Tau     &    Two rings + two arcs   &    $10.334^{+0.002}_{-0.002}$    &    $366.0^{+0.2}_{-0.2}$ & $41.8^{+0.2}_{-0.2}$   &    $10.722^{+0.001}_{-0.001}$    &    $252.2^{+0.2}_{-0.2}$    &    $62.8^{+0.2}_{-0.2}$    &    $0.002^{+0.003}_{-0.001}$    &    $90.2^{+0.3}_{-0.2}$ \\
   &       &    $9.856^{+0.005}_{-0.005}$    &    $387^{+2}_{-2}$ & $133.2^{+0.6}_{-0.6} $ & $10.7518^{+0.0007}_{-0.0007}$    &    $247.1^{+0.1}_{-0.1}$    &    $58.6^{+0.1}_{-0.2}$    &    $165.86^{+0.04}_{-0.04}$    &    $39.21^{+0.07}_{-0.07}$ \\
\hline
HD 135344B     &    One rings + one arc   &    $10.4123^{+0.0003}_{-0.0003}$    &    $371.44^{+0.04}_{-0.04}$ & $64.98^{+0.06}_{-0.06}$   &    $10.5283^{+0.0003}_{-0.0003}$    &    $588.90^{+0.06}_{-0.06}$    &    $69.48^{+0.05}_{-0.05}$    &    $146.42^{+0.09}_{-0.07}$    &    $61.51^{+0.03}_{-0.03}$ \\
\hline
HD 143006     &   Central Gaussian    &    $10.451^{+0.004}_{-0.003}$    &    Fixed at 0 & $51.3^{+0.3}_{-0.3}$   &    $10.219^{+0.003}_{-0.003}$    &    $445.3^{+0.5}_{-0.4}$    &    $43.8^{+0.3}_{-0.3}$    &    $129.6^{+0.4}_{-0.3}$    &    $20.65^{+0.08}_{-0.07}$ \\
&  +  two rings + one arc     &    $10.057^{+0.002}_{-0.002}$    &    $237.8^{+0.3}_{-0.3}$ & $35.7^{+0.4}_{-0.3} $ &    &        &       &        &     \\
&       &    $10.0374^{+0.0008}_{-0.0007}$    &    $382.7^{+0.4}_{-0.4}$ & $64.6^{+0.3}_{-0.3} $ &    &        &       &        &     \\
\hline
HD 34282     &    Three rings + one arc   &    $9.907^{+0.005}_{-0.004}$    &    $362.1^{+0.4}_{-0.3}$ & $32.2^{+0.7}_{-0.7}$   &    $10.288^{+0.002}_{-0.002}$    &    $451.2^{+0.4}_{-0.4}$    &    $50.8^{+0.3}_{-0.2}$    &    $20.3^{+0.1}_{-0.1}$    &    $28.2^{+0.1}_{-0.1}$ \\
&       &    $9.375^{+0.007}_{-0.006}$    &    $407^{+4}_{-4}$ & $386^{+2}_{-2} $ &   &        &       &       &     \\
&       &    $10.305^{+0.001}_{-0.001}$    &    $470.2^{+0.5}_{-0.5}$ & $129.0^{+0.3}_{-0.4} $ &   &        &       &       &     \\
\hline
MWC 758     &    Two rings + two arcs   &    $10.902^{+0.006}_{-0.005}$    &    $328.7^{+0.1}_{-0.1}$ & $2.30^{+0.04}_{-0.05}$   &    $11.79^{+0.01}_{-0.01}$    &    $277.37^{+0.04}_{-0.05}$    &    $1.24^{+0.05}_{-0.04}$    &    $122.7^{+0.2}_{-0.2}$    &    $39.2^{+0.2}_{-0.2}$ \\
&       &    $9.8855^{+0.0007}_{-0.0007}$     &    $432.5^{+0.4}_{-0.4}$ & $99.4^{+0.2}_{-0.2} $ & $10.551^{+0.001}_{-0.001}$    &    $550.7^{+0.3}_{-0.3}$    &    $45.4^{+0.2}_{-0.2}$    &    $256.1^{+0.1}_{-0.1}$    &    $20.07^{+0.02}_{-0.02}$ \\
	\enddata
	\tablecomments{Column (1): target name. Column (2): parametric model assumed for the \galario fit. Columns (3) - (5): best-fit parameters for ring emission. Columns (6)-(10): best-fit parameters for arc emission. The median of the marginalized posterior distribution is shown, along with the associated statistical uncertainties from the 16th and 84th percentiles of the MCMC marginalized distribution.}
 \label{tab:galario_results_2D}
\end{deluxetable}

\section{Comparing geometrical parameters obtained from continuum and gas} \label{sect:comparing_gas_dust_geom_param}

\begin{figure}[h]
\centering
\includegraphics[width=0.9\hsize]{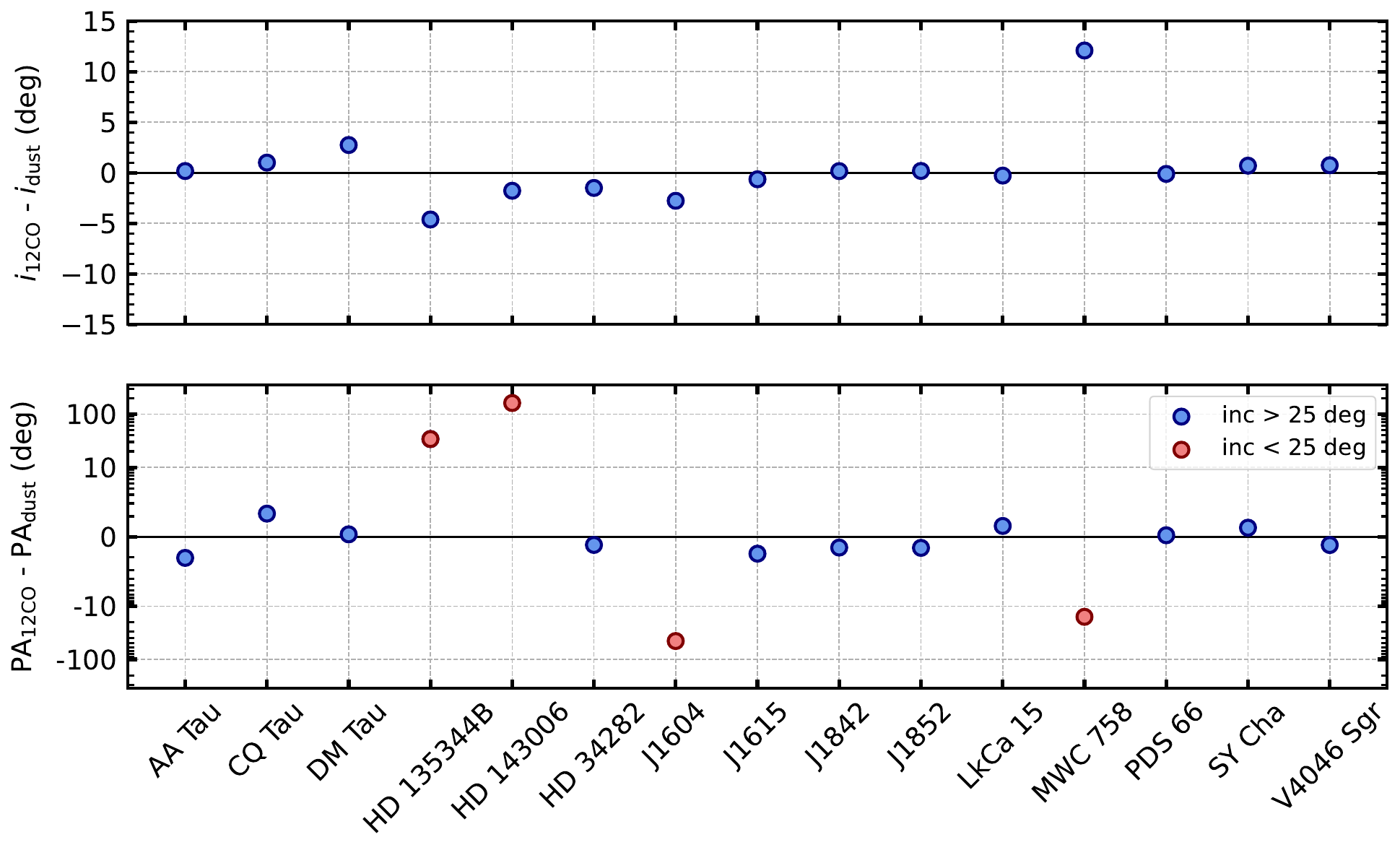}
\caption{Difference between the $i$ and PA values obtained by analyzing the continuum with \galario and the $^{12}$CO data with \discminer \citep{Izquierdo_exoALMA}. An asinh stretch has been applied to the y-axis of the bottom panel to include the disks with a large difference in PA due to a low inclination (red circles).}
\label{fig:inc_PA_comparison}
\end{figure}

\begin{figure}[h]
\centering
\includegraphics[width=0.9\hsize]{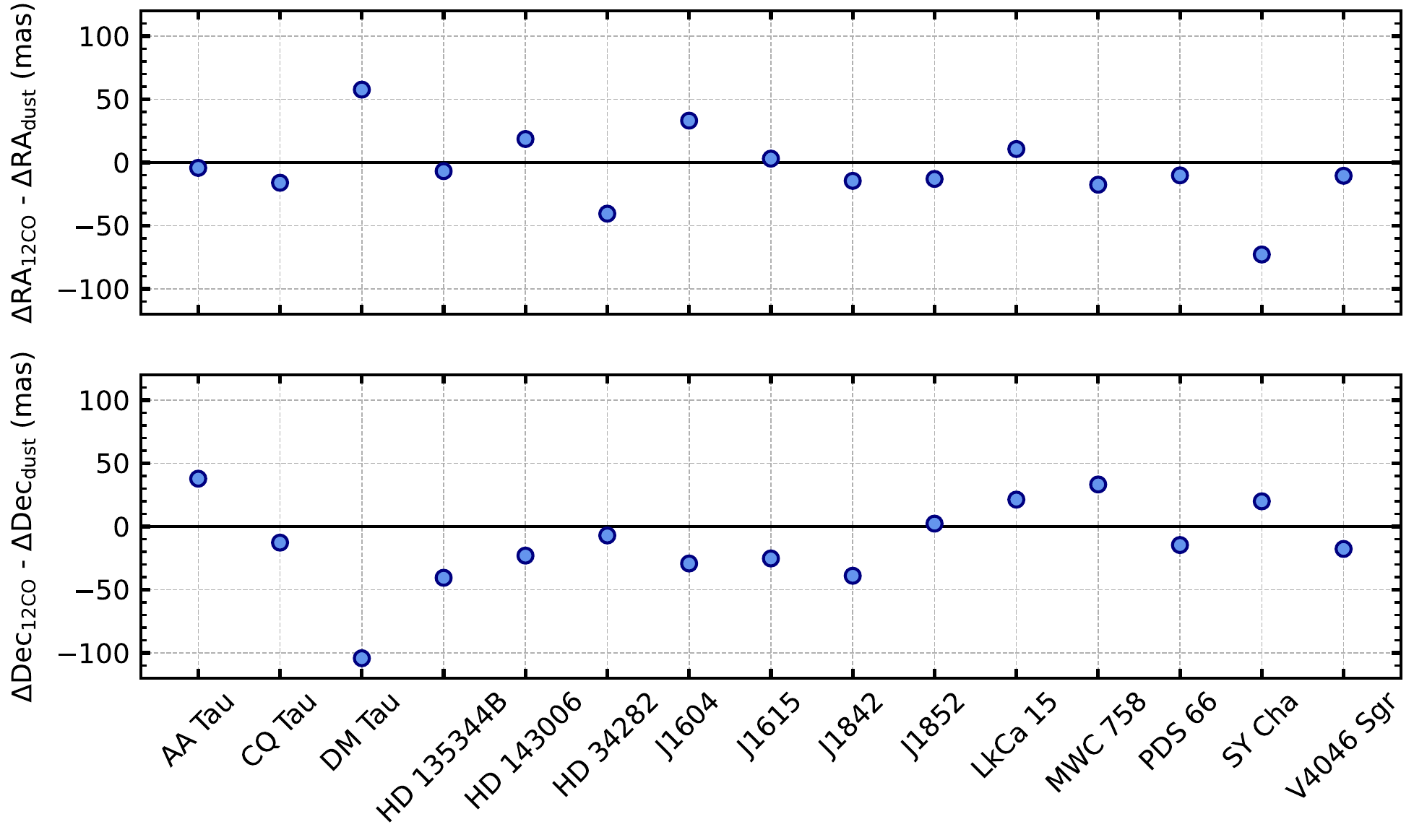}
\caption{Difference between the offset in R.A. and decl. obtained by analyzing the continuum with \galario and the $^{12}$CO data with \discminer \citep{Izquierdo_exoALMA}.}
\label{fig:dRA_dDec_comparison}
\end{figure}

In Fig.~\ref{fig:inc_PA_comparison} and~\ref{fig:dRA_dDec_comparison} are shown the comparisons between the geometrical parameters ($i$, PA, $\Delta$R.A., and $\Delta$decl.) derived by analyzing the continuum data with \galario and the $^{12}$CO channel maps with \discminer \citep{Izquierdo_exoALMA}. We note that most inclination values are within 5 deg, with the only exception being MWC~758 with about 12 deg. PAs are all within 10 deg for disks with relevant inclination ($>$25 deg), whereas the two methods do not agree for low-inclination disks ($<$25 deg). Most of the differences in R.A. and decl. offsets are within 50~mas (half of the synthesized beam), with only three cases between 50 and 110 mas. 

\section{Accretion rate and NIR excess values} \label{sect:Macc_NIR}

In Table~\ref{tab:Macc_NIR} we report the mass accretion rate $\dot{M}$ and NIR excess for the disks in the exoALMA sample. NIR excess values are obtained as explained in Sect.~\ref{subsect:correlations_dust_star}.

\begin{deluxetable}{llll} 
	\tablewidth{1\textwidth} 
	\tablecaption{Accretion Rate and NIR Excess Values of the exoALMA Sample}
	\tablehead{
		\colhead{Source} & 
		\colhead{$\log_{10}\dot{M}$} &  
		\colhead{References $\dot{M}$} &  
		\colhead{NIR Excess} \\
        \colhead{} &
		\colhead{($M_\odot\,\mathrm{yr}^{-1}$)} &
		\colhead{} &   
		\colhead{(\%)} 
		} 
	\colnumbers
	\startdata 
AA Tau     &    -8.1   &    \cite{Bouvier2013}   &   $4.7\pm3.6$  \\
CQ Tau     &    -7.0   &    \cite{Donehew2011}   &  $25.4\pm2.5$   \\
DM Tau     &    -8.2   &    \cite{Manara2014}   &  $<0.6$   \\
HD 135344B     &    -8.0   &    \cite{Sitko2012}   &  $27.2\pm3.1^{(a)}$  \\
HD 143006     &    -8.1   &    \cite{Rigliaco2015}   &   $21.3\pm1.4^{(a)}$  \\
HD 34282    &    -7.7   &    \cite{Fairlamb2015}   &   $9.2\pm1.0^{(a)}$    \\
J1604    &    -10.5   &    \cite{Bouvier2013}   &  $17.5\pm3.6^{(a)}$   \\
J1615     &    -8.5   &    \cite{Manara2014}   &  $<0.9^{(a)}$     \\
J1842   &    -8.8   &    \cite{Manara2014}   &   $12.3\pm1.1$    \\
J1852   &    -8.7   &    \cite{Manara2014}   &   $<1.1$    \\
LkCa 15    &    -8.4   &    \cite{Manara2014}   &  $13.4\pm1.0$  \\
MWC 758    &    -8.0   &    \cite{Boehler2018}   &  $27.5\pm2.9$    \\
PDS 66     &    -9.9   &    \cite{Ingleby2013}   &  $7.3\pm1.4^{(a)}$ \\
SY Cha     &    -9.2   &    \cite{Manara2023}   &  $7.6\pm1.1^{(a)}$   \\
V4046 Sgr     &    -9.3   &    \cite{Donati2011}   &  $<0.9^{(a)}$  \\
	\enddata
	\tablecomments{Column (1): target name. Column (2): mass accretion rate. The uncertainty associated with each value is 0.35~dex, following what is reported in Sect.~2.1.3 of \cite{Manara2023}. Column (3): reference paper for the mass accretion rate values. Column (4): NIR excess. Values with~$^{(a)}$ are from \cite{Garufi2018}, while for disks not included in that work, the NIR excess was calculated following the same procedure (see Appendix~\ref{sect:Macc_NIR}).}
 \label{tab:Macc_NIR}
\end{deluxetable}

\section{Other correlations with the NAI} 
\label{sect:correlations_NAI}

\begin{figure}[t]
\centering
\includegraphics[width=0.5\columnwidth]{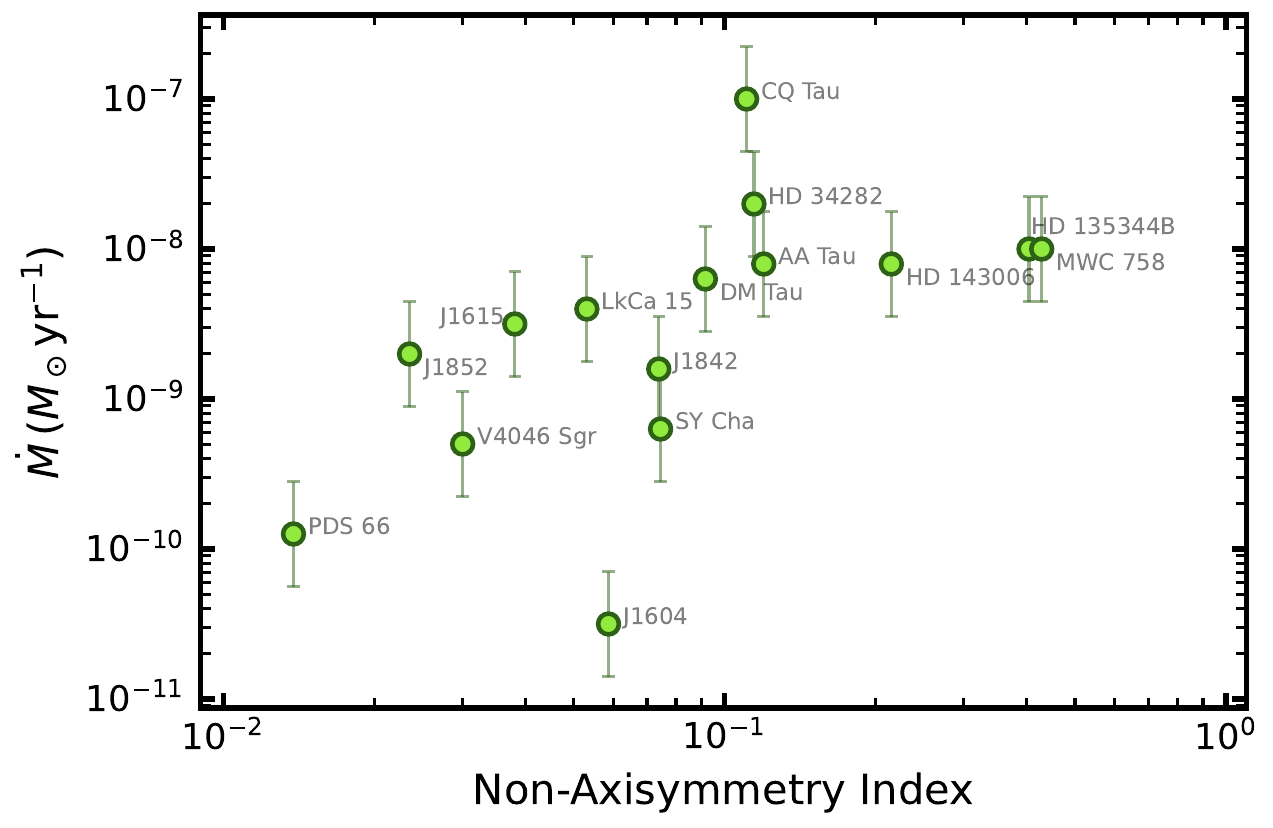}
\caption{Same as the left panel in Fig.~\ref{fig:Macc_NIR_vs_NAI}, but without normalizing the mass accretion rate for its dependence on stellar mass.}
\label{fig:Macc_NOnorm_vs_NAI}
\end{figure}

\begin{figure}[t]
\centering
\includegraphics[width=1\columnwidth]{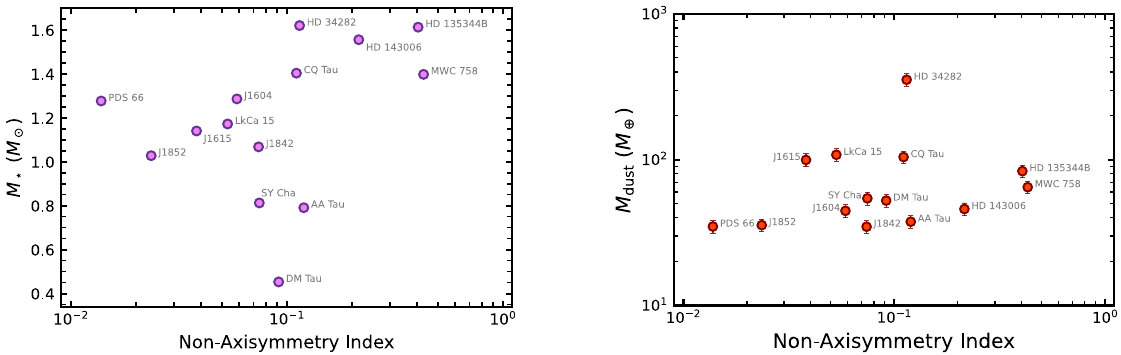}
\caption{Stellar mass from \discminer \citep{Izquierdo_exoALMA} and dust disk mass, calculated in Sect.~\ref{sect:data}, as functions of the NAI. The uncertainties in stellar mass are too small to be shown, while for the dust disk mass, we show the 10\% absolute flux calibration error, which dominates over the statistical uncertainty.}
\label{fig:StellaMass_and_dust_mass_vs_NAI}
\end{figure}

Figure~\ref{fig:Macc_NOnorm_vs_NAI} shows the correlation between the unscaled accretion rate and the NAI. Figure~\ref{fig:StellaMass_and_dust_mass_vs_NAI} presents the stellar mass and dust disk mass as functions of the NAI.

\section{External sources in the FOV} \label{sect:Appendix_fullFOV}

\begin{figure}[h]
\centering
\includegraphics[width=0.76\hsize]{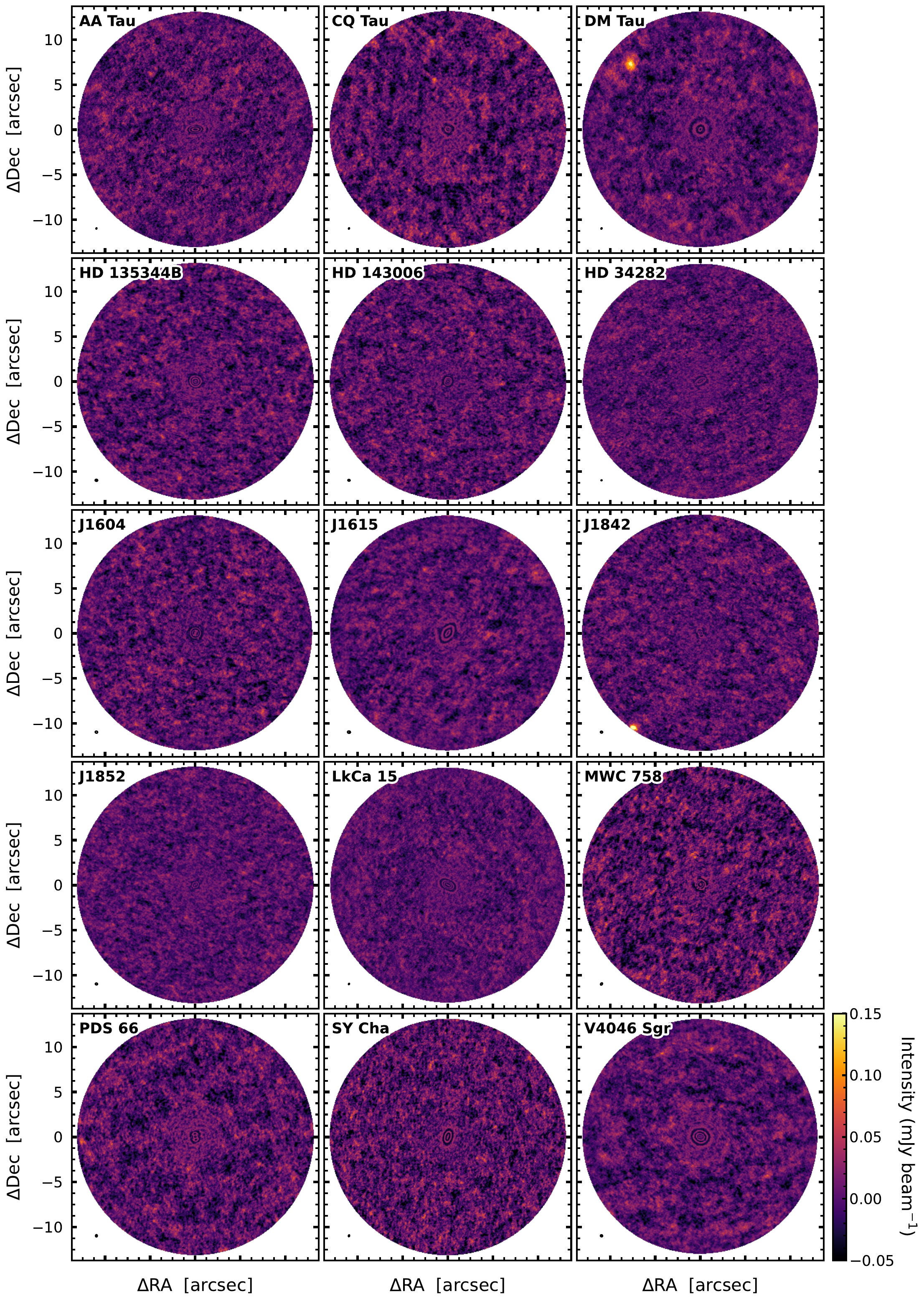}
\caption{Gallery of CLEAN residuals of the full FOV after deconvolution using a central spherical mask $3''$ wide and a $2\sigma$ stopping threshold. Each panel shares the same color bar ranging from $-0.05$ to $0.15$~mJy/beam. The robust value of $2.0$ has been employed, and the associated beam is indicated by the black ellipse in the lower left corner of each panel.}
\label{fig:fullFOV}
\end{figure}

To evaluate the presence of external sources in the FOV, we generated a gallery of CLEAN residuals (Fig.~\ref{fig:fullFOV}). The CLEAN algorithm was applied using a central mask $3''$ wide, a robust parameter of $2.0$, and a stopping threshold of $2\sigma$. Notable external sources are apparent within the FOVs of a few targets. For DM~Tau, an external source with an integrated flux of approximately 4.0~mJy is located $10.5''$ northeast of the central disk. In the FOV of J1842, an external source with an integrated flux of approximately 2.6~mJy is positioned $12.8''$ southeast of the central disk, right at the edge of the FOV. Additionally, there is a tentative detection of an external source with an integrated flux of approximately 0.5~mJy within the CQ~Tau FOV, $5.7''$ north of the central disk. The reported flux density have been computed from the primary-beam-corrected images. We did not find any correspondence of these external sources in the SIMBAD catalog \citep{Wenger2000}, the VLA Sky Survey \citep{Lacy2020}, or the ALMA continuum source catalogs from the A$^3$COSMOS and A$^3$GOODSS projects \citep{Adscheid2024}.

\end{document}